\documentclass[numberedappendix]{emulateapj}
\usepackage{graphicx}

\newcommand{ \St}{ {\rm St} }
\newcommand{ \Rep}{ {\rm Re_p} }
\newcommand{ \Ret}{ {\rm Re_t} }
\newcommand{ \rhoint}{ \tilde{\varrho} }

\begin{document}

\title{Aggregate Growth and Internal structures of Chondrite Parent Bodies Forming from Dense Clumps}
\shorttitle{Chondrule Mass Fraction in Chondrite Parent Bodies via Aggregate Growth}
\author{Yuji Matsumoto,\altaffilmark{1}
		Shigeru Wakita,\altaffilmark{2,3,4}
		Yasuhiro Hasegawa,\altaffilmark{5}
		\& Shoichi Oshino\altaffilmark{6,7}
		}
\email{ymatsumoto@asiaa.sinica.edu.tw}

\altaffiltext{1}{Institute of Astronomy and Astrophysics, Academia Sinica, Taipei 10617, Taiwan}
\altaffiltext{2}{Earth-Life Science Institute, Tokyo Institute of Technology, Ookayama, Meguro-ku, Tokyo, 152-8550, Japan}
\altaffiltext{3}{Department of Earth, Environmental and Planetary Sciences, Brown
University, Providence, RI 02912, USA}
\altaffiltext{4}{Department of Earth, Atmospheric, and Planetary Sciences, Purdue
University, West Lafayette, IN 47907, USA}
\altaffiltext{5}{Jet Propulsion Laboratory, California Institute of Technology, Pasadena, CA 91109, USA}
\altaffiltext{6}{Center for Computational Astrophysics, National Astronomical Observatory of Japan, Osawa, Mitaka, Tokyo, 181-8588, Japan}
\altaffiltext{7}{Institute for Cosmic Ray Research, University of Tokyo, Hida, Gifu 506-1205, Japan}

\begin{abstract}

Major components of chondrites are chondrules and matrix. 
Measurements of the volatile abundance in Semarkona chondrules suggest that chondrules formed in a dense clump that had a higher solid density than the gas density in the solar nebula.
We investigate collisions between chondrules and matrix in the surface region of dense clumps using fluffy aggregate growth models.
Our simulations show that the collisional growth of aggregates composed of chondrules and matrix takes place in the clumps well before they experience gravitational collapse.
The internal structure of chondrite parent bodies (CPBs) can be thereby determined by aggregate growth.
We find that the aggregate growth generates two scales within CPBs.
The first scale is involved with the small scale distribution of chondrules and determined by the early growth stage, where chondrules accrete aggregates composed of matrix grains.
This accretion can reproduce the thickness of the matrix layer around chondrules found in chondrites.
The other scale is related to the large scale distribution of chondrules.
Its properties (e.g., the abundance of chondrules and the overall size) depend on the gas motion within the clump, which is parameterized in this work.
Our work thus suggests that the internal structure of CPBs may provide important clues about their formation conditions and mechanisms.

\end{abstract}


\section{Introduction}

The origin of the solar system is one of the fundamental questions and can be explored by investigating primitive materials currently remaining in the solar system.
This is because such materials and their components formed during the birth of the solar system, providing clues about its early evolution.

Chondrites are one of the primitive bodies.
They make up over 80\% of meteorite falls \citep{Scott&Krot2005,Weisberg+2006}.
Except for metal-rich chondrites, their major components are chondrules and matrix materials.
Chondrules are mm-sized spherical to semi-spherical objects, which originated from molten silicate droplets in the solar nebula \citep[e.g.,][]{Scott&Krot2005, Scott2007, Krot+2009, Bizzarro+2017}.
Isotope measurements of chondrules suggest that chondrules formed in the first $\sim 5$~Myr after the formation time of Ca-Al-rich inclusions (CAIs) \citep{Connelly+2012, Bollard+2017}.
Chondrules and other components (such as refractory inclusions and metallic grains) are coated by matrix grains.
The sizes of matrix grains are 1~nm to 10~$\mu$m \citep{Toriumi1989,Scott&Krot2005}.
The fraction of chondrules in chondrites is different among each group of chondrites \citep{Scott&Krot2005}.
The chondrule abundances in ordinary chondrites and enstatite chondrites are 60 -- 80~vol.\%. 
Carbonaceous chondrites tend to have fewer chondrules from other types of chondrites. 

The measurements of Na in the chondrules of Semarkona ordinary chondrite suggest that the dust density of the chondrule-forming region was between $10^{-6}\mbox{~g~cm}^{-3}$ and $10^{-2}\mbox{~g~cm}^{-3}$ \citep{Alexander+2008}.
This mass density is several orders of magnitude higher than the gas mass density of the minimum mass solar nebula model around a few au \citep{Hayashi1981}.
Such a high-density clump might have formed through streaming instability \citep[e.g.,][]{Youdin&Goodman2005, Johansen+2012} or turbulent concentration \citep{Cuzzi+2001}, which would be the birthplace of planetesimals and chondrite parent bodies (CPBs).
Since the dust mass density in such a dense clump is higher than the Roche density, gravitational instability would play an important role in the subsequent evolution of the clump and planetesimal formation there \citep[e.g.,][]{Safronov1972,Sekiya1983, Shi&Chiang2013}.

When CPBs formed from high-density dust clumps, collisions between dust particles (e.g., chondrules and matrix grains) should have occurred as well. 
In fact, there are previous studies that investigate collisions between chondrules and matrix in the solar nebula.
For instance, \cite{Ormel+2008} and \cite{Xiang+2018} showed that a layer composed of porous dust particles forms around the surface of chondrules.
Since such a layer is capable of absorbing the collisional energy \citep{Beitz+2012, Gunkelmann+2017}, 
aggregates can grow via collisions \citep[e.g.,][]{Ormel+2008,Arakawa2017}.
As will be shown below, these collisions generate two types of aggregates: the ones are composed of chondrules and matrix grains, and the other ones consist purely of matrix grains.
In this paper, the former ones are referred to as compound/chondrule aggregates (CAs) and the latter ones are as matrix aggregates (MAs).
\cite{Ormel+2008} investigate the growth of CAs using the studies of particle-cluster aggregation. 
This corresponds to the situation that CAs (or chondrules) collide with matrix grains.
When CAs (or chondrules) collide with MAs, which is called cluster-cluster aggregation, the internal densities of aggregates become smaller than those derived from particle-cluster aggregation.
The resulting formed aggregates via cluster-cluster aggregation have internal densities that can be several orders of magnitude smaller than the bulk density of monomers.
These aggregates are called fluffy aggregates.
This fluffy aggregate growth occurs when aggregates are composed of small dust monomers \citep[e.g., ][]{Okuzumi+2012,Kataoka+2013, Arakawa&Nakamoto2016}.

The above studies, however, do not consider dense clumps as a natal place of planetesimals/CPBs.
As demonstrated below (see Section \ref{sect:outline}), the clumps host collisions between chondrules and matrix, which can occur well before the clumps experience gravitational collapse for most cases.
In this paper, we calculate the collisional evolution of both CAs and MAs in dense clumps using the fluffy aggregate growth model.
We also estimate their internal structure that is determined by their collisional history.

The plan of this paper is as follows.
We give the outline of our model in Section \ref{sect:outline}.
We describe the detail of our model in Section \ref{sect:model}.
In Sections \ref{sect:resut_weddy} and \ref{sect:resut_leddy}, 
we show how aggregates form and grow and how the chondrule fraction evolves in aggregates using two models.
In Section \ref{sect:discussion}, the influences of our assumptions on the results are discussed.
Conclusions are given in Section \ref{sect:conclusion}.

\section{Outline}\label{sect:outline}

\begin{deluxetable*}{ll}
	\tablenum{1}
	\tablecaption{Key quantities}\label{table:quantities}
	\tablehead{
		Quantities	&	Meaning	
	}
	\startdata
	$M_{\rm clump}$		&	The total dust mass of the dense clump\\
	$R_{\rm clump}$		&	The size of the dense clump (6000~km)\\
	$\chi_{\rm clump}$	&	The chondrule mass fraction in the whole dense clump\\
	$r$					&	The orbital radius of the dense clump\\
	$\alpha$			&	The strength of the turbulence ($10^{-4}$)\\
	$c_{\rm s}$			&	The sound speed in the clump\\
	$\rho_{\rm d,ch}$	&	The mass density of chondrules in the dense clump\\
	$m_{\rm ch}$		&	The mass of chondrules\\
	$a_{\rm ch}$		&	The radius of chondrules\\
	$\rho_{\rm d,mat}$	&	The mass density of matrix grains in the dense clump at the onset of simulations\\	
	$m_{\rm mat}$		&	The mass of matrix monomers\\
	$a_{\rm mat}$		&	The radius of matrix monomers\\
	$t_{\rm GC}$		&	The timescale of gravitational collapse of the clump\\
	$t_{\rm gr,1-2}$ 	&	The growth timescale via collisions between aggregates 1 and 2\\
	$\St$				&	Stokes number of solid aggregates (chondrules, matrix grains, CAs and/or MAs) \\
	$t_{\rm s}$			&	The stopping time of aggregates\\
	$\Delta v_{1-2}$	&	The collision velocity between aggregates 1 and 2\\
	$v_{\rm tur}$		&	The turbulence-induced relative velocity of the aggregate to the surrounding gas\\
	$v_{\rm tur,L}$		&	The velocity induced by the eddies whose turnover timescales are longer than $t_{\rm s}$\\
	$v_{\rm tur,S}$		&	The velocity induced by the eddies whose turnover timescales are shorter than $t_{\rm s}$\\
	$v_{\rm frag,1-2}$	&	The fragmentation velocity for collisions between aggregates 1 and 2\\
	$M_{\rm CA}$		&	The mass of CAs\\
	$a_{\rm CA}$		&	The radius of CAs\\
	$\rhoint_{\rm CA}$	&	The internal density of CAs\\
	$v_{\rm CA}$		&	The relative velocity between CAs and gas\\
	$M_{\rm MA}$		&	The mass of MAs\\
	$a_{\rm MA}$		&	The radius of MAs\\
	$\rhoint_{\rm MA}$	&	The internal density of MAs\\
	$\rho_{\rm d,MA}$	&	The mass density of MAs in the dense clump\\	
	$\chi$				&	The chondrule mass fraction in CAs\\
	$1-\chi$			&	The matrix mass fraction in CAs
	\enddata 
\end{deluxetable*}

\begin{deluxetable*}{lll}
	\tablenum{2}
	\tablecaption{Summary of parameters}\label{table:parameters}
	\tablehead{
		parameter	&	values	&	fiducial value
	}
	\startdata
	$\rho_{\rm d,ch}$ [$\mbox{g~cm}^{-3}$]	&	$10^{-6},\,10^{-5},\,10^{-4},\,10^{-3},\, 10^{-2}$	&	$10^{-4} $	\\
	$\chi_{\rm clump}$	&	1/9, 1/5, 1/3, 1/2, 2/3, 4/5	&	1/2\\
	$a_{\rm ch}$ [cm]		&	$10^{-3},\ 10^{-2}$, $10^{-1}, 10^{0}$	&	$10^{-1}$ \\
	$a_{\rm mat}$ [cm]	&	$2.5\times10^{-7}$, $2.5\times10^{-6},\ 2.5\times10^{-5},\ 2.5\times10^{-4}$	&	$2.5\times10^{-7}$
	\enddata 
\end{deluxetable*}

In this section, we describe the outline of our models.
Key quantities and parameters are summarized in Tables \ref{table:quantities} and \ref{table:parameters}, respectively.

\subsection{Dense clump}
\label{sect:dense_clump}

\begin{figure}[hbt]
	\plotone{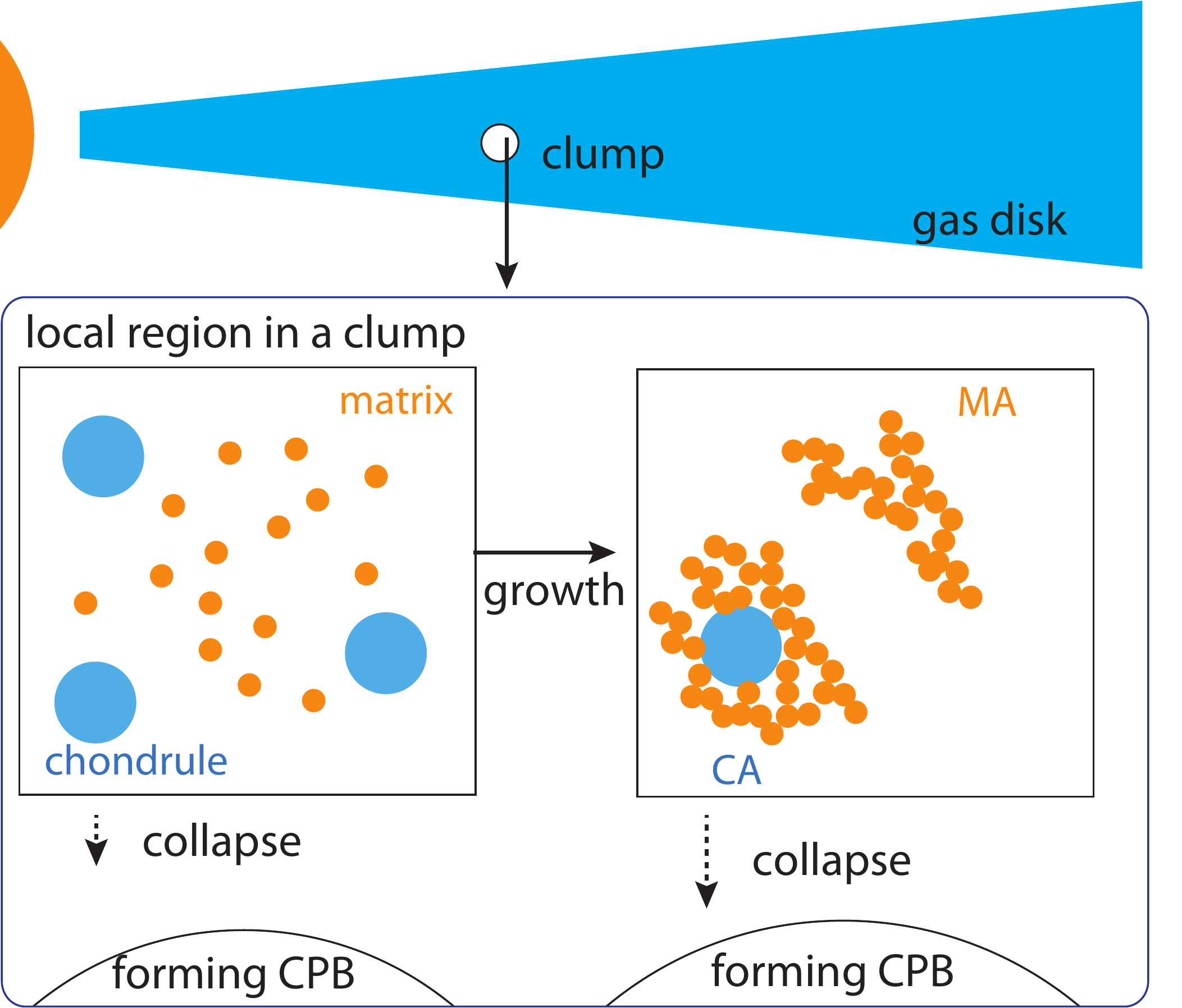}
	\caption{
		The schematic picture of our model.
		A dense clump is embedded in the gas disk around the Sun.
		We focus on the local surface region in the clump that is initially composed of chondrules and matrix grains.
		We will examine whether these particles grow via collisions or sediment to the clump center via gravitational collapse, where a CPB is forming.
		}
	\label{Fig:overview}
\end{figure}

We consider that a dense clump is embedded in the gas disk.
It is assumed that the clump is composed of chondrules and matrix grains, which is located at $r=2$ au unless otherwise mentioned (Figure \ref{Fig:overview}).

Under the above assumption,
the dense clump can be characterized by three quantities: 
the initial mass densities of chondrules and matrix grains ($\rho_{\rm d,ch}$, $\rho_{\rm d,mat}$) and the radius of the clump ($R_{\rm clump}$).
Then, the initial mass fraction of chondrules in the clump ($\chi_{\rm clump}$) is given as
\begin{eqnarray}
	\chi_{\rm clump}=\frac{\rho_{\rm d,ch}}{\rho_{\rm d,mat}+\rho_{\rm d,ch} },
	\label{eq:chi_GI}
\end{eqnarray}
and the total mass of the clump is written as
\begin{eqnarray}
 \label{eq:M_clump}
	M_{\rm clump} 
	& = & \frac{4 \pi}{3} (\rho_{\rm d,mat}+\rho_{\rm d,ch}) R_{\rm clump}^3 \\ \nonumber
	& = & \frac{4 \pi}{3} \frac{\rho_{\rm d,ch}}{\chi_{\rm clump}} R_{\rm clump}^3,
\end{eqnarray}
where the clump is assumed to be spherical.

The main purpose of this work is to examine how collisional growth of chondrules and matrix grains is important 
for determining the internal structure of CPBs forming out of self-gravitating dense clumps.
To achieve such a situation, the clump should have the initial mass density that is higher than the Roche density \citep[$\rho_{\rm R}$, e.g.,][]{Safronov1972, Shi&Chiang2013}:
\begin{eqnarray}
	\rho_{\rm R}\simeq 3.5M_{\odot}/r^3 = 2.6\times 10^{-7} (r/\mbox{2 au})^{-3} \mbox{~g~cm}^{-3},
	\label{eq:rho_roche}
\end{eqnarray}
where $r$ is the orbital radius.
Furthermore, the experimental study of chondrules in Semarkona ordinary chondrite suggests that 
its parent clump should have had $10^{-6} \mbox{~g~cm}^{-3} \leq \rho_{\rm d,ch} \leq 10^{-2} \mbox{~g~cm}^{-3}$ \citep{Alexander+2008} and 150 km $ \la R_{\rm clump} \la$ 6000 km \citep{Cuzzi&Alexander2006}. 
We thus adopt that $\rho_{\rm d,ch} = 10^{-4} \mbox{~g~cm}^{-3}$ and $R_{\rm clump} =$ 6000 km for our fiducial case (see Table \ref{table:parameters} and Section \ref{sect:params}).
Assuming that $\chi_{\rm clump}=1/2$, the total mass of the clump is written as
\begin{eqnarray}
	M_{\rm clump}=1.8\times 10^{23} (\rho_{\rm d,ch}/10^{-4} \mbox{~g~cm}^{-3}) (\chi_{\rm clump}/0.5)^{
		-1}~\mbox{g}.
	\nonumber\\
\end{eqnarray}
The corresponding planetesimal sizes are about 50~km -- 1100~km, if all dust components are accreted onto the same planetesimal.
\footnote{
Note that the clump size inferred by \citet{Cuzzi&Alexander2006} represents chondrule-forming regions, not planetesimal-forming regions.
The planetesimal-forming regions may have different concentrations and spatial scales from the chondrule-forming regions. 
All of the chondrules formed together in the region may not necessarily form into the same planetesimal together.
}

\subsection{Growth modes}
\label{sect:modes}

We now consider what kinds of growth can occur in the clump described above.
There are two possible growth processes: the gravitational collapse and collisional growth.
We here discuss the importance of these processes by estimating the corresponding timescales.

We first consider gravitational collapse.
As discussed above, the clump is self-gravitating.
Hence, it is natural to expect that the collapse of solid particles (chondrules and/or matrix) occurs on the free-fall timescale, which is given as
\begin{eqnarray}
	t_{\rm ff} &=& \frac{\pi}{2} \left( \frac{3}{4\pi {\rm G} \rho_{\rm d} } \right)^{1/2} \equiv \frac{\pi}{2} \omega_{\rm ff}^{-1}
	\nonumber \\
	&=& 9.5\times10^{-3} \left( \frac{ \rho_{\rm d} }{10^{-4}\mbox{~g~cm}^{-3}} \right)^{-1/2} \mbox{~yr}
	,
	\label{eq:t_ff}
\end{eqnarray}
where $\rho_{\rm d}$ is the mass density of solid in the clump and ${\rm G}$ is the gravitational constant.
The above value corresponds to 3.5~days in this specific case.
At the same time, the clump is embedded in the gas disk in our setup (see below), and thereby the dynamics of solid particles may be affected by the gas within the clump.
In particular, if the size of solid particles is small enough, they can be coupled well with the local gas motion.
In this case, the collapse of these particles is regulated by sedimentation toward the clump center, which is much slower than free-fall.
Consequently, the collapse timescale ($t_{\rm GC}$) of solid particles in the clump can be computed by
\begin{equation}
t_{\rm GC} = \mbox{max}[t_{\rm ff}, t_{\rm sed}],
\label{eq:t_gc}
\end{equation}
where the sedimentation timescale ($ t_{\rm sed}$) is estimated from the terminal velocity of solid particles under the gas drag \citep{Cuzzi+2008}:
\begin{eqnarray}
	t_{\rm sed} = \omega_{\rm ff}^{-2} t_{\rm s}^{-1} \propto \rho_{\rm d}^{-1} t_{\rm s}^{-1}.
	\label{eq:t_sed}
\end{eqnarray}
where $t_{\rm s}$ is the stopping time of solid particles.
Thus, the gravitational collapse will occur on the timescale of 3.5~days or longer in our setup.
\footnote{
In the dense clump, gravitational collapse can be frustrated by the gas pressure that is enhanced by the back reaction of dust on the gas motion \citep[e.g.,][]{Shariff&Cuzzi2015}.
We will discuss this effect in Section \ref{sect:discuss_time}.
}

We now discuss the growth timescale of solid particles via collisions in the clump.
As an example, we consider collisions between matrix grains.
To proceed, we assume that the size of matrix grains is $a_{\rm mat}=2.5\mbox{ nm}$, 
their material density is $\rhoint_{\rm mat}=3 \mbox{~g~cm}^{-3}$, 
and their mass is $m_{\rm mat} = 1.96 \times10^{-19} (a_{\rm mat}/\mbox{2.5~nm} )^3 \mbox{~g}$ (see Table \ref{table:parameters} and Section \ref{sect:params}).
Then, the growth timescale of matrix-matrix collisions in the clump is given by 
\begin{eqnarray}
	t_{\rm gr,mat-mat}
	&=& \left( \frac{ \rho_{\rm d,mat} \sigma_{\rm mat-mat} \Delta v_{\rm mat-mat}}{m_{\rm mat} }\right)^{-1}
	\nonumber \\
	&\sim&
	3\times10^{-6}
	\left( \frac{ \rho_{\rm d,mat} }{ 10^{-4}~\mbox{g~cm}^{-3} }\right)^{-1}
	\left( \frac{ \rhoint_{\rm mat} }{ 3~\mbox{g~cm}^{-3} }\right)
	\nonumber \\ && \times
	\left( \frac{ a_{\rm mat} }{ 2.5~\mbox{nm} }\right)
	\left( \frac{ \Delta v_{\rm mat-mat} }{ 10^3~\mbox{cm~s}^{-1} }\right)^{-1}
	~\mbox{s},
	\label{eq:t_gr_mat_mat_ex}
\end{eqnarray} 
where $\sigma_{\rm mat-mat}$ and $\Delta v_{\rm mat-mat}$ are the cross-section and the relative velocity between matrix grains, respectively.
For simplicity, we here adopt the Brownian motion for computing $\Delta v_{\rm mat-mat}$ (see Section \ref{sect:dynamics} for the detail).
Equation (\ref{eq:t_gr_mat_mat_ex}) explicitly shows that the growth timescale of matrix-matrix collisions is significantly shorter than the collapse timescale.
This suggests that chondrules and matrix grains can grow via collisions in the dense clump well before they experience gravitational collapse.

In summary, there are two growth modes in the dense clump (see Figure \ref{Fig:overview}).
The first one is the collisional growth of chondrules and matrix grains.
This mode occurs when the growth timescale via collisions is shorter than the collapse timescale.
The expected outcome is the formation of aggregates of chondrules and matrix in the clump.
We refer to this mode as the accretion mode in this paper.
The other one is the collapse mode.
In this mode, chondrules, matrix grains, and/or the subsequently formed CAs and MAs experience gravitational collapse,
which leads to sedimentation toward the clump center.

In the following sections, we will compute the formation and growth of CAs and MAs via collisions in the dense clump 
and investigate when the accretion mode dominates over the collapse one.
Also, we will explore the effect of the accretion mode on the internal structure of CPBs forming out of the dense clump.

\subsection{Accretion mode}

Before describing the detail of our model for the accretion mode, we here provide its overview.

\subsubsection{Basic model and assumptions}
\label{sect:assump}

We extend the fluffy aggregate growth model \citep{Okuzumi+2012, Kataoka+2013, Arakawa&Nakamoto2016, Arakawa2017}, 
by considering both the self-gravitating environment and two kinds of solid particles (chondrules and matrix grains).
These two new ingredients make the model very complicated.
We, therefore, adopt the following assumptions to simplify the problem.

{\it Assumption 1.} Both the mass density and the temperature of the gas within the dense clump are the same as those of the surrounding disk gas.
It can be expected that the dynamics of particles in the clump would affect these gas quantities, and hence numerical simulations are needed to verify this assumption.

{\it Assumption 2.} The existence of the dense clump does not affect the surrounding gas motion.
This assumption is justified as follows.
The Bondi radius ($r_{\rm B}$) of the clump is given as (using equations (\ref{eq:M_clump}) and (\ref{eq:rho_roche})),
\begin{eqnarray}
	r_{\rm B} = \frac{GM_{\rm clump}}{c_{\rm s}^2}
	= \frac{14\pi}{3} \left( \frac{R_{\rm clump}}{h_{\rm g}} \right)^2 \left( \frac{\rho_{\rm d}}{\rho_{\rm R}} \right) R_{\rm clump}
	,
	\label{eq:r_B}
	\nonumber\\
\end{eqnarray}
where $c_{\rm s}$ is the sound speed and $h_{\rm g}$ is the gas scale height.
Since $h_{\rm g} \simeq 0.04r\sim 10^7$~km at 2~au for the optically thin disk (see Section \ref{sect:disk_gas} for the detail), $r_{\rm B} \ll R_{\rm clump}$.
Thus, the disk gas motion is not altered even in the vicinity of the clump significantly.
Note that the disk gas can affect the motion of particles in the clump.

{\it Assumption 3.} With Assumptions 1 and 2, it would be consistent to assume that the gas in the clump is turbulent, which is similar to the one in the surrounding gas disk.
This assumption would be reasonable, especially for the surface region of the clump, and we focus on such a local region in this paper (Figure \ref{Fig:overview}). 
Turbulence plays an important role in determining the relative velocity of colliding solid particles and hence the growth of CAs and MAs.
It is assumed that the motions of particles are regulated by their stopping times.
The resulting turbulence-induced velocity ($v_{\rm tur}$) of particles can be written as the summation of the velocities induced by eddies with different scales \citep{Ormel&Cuzzi2007}. 
It is, however, unclear how these eddies behave in the clump and which eddies contribute to the velocity of particles.
Therefore, we divide $v_{\rm tur}$ into two components:
\begin{equation}
	v_{\rm tur}^2 = v_{\rm tur,L}^2+v_{\rm tur,S}^2,
\end{equation} 
where $v_{\rm tur,L}$ is induced by the eddies whose turnover timescales are longer than $t_{\rm s}$; and $v_{\rm tur,S}$ is by the shorter timescale eddies.
In this paper, we perform two kinds of simulations: 
one is called the whole eddy model where $v_{\rm tur}^2=v_{\rm tur,L}^2+v_{\rm tur,S}^2$ 
and the other is the large eddy model where $v_{\rm tur}^2 = v_{\rm tur,L}^2$.
Note that it would be ideal to realistically determine what sizes of eddies can survive in the {\rm dense} clump and contribute to $v_{\rm tur}$ in the clump.
However, such determination requires detailed numerical simulations, which is beyond the scope of this work, and would depend on a number of parameters (see Table \ref{table:parameters}).
We, therefore, simplify the problem, assuming that the smallest size of eddies that contribute to $v_{\rm tur,L}$ is determined by the particle size (equivalently $t_{\rm s}$).
Since the particle sizes increase according to their growth, our assumption effectively mimics the situation that smaller-sized eddies will {\rm be damped by particles} with time and their effects on particle growth will become negligible.
Thus, we here attempt to examine the effect of $v_{\rm tur}$ on particle growth by considering these two extreme cases:
In the whole eddy model, all the eddies affect the motion of particles with various sizes all the time;
In the large eddy model, only the eddies whose turnover time is longer than the stopping times of a particle affect the velocity of a particle.
The collision velocities of particles are also different in these models.
\footnote{
See also Equations (B2) and (B3) in	\cite{Ormel&Cuzzi2007}, where $\Delta V_{\rm I}$ is the collision velocity induced by $v_{\rm tur,L}$ and $\Delta V_{\rm II}$ is the collision velocity induced by $v_{\rm tur,S}$.
In the whole eddy model, both $\Delta V_{\rm I}$ and $\Delta V_{\rm II}$ contribute to the collision velocities.
In the large eddy model, we simply assume there is no contribution from $\Delta V_{\rm II}$ to act on particles and only consider the contribution from $\Delta V_{\rm I}$ for any stopping times of particles.
}
The detailed expression of $v_{\rm tur}$ is given in Section \ref{sect:dynamics}.

{\it Assumption 4.} There is no inflow and outflow of solid particles in the clump. 
This assumption greatly simplifies the problem and is related to the formation mechanism of dense clumps,
which is out of the scope of this work.
Thus, its validity should be examined through numerical simulations.

{\it Assumption 5.} We assume that collisions of solid particles lead to perfect mergers unless the collision velocity exceeds the critical velocity for the collisional growth of aggregates, which is called the fragmentation velocity ($v_{\rm frag}$, see \ref{sect:v_crit} for the detail).
If exceeds, the corresponding collisions are neglected in aggregate growth calculations.
Thus, the outcome of collisions is idealized, where only perfect sticking and bouncing without the mass loss are effectively considered.

{\it Assumption 6.} We assume that the resulting formed CAs are identical with each other.
Equivalently, their size distribution is not computed.
The same assumption is adopted to MAs.
This greatly simplifies the problem.
For instance, the numerical cost can be reduced significantly;
about $10^{25}$ chondrules and about $10^{42}$ matrix grains are initially present in the clump for our fiducial case (Section \ref{sect:params}).
Under this assumption, there is no need for computing the growth of these particles individually. 
In addition, it may be reasonable to adopt this assumption as a first extension of the fluffy aggregate growth model.

{\it Assumption 7.} With Assumption 6, 
it would be natural to neglect both the spatial distributions of CAs and MAs and the spatial variation of gas quantities within the clump.
These are the mass density, temperature, and turbulence properties of the gas in the clump.

\subsubsection{Resulting collisional growth}
\label{sect:col_growth}

\begin{figure}[hbt]
	\plotone{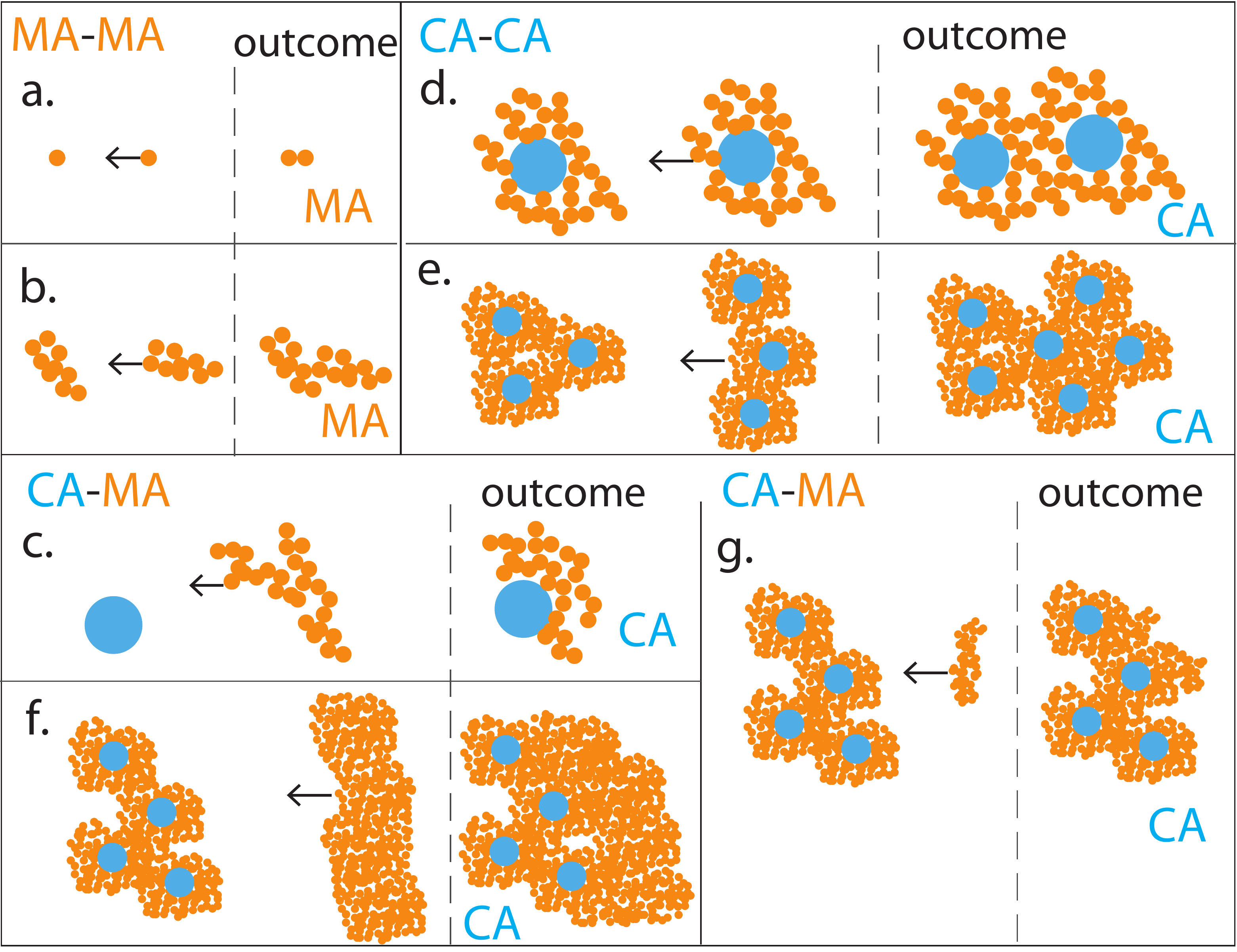}
	\caption{
		The schematic picture of the expected outcome in the accretion mode.
		Each panel from (a) to (g) shows solid particles both before and after collisions.
		Chondrules are represented by the blue circles, matrix grains by the orange circles, 
		CAs by the mixture of chondrules and matrix, and MAs by the pure collection of matrix grains.
		Panel (a) shows the picture of the collision between matrix grains.
		Panel (b) is between MAs.
		Panel (c) is between a chondrule and a MA.
		Panel (d) is between CAs. These aggregates are composed of single chondrules covered by matrix grains.
		Panel (e) is between CAs that contain multiple chondrules and matrix components.
		Panel (f) is between a CA and a MA. These two aggregates are similar in size.
		Panel (g) is between a CA and a MA. The CA is much larger than the MA in this case.
		}
	\label{Fig:collisions}
\end{figure}

With the above model and assumptions, we conduct detailed aggregate growth calculations in the following sections.
Here, we discuss the expected outcome of collisions. 
The schematic picture is shown in Figure \ref{Fig:collisions}.
Our results are provided in Sections \ref{sect:resut_weddy} and \ref{sect:resut_leddy}, and Appendices \ref{sec:results_evolution} and \ref{sec:results_evolution_v1}.

First, it is expected that matrix-matrix collisions occur and matrix grains grow up to be MAs (see Figure \ref{Fig:collisions} (a) and (b), also see Section \ref{sect:modes}).
In this paper, MAs are regarded as the ones composed purely of two or more matrix grains.
As MAs form and grow, their radii become comparable to those of chondrules.
At that time, the number density, cross-section, and relative velocity of MAs can also be comparable to those of chondrules.
Then, chondrule-MA collisions begin.
Note that the growth timescale of MAs is affected by their internal density evolution as well;
the internal densities of MAs become several orders of magnitude smaller than those of matrix grains due to the hit-and-stick growth process (see Section \ref{sect:material_density}). 

Subsequently, chondrules would collide with MAs and become CAs (Figure \ref{Fig:collisions} (c)).
These CAs are regarded as chondrules covered by the matrix surface layer. 
Assuming that $N_{\rm ch}$ chondrules are contained in a CA with a mass of $M_{\rm CA}$,
the chondrule mass fraction ($\chi$) of the CA is given as\footnote{Note that our definition of $\chi$ is different from that of \cite{Arakawa2017}.}
\begin{eqnarray}
	\chi=\frac{N_{\rm ch} m_{\rm ch}}{M_{\rm CA}}.
	\label{eq:chi_def}
\end{eqnarray}
The value of $\chi$ is initially unity since any chondrules do not have matrix components yet.
As time goes on, the value decreases because chondrules collide with MAs.
If the resulting formed CAs accrete all MAs in the clump, then $\chi = \chi_{\rm clump}$ (see equation (\ref{eq:chi_GI})).
Note that due to Assumption 6 (see Section \ref{sect:assump}), 
$\chi$ does not change via CA-CA collisions.
This is simply because a collision between two identical CAs forms a newly formed CA, 
but the new CA doubles both the chondrule mass and total aggregate mass (Figure \ref{Fig:collisions} (d) and (e)).

Eventually, CAs collide with other CAs and/or MAs and grow in mass.
As discussed above, the value of $\chi$ decreases when CA-MA collisions occur (Figure \ref{Fig:collisions} (f) and (g)).

In the following sections, we will investigate how $\chi$ evolves with time by tracing the growth paths of CAs in the clump.

\subsubsection{Model parameters}
\label{sect:params}

There are four key parameters in our calculations: $\rho_{\rm d, ch}$, $a_{\rm ch}$, $\chi_{\rm clump}$, and $a_{\rm mat}$ (see Table \ref{table:parameters}).
We here discuss the motivation of values adopted in this paper.

We consider that $10^{-6} \mbox{~g~cm}^{-3} \leq \rho_{\rm d,ch} \leq 10^{-2} \mbox{~g~cm}^{-3}$, following \cite{Alexander+2008} as discussed in Section \ref{sect:dense_clump}.
The radius of chondrules is also taken as a parameter ($a_{\rm ch}=1, 10^{-1}, 10^{-2}, 10^{-3}$~cm), and $a_{\rm ch}=10^{-1}$~cm is our fiducial value.
We choose these values because they cover the typical size range of chondrules \citep[e.g.,][]{Scott&Krot2005,Scott2007,Krot+2009,Friedrich+2015,Simon+2018}.
Assuming that the material density of chondrules is $\rhoint_{\rm ch}=3 \mbox{~g~cm}^{-3}$,
their mass is written as $m_{\rm ch}\simeq 1.26\times10^{-2} (a_{\rm ch}/\mbox{10$^{-1}$ cm} )^3 \mbox{ g}$.

We treat $\chi_{\rm clump}$ as a parameter (rather than $\rho_{\rm d, mat}$), 
in order to examine how the matrix abundance affects the growth of CAs in the clump (see Equation (\ref{eq:chi_GI})).
We consider $\chi_{\rm clump}=4/5,\,2/3,\,1/2,\,1/3,\,1/5,\,1/9$, which covers the present chondrule volume fraction in chondrites \citep{Scott&Krot2005}.
The fiducial value is $\chi_{\rm clump}=1/2$.

Finally, we parameterize the size of matrix grains ($a_{\rm mat}$) with the range from 2.5 nm ($2.5\times10^{-7}$~cm) to 2.5 $\mu$m ($2.5\times10^{-4}$~cm).
In our fiducial case, we adopt $a_{\rm mat}=2.5\mbox{ nm}$, following \cite{Arakawa&Nakamoto2016}. 
Note that this size is the peak value in the size distribution of matrix grains in Allende chondrite \citep{Toriumi1989}.
Assuming that the material density of matrix grains is $\rhoint_{\rm mat}=3 \mbox{~g~cm}^{-3}$, 
the mass of matrix grains is given as $m_{\rm mat} = 1.96 \times10^{-19} (a_{\rm mat}/\mbox{2.5~nm} )^3 \mbox{~g}$.

\subsubsection{Simulation procedure}\label{sect:calculation_summary}

We perform simulations of fluffy aggregate growth with the above setups.
We here describe the procedure of our simulations.

First, we set up a dense clump that is composed of chondrules and matrix grains (Section \ref{sect:dense_clump}) in the gas disk (Section \ref{sect:disk_gas}).
In this setup, we pick up one set of model parameters from Table \ref{table:parameters} (Section \ref{sect:params}).
we calculate the Stokes numbers of chondrules and matrix, where the Stokes number is the normalized stopping time ($t_{\rm s}$) and given by $\St=t_{\rm s}\Omega_{\rm K}$ (Section \ref{sect:dynamics}).
This allows us to compute their velocities relative to the gas motion and collision velocities in a timestep.

Second, we numerically calculate the growth timescales (Section \ref{sect:timescale}), the collapse timescale ($t_{\rm GC}$, Section \ref{sect:modes}), and the fragmentation velocity of aggregates (Section \ref{sect:v_crit}) in this timestep, under the above assumptions (Section \ref{sect:assump}).
The growth timescale of matrix-matrix collisions and MA-MA collisions is denoted by $t_{\rm gr,MA-MA}$, that of chondrule-MA collisions and CA-MA collisions by $t_{\rm gr,CA-MA}$, 
and that of CA-CA collisions by $t_{\rm gr,CA-CA}$, in the following.
 
Third, we judge which growth mode (accretion vs collapse) will occur in this timestep, based on the computed timescales.
If the accretion mode will be realized, 
we compare the growth timescales of all the collisions ($t_{\rm gr,MA-MA}$, $t_{\rm gr,CA-MA}$, and $t_{\rm gr,CA-CA}$)
and find out a collision that satisfies two conditions: 
1) the corresponding growth timescale is the shortest among other collisions, 
and 2) the collision velocity is slower than the fragmentation velocity.

Fourth, if the collision satisfying the above two conditions is identified, we increase the mass of the particles that experience the collision.
We also calculate their internal densities (Section \ref{sect:material_density}) and Stokes numbers.
The internal densities of CAs and MAs are denoted by $\rhoint_{\rm CA}$ and $\rhoint_{\rm MA}$, respectively.
We also update the mass densities of CAs and MAs in the clump (Section \ref{sec:model_volumedensity}).

Fifth, the above steps (from second to fourth) are repeated until either the numbers of CAs or MAs become less than two, or the collapse mode is realized.

\section{Model}\label{sect:model}

In this section, we provide the detail of our fluffy aggregate growth model.
The general readers may proceed to Sections \ref{sect:resut_weddy} and \ref{sect:resut_leddy} for the main results of this paper.

\subsection{Disk gas}\label{sect:disk_gas}

We here describe a model of the surrounding gas disk.
We adopt a power-law model similar to the minimum-mass solar nebula \citep{Hayashi1981}.

The gas surface density ($\Sigma_{\rm g}$) is given by $\Sigma_{\rm g}=2400 (r/1\mbox{ au})^{-3/2} \mbox{~g~cm}^{-2}$. 
Note that $\Sigma_{\rm g}$ is increased by 1.5 following previous studies \citep[e.g.,][]{Kokubo&Ida2000,Ida&Lin2004a,Hasegawa+2016a}.
Considering an optically thin disk around a one solar mass star ($M_{\odot}$), the disk temperature is given by $T=280 (r/1\mbox{ au})^{-1/2} \mbox{ K}$.
The sound speed is computed as $c_{\rm s}=\sqrt{k_{\rm B} T/m_{\rm g}}$, where $k_B$ is the Boltzmann constant; and $m_{\rm g}=3.9\times10^{-24}\mbox{ g}$ is the mean molecular mass.
The vertical structure of the disk is assumed to be in hydrostatic equilibrium. 
Then the gas mass density at the midplane ($\rho_{\rm g}$) is written as $\rho_{\rm g} = \Sigma_{\rm g}/\sqrt{2\pi} h_{\rm g}$, where $h_{\rm g}=c_{\rm s}/\Omega_{\rm K}$ is the gas scale height and $\Omega_{\rm K}$ is the Kepler frequency. 
When the clump is located at 2~au in our fiducial, $\rho_{\rm g} =2.9\times10^{-10} \mbox{~g~cm}^{-3}$, which is smaller than $\rho_{\rm d,ch}$ (Section \ref{sect:dense_clump}).
The disk gas moves at a sub-Keplerian velocity, $(1-\eta) v_{\rm K}$, where $v_{\rm K}$ is the Kepler velocity, and $\eta$ can be written as
\begin{eqnarray}
	\eta = \frac{1}{2}\left( \frac{c_{\rm s}}{v_{\rm K}} \right)^2 \frac{\partial \ln{(\rho_{\rm g} c_{\rm s}^2)}}{\partial r}, 
\end{eqnarray}
\citep{Adachi+1976}.
In our model, the strength of gas turbulence is prescribed by the $\alpha$-parameter \citep{Shakura&Sunyaev1973}.
\cite{Fu+2014} conducted paleomagnetic measurements of chondrules in Semarkona ordinary chondrite and 
suggest that the magnetic field strength was 5 to 54 microteslas at the chondrule-forming region of the solar nebula.
These values correspond to $\alpha\sim 10^{-4}$ \citep{Wardle2007,Okuzumi&Hirose2011,Hasegawa+2016b}, and we adopt $\alpha= 10^{-4}$.

\subsection{Mass densities of MAs in the clump} \label{sec:model_volumedensity}

The mass densities of chondrules and matrix grains in the dense clump are important quantities for aggregate growth.
In this section, we derive a correlation between the mass densities of chondrules ($\rho_{\rm d,ch}$) and MAs ($\rho_{\rm d,MA}$) in the clump at a certain time.

The value of $\rho_{\rm d,ch}$ does not change with time under Assumption 4 (Section \ref{sect:assump}).
While the total abundance of matrix grains also does not vary due to the same reason, (i.e., $\rho_{\rm d, mat}$ is constant, see Equation (\ref{eq:chi_GI})),
the mass density of MAs ($\rho_{\rm d,MA}$) will evolve with time, following aggregate growth.
Considering a CA with the mass of $M_{\rm CA}$ and the chondrule mass fraction of $\chi$,
the total mass of matrix within this aggregate is $(1-\chi)M_{\rm CA}$.
With Assumption 6, the number density of CAs in the clump is given as $\rho_{\rm d,ch}/(\chi M_{\rm CA})$.
Consequently, the mass density of matrix that is contained in all the CAs is written as $(1-\chi)M_{\rm CA}\rho_{\rm d,ch}/(\chi M_{\rm CA})$.
Then, $\rho_{\rm d,MA}$ in the clump can be calculated as due to Assumption 4,
\begin{eqnarray}
	\rho_{\rm d,mat} 
	= \rho_{\rm d,MA} 
	+ (1-\chi) M_{\rm CA} \left( \frac{\rho_{\rm d,ch}}{ \chi M_{\rm CA} } \right).
	\nonumber\\
\end{eqnarray}
Equivalently,
\begin{eqnarray}
	\rho_{\rm d, MA}
	&=&\rho_{\rm d, mat}- \frac{1-\chi}{\chi}\rho_{\rm d,ch}.
	\label{eq:rhod_MA}
\end{eqnarray}
Note that under Assumption 6, the value of $\rho_{\rm d,MA}$ does not change due to MA-MA collisions. 
We adopt an extremely small value for $\rho_{\rm d,MA}$ when $\chi=\chi_{\rm clump}$.

\subsection{Growth timescales of CAs and MAs}\label{sect:timescale}

In this section, we describe the growth timescales of CAs and MAs.

The growth timescale via collisions is estimated by the mass doubling timescale, which is written as
\begin{equation}
	t_{\rm gr,1-2}=\frac{M_1}{M_2/t_{\rm col,1-2}}, 
\label{eq:t_gr_1-2}
\end{equation}
where an aggregate with a mass of $M_1$ grows via perfect mergers with the other aggregates with a mass of $M_2$.
In this equation, $t_{\rm col,1-2}$ is the collision timescale between these two aggregates, and $M_2/t_{\rm col,1-2}$ is the mass growth rate of the aggregate 1 due to a collision with the aggregate 2.
Note that the {\it total} growth timescale is a couple of tens times longer than the above mass doubling timescale (see Section \ref{sect:discuss_time}).

We here consider three growth timescales ($t_{\rm gr,MA-MA}$, $t_{\rm gr,CA-MA}$, and $t_{\rm gr,CA-CA}$, 
see Section \ref{sect:calculation_summary}, also see Figure \ref{Fig:collisions}). 
Due to Assumption 6 (Section \ref{sect:assump}),
$t_{\rm gr,MA-MA}$ and $t_{\rm gr,CA-CA}$ become equal to their collision timescales.
Mathematically, $t_{\rm gr,MA-MA}$ is given by 
\begin{eqnarray}
	t_{\rm gr,MA-MA} &=& \left( \frac{ \rho_{\rm d,MA} }{M_{\rm MA} } \sigma_{\rm MA-MA} \Delta v_{\rm MA-MA}\right)^{-1}
	\nonumber\\
	&\propto & \rho_{\rm d,MA}^{-1} M_{\rm MA}^{1/3} \rhoint_{\rm MA}^{\,2/3} \Delta v_{\rm MA-MA}^{-1} (1+\Theta_{\rm MA-MA})^{-1}
	,\label{eq:t_gr_MA-MA}
	\nonumber \\
\end{eqnarray}
where $\rhoint_{\rm MA} = 3M_{\rm MA} / (4 \pi a_{\rm MA}^3)$; $M_{\rm MA}$ is the mass of colliding MAs, $a_{\rm MA}$ is their radius, $\sigma_{\rm MA-MA}$ is their cross-section, and $\Delta v_{\rm MA-MA}$ is the collision velocity.
In general, the cross-section is written as $\sigma_{1-2}=\pi(a_1+a_2)^2(1+\Theta_{1-2})$, where $a_1$ and $a_2$ are the radii of the collision bodies; and $\Theta_{1-2}$ is the Safronov parameter \citep{Safronov1972}.
The Safronov parameter is also known as the gravitational focusing factor and given by the square of the ratio of the escape velocity and the collision velocity.
When the Safronov parameter is larger than unity, the corresponding collision leads to runaway growth.
This accretion occurs only at the final stage of aggregate growth in our simulations (see Appendices \ref{sec:res_stage7_v12} and \ref{sec:res_stage5_v1}), 
and the Safronov parameter is not effective until then.
Recent fluffy aggregate calculations suggest that the condition of the runaway growth is satisfied when the aggregate mass exceeds $\sim 10^{15}$ g \citep{Kataoka+2013,Arakawa&Nakamoto2016}.
This value is significantly smaller than the total dust mass of the clump in our setup, and hence the runaway growth is realized in our calculations.
Note that it is unclear whether the detailed behavior of this runaway growth, especially in the self-gravitating dense clump, is similar to that in the runaway growth of planetesimals \citep[e.g.,][]{Wetherill&Stewart1989, Ohtsuki+1993, Kokubo&Ida1996}.

The growth timescale of CAs via CA-CA collisions is computed in the same manner, 
\begin{eqnarray}
	t_{\rm gr,CA-CA}&=& \left( \frac{\rho_{\rm d,ch}}{\chi M_{\rm CA}} \sigma_{\rm CA-CA} \Delta v_{\rm CA-CA} \right)^{-1}
	\nonumber\\
	&\propto & \rho_{\rm d,ch}^{-1} M_{\rm CA}^{1/3} \rhoint_{\rm CA}^{\,2/3} \Delta v_{\rm CA-CA}^{-1} (1+\Theta_{\rm CA-CA})^{-1}
	,\label{eq:t_gr_CA-CA}
	\nonumber \\
\end{eqnarray}
where $\rhoint_{\rm CA} = (3M_{\rm CA} / (4 \pi a_{\rm CA}^3))$; 
$a_{\rm CA}$ is their radius; and $\sigma_{\rm CA-CA}$ and $\Delta v_{\rm CA-CA}$ are the cross-section and relative velocity between two colliding CAs, respectively.

In contrast to MAs, CAs can grow via CA-MA collisions as well.
The growth timescale of CA-MA collisions is computed as $t_{\rm gr,CA-MA}=M_{\rm CA}/( M_{\rm MA}/t_{\rm col,CA-MA} )$ (see Equation (\ref{eq:t_gr_1-2})).
Given that there is a (large) mass difference between CAs and MAs as chondrules are significantly more massive than matrix grains, $t_{\rm gr,CA-MA} \geq t_{\rm col,CA-MA}$.
To avoid the case that $t_{\rm gr,CA-MA} < t_{\rm col,CA-MA}$, where $M_{\rm CA}<M_{\rm MA}$, we adopt
\begin{eqnarray}
	t_{\rm gr,CA-MA}&=&\max{\left( \frac{M_{\rm CA}}{M_{\rm MA}}t_{\rm col,CA-MA},t_{\rm col,CA-MA} \right)}
	,\label{eq:t_gr_CA-MA}
	\nonumber \\
	t_{\rm col,CA-MA} &=& \left( \frac{\rho_{\rm d,MA} }{M_{\rm MA} } \sigma_{\rm CA-MA} \Delta v_{\rm CA-MA}\right)^{-1},
\end{eqnarray}
where $\sigma_{\rm CA-MA}$ and $\Delta v_{\rm CA-MA}$ are the cross-section and relative velocity between CAs and MAs, respectively.
In Equation (\ref{eq:t_gr_CA-MA}), we have assumed that the target is a CA and hence that $t_{\rm gr,CA-MA}$ does not depend on $\rho_{\rm d,ch}$ but on $\rho_{\rm d,MA}$.

We will compute these three growth timescales (see Equations (\ref{eq:t_gr_MA-MA}), (\ref{eq:t_gr_CA-CA}), and (\ref{eq:t_gr_CA-MA})) to find out the shortest one and to follow the mass evolution of CAs and MAs.
Practically, we calculate $t_{\rm gr,MA-MA}$ and $t_{\rm gr,CA-CA}$ at every numerical step.
For $t_{\rm gr,CA-MA}$, we will adopt the following numerical technique to speed up calculations: 
suppose that a series of CA-MA collisions have been realized and the mass of CAs has been increased by $10^{p_{\rm CM}} M_{\rm MA}$ over the previous numerical steps, 
where $p_{\rm CM}$ is an integer and we normally take $p_{\rm CM}=3$.
If we can confirm that a CA-MA collision still becomes the shortest growth timescale at the current numerical step, 
then we increase the mass of CAs by $10^{p_{\rm CM}-2} M_{\rm MA}$ during this single step. 
For example, we numerically add $10 M_{\rm MA}$ to a CA in a numerical step if the CA has accreted $10^3 M_{\rm MA}$ so far, 
and $10^2 M_{\rm MA}$ is added after CAs have obtained $10^4 M_{\rm MA}$.
Thus, we take account of multiple CA-MA collisions in a single numerical step in order to reduce numerical cost.

\subsection{Free fall vs sedimentation in the collapse mode}\label{sect:st_ff}

As discussed in Section \ref{sect:modes}, 
the collapse timescale is determined by Equation (\ref{eq:t_gc}).
Here, we estimate the characteristic value of the Stokes number ($\St_{\rm ff}$) of aggregates that satisfy $t_{\rm ff} = t_{\rm sed}$.

We find that $\St_{\rm ff}$ is written as
\begin{eqnarray}
	\St_{\rm ff} &=& 
		0.95\times10^{-2} \left( \frac{a}{\mbox{2 au} }\right)^{-3/2} 
			\left( \frac{\rho_{\rm d,ch}}{10^{-4} \mbox{~g~cm}^{-3}} \right)^{-1/2} \nonumber \\ &&\times 
			\left( \frac{\chi_{\rm clump} }{0.5} \right)^{1/2} .
\end{eqnarray}
This means that aggregates are decoupled from gas when $\St \sim$~0.1.
Considering the dependence of $\rho_{\rm d}$ on the free-fall timescale (Equation (\ref{eq:t_ff})), sedimentation timescale (Equation (\ref{eq:t_sed})), and growth timescales (Equations (\ref{eq:t_gr_MA-MA}), (\ref{eq:t_gr_CA-CA}), and (\ref{eq:t_gr_CA-MA})), it is expected that the collapse mode becomes effective when $\rho_{\rm d}$ is small and $\St>\St_{\rm ff}$.

\subsection{Internal densities of growing aggregates}
\label{sect:material_density}

The internal densities of aggregates ($\rhoint_{\rm CA}, \rhoint_{\rm MA}$) are important quantities for computing their growth timescales.
In this section, we describe how we derive the internal densities of aggregates.

We assume that the internal density of MAs evolves due to hit-and-stick, collisional compression, gas compression, and self-gravitational compression,
following the previous studies of fluffy aggregate growth \citep[e.g.,][]{Okuzumi+2012, Kataoka+2013,Arakawa&Nakamoto2016}.

The hit-and-stick growth occurs at first. 
In this case, the internal density of MAs ($\rhoint_{\rm hit}$) is given as \citep{Wada+2008}
\begin{eqnarray}
	\rhoint_{\rm hit}=(3/5)^{2/3} \left( M_{\rm MA}/m_{\rm mat}\right)^{-1/2}\rhoint_{\rm mat}. \label{eq:rho_hit}
\end{eqnarray}
The hit-and-stick growth decreases the internal density of aggregates.
This process becomes effective when colliding aggregates merge together without restructuring of their internal structures.
Equivalently, the impact energy ($E_{\rm imp}$) is smaller than the rolling energy ($E_{\rm roll}$).
Note that $E_{\rm imp}$ is the kinetic energy of two colliding bodies and hence proportional to $M_{\rm MA}$.
On the other hand, the rolling energy is given as $E_{\rm roll}=6\pi^2\gamma a_{\rm mat} \xi_{\rm crit}$, 
where $\gamma=25 \mbox{~erg~cm}^{-2}$ is the surface energy per unit contact area between two particles; and $ \xi_{\rm crit}=0.3\mbox{ nm}$ is the critical displacement of rolling \citep{Dominik&Tielens1997, Wada+2007}.
These two specific values are measured for silicate dust aggregates.

As MAs become more massive, $E_{\rm imp}$ increases and eventually reaches $\sim E_{\rm roll}$.
Then, the internal density of MAs is compressed through collisions.
The compaction rate through collisions was studied by \cite{Wada+2008} and the resulting internal density of MAs ($\rhoint_{\rm col}$) is given as
\begin{eqnarray}
	\rhoint_{\rm col}
		&=&(E_{\rm imp}/0.15E_{\rm roll})^{3/10} \rhoint_{\rm hit} \nonumber \\
		&\propto& M_{\rm MA}^{-1/5} \Delta v_{\rm MA-MA}^{3/5} 
		.\label{eq:rho_col}
\end{eqnarray}

Furthermore, aggregates can be compressed by the ram pressure and self-gravitational pressure.
Following \cite{Kataoka+2013a}, the internal density of MAs due to the ram pressure of the disk gas ($\rhoint_{\rm ram}$) is given by 
\begin{eqnarray}
	\rhoint_{\rm ram}&=&\left(\frac{a_{\rm mat}^3}{E_{\rm roll}} \frac{M_{\rm MA}v_{\rm MA}}{\pi a_{\rm MA}^2 t_{\rm s}} \right)^{1/3} \rhoint_{\rm mat} \nonumber\\
	&\propto& M_{\rm MA}^{1/7} v_{\rm MA}^{3/7} t_{\rm s}^{-3/7},
	\label{eq:rho_ram}
\end{eqnarray}
where $v_{\rm MA}$ is the relative velocity between a MA and the surrounding gas.
Note that $\rhoint_{\rm ram}$ depends on $v_{\rm MA}/t_{\rm s}$ because ram pressure is caused by the relative motion between the disk gas and aggregates.
The internal density of MAs due to the self-gravitational pressure ($\rhoint_{\rm grav}$) is \citep{Kataoka+2013a}
\begin{eqnarray}
	\rhoint_{\rm grav}&=&\left(\frac{a_{\rm mat}^3}{E_{\rm roll}} \frac{{\rm G}M_{\rm MA}^2}{\pi a_{\rm MA}^4} \right)^{1/3} \rhoint_{\rm mat} 
	\ \propto M_{\rm MA}^{2/5}
	.
	\label{eq:rho_grav}
\end{eqnarray}
These two equations show that the ram pressure ($M_{\rm MA}v_{\rm MA}/\pi a_{\rm MA}^2 t_{\rm s}$) and self-gravitating pressure (${\rm G}M_{\rm MA}^2/\pi a_{\rm MA}^4$) 
are normalized by the pressure from the rolling energy ($E_{\rm roll}/a_{\rm mat}^3$), and this normalized pressure determines the filling factor.

Thus, the internal density of MAs is computed by the above four equations and changes in the order of $\rhoint_{\rm hit}$, $\rhoint_{\rm col}$, $\rhoint_{\rm ram}$, and $\rhoint_{\rm grav}$ as $M_{\rm MA}$ increases \citep{Kataoka+2013,Arakawa&Nakamoto2016}.
We find that collisional compression is not effective in weak turbulent cases where collision velocities are small.
In addition, our preliminary results suggest that the internal density can become a decreasing function of $M$ in the $\rhoint_{\rm ram}$ regime.
It, however, can be expected that 
aggregates should be compressed and their internal densities should not decrease after the hit-and-stick regime.
We, therefore, neglect the density reduction once the Stokes number of aggregates exceeds unity.
We compute the filling factor of a MA with the equation that $\phi_{\rm MA}=\rhoint_{\rm MA}/\rhoint_{\rm mat}$.

The internal density of CAs ($\rhoint_{\rm CA}$) is determined by their internal structure and the mass fraction of matrix grains in the CAs ($1-\chi$).
This is because CAs are composed of both chondrules and matrix grains.
By denoting $\phi_{\rm mat,CA}$ as the filling factor of matrix grains in CAs, 
$\rhoint_{\rm CA}$ can be written as \citep{Arakawa2017}
\begin{eqnarray}
	\rhoint_{\rm CA} 
	&=& \frac{M_{\rm CA}}{ (M_{\rm CA} (1-\chi) )/(\phi_{\rm mat,CA}\rhoint_{\rm mat}) + \chi M_{\rm CA}/\rhoint_{\rm ch} }
	\nonumber\\
	&=& \left( \frac{1-\chi}{\phi_{\rm mat,CA}\rhoint_{\rm mat}/\rhoint_{\rm ch}} + \chi \right)^{-1} \rhoint_{\rm ch}
	,
	\label{eq:rho_CA}
\end{eqnarray}
where $(M_{\rm CA}(1-\chi) ) / ( \phi_{\rm mat,CA}\rhoint_{\rm mat})$ comes from the volume contribution of matrix grains; and $\chi M_{\rm CA}/\rhoint_{\rm ch}$ is from that of chondrules.
The derivation of $\phi_{\rm mat,CA}$ is the same as $\phi_{\rm MA}$.
This equation shows that $\rhoint_{\rm CA}$ is always larger than $\rhoint_{\rm MA}$ when CAs and MAs have the same mass.

\subsection{Dynamics of aggregates}
\label{sect:dynamics}

We here describe the dynamics of aggregates, which is important in computing their growth timescales (Section \ref{sect:timescale}) and judging a growth mode (Section \ref{sect:modes}).
According to Assumptions 1, 2, and 3, we consider the local surface region of the clump, where the motion of each aggregate may be controlled by the surrounding gas motion.
We consider that the dust motion is induced by the Brownian motion, turbulence, radial drift, and azimuthal drift.
For the purpose of clear presentation, the internal density, mass, and radius of dust aggregates are denoted by $\rhoint$, $M$, and $a$, respectively, and the relative velocity between gas and dust aggregates is by $v$ in this section.
We also use both the stopping time ($t_{\rm s}$) and Stokes number ($\St$) interchangeably.

The motion of dust particles is described as a function of $t_{\rm s}$.
Several regimes are introduced, 
depending on $a$ and the particle Reynolds number ($\Rep=2av/\nu_{\rm mol}$), 
where $\nu_{\rm mol}=v_{\rm th} \lambda_{\rm m}/2$; $v_{\rm th}=\sqrt{8/\pi}c_{\rm s}$ is the thermal velocity, 
and $\lambda_{\rm m}$ is the free path of gas particle \citep{Adachi+1976, Weidenschilling1977}.
Then, $t_{\rm s}$ obeys the Epstein's law for $a<(9/4)\lambda_{\rm m}$, 
\begin{eqnarray}
	t_{\rm s}^{\rm Ep}&=&3M/(4\pi \rho_{\rm g}v_{\rm th} a^2)
		\ \propto M^{1/3} \rhoint^{\,2/3}
	.\label{eq:t_s_Ep}
\end{eqnarray}
For $a>(9/4)\lambda$ and $\Rep<1$, in the Stokes' law,
\begin{eqnarray}
	t_{\rm s}^{\rm St}=M/(6\pi \rho_{\rm g}\nu_{\rm mol} a)
		\ \propto M^{2/3} \rhoint^{\,1/3}
	.\label{eq:t_s_St}
\end{eqnarray}
For $a>(9/4)\lambda$ and $1\leq\Rep<54^{5/3}$, the Allen's law, 
\begin{eqnarray}
	t_{\rm s}^{\rm Al}&=& 2^{3/5} M / (12\pi\rho_{\rm g}\nu_{\rm mol}^{3/5} v^{2/5} a^{7/5})
		\nonumber\\
		&\propto& M^{8/15} \rhoint^{\,7/15} v^{-2/5}
	.\label{eq:t_s_Al}
\end{eqnarray}
For $a>(9/4)\lambda$ and $54^{5/3}<\Rep$, the Newton's law, 
\begin{eqnarray}
	t_{\rm s}^{\rm Ne}= 9M/(2\pi \rho_{\rm g} v a^2)
		\ \propto M^{1/3} \rhoint^{\,2/3} v^{-1}
	.\label{eq:t_s_Ne}
\end{eqnarray}

The relative velocity between dust aggregates and gas is given as 
\begin{eqnarray}
	v = \sqrt{ v_{\rm B}^2 + v_{\rm tur}^2 + v_r^2 + v_{\phi}^2 }, 
\end{eqnarray}
where $v_{\rm B},\ v_{\rm tur},\ v_r,\ v_{\phi}$ are those induced by the Brownian motion, turbulence, radial drift, and azimuthal drift. 
The collision velocity between two dust aggregates is
\begin{eqnarray}
	\Delta v = \sqrt{\Delta v_{\rm B}^2 + \Delta v_{\rm tur}^2 + \Delta v_r^2 + \Delta v_{\phi}^2 }.
\end{eqnarray}
Each component of collision velocities depends on the properties of two colliding aggregates.

The velocities induced by the Brownian motion ($v_{\rm B}$ and $\Delta v_{\rm B}$) are given by
\begin{eqnarray}
	v_{\rm B} &=& \sqrt{\frac{8 k_{\rm B}T }{ \pi M } }=\sqrt{\frac{8 m_{\rm g} }{ \pi M } }c_{\rm s}, \\
	\Delta v_{\rm B} &=& \sqrt{\frac{8 (M_1+M_2) k_{\rm B}T }{ \pi M_1 M_2 } }= \sqrt{\frac{8m_{\rm g} }{ \pi\mu_{12} } } c_{\rm s},
	\label{eq:v_B}
\end{eqnarray}
where $M_1$ and $M_2$ are the masses of two colliding aggregates; and $\mu_{12}=M_1M_2/(M_1+M_2)$ is the reduced mass.

The radial drift velocity ($v_r$) is
\begin{eqnarray}
	v_{r} &=& 
	-2 
	\left( \frac{ \St/\left(1+\rho_{\rm d}/\rho_{\rm g} \right) }{1 + (\St/\left(1+\rho_{\rm d}/\rho_{\rm g} \right) )^2 } \right)\eta v_{\rm K} ,
	\label{eq:v_r}
\end{eqnarray}
where $\rho_{\rm d}=\rho_{\rm d,ch} + \rho_{\rm d,MA}$ \citep{Nakagawa+1986}. 
The value of $v_r$ is almost equal to 0 until dust aggregates become large enough to satisfy $\St\sim \rho_{\rm d}/\rho_{\rm g}$.
The collision velocity due to radial drift is the velocity difference between two colliding bodies, $\Delta v_r = |v_r(\St_1) - v_r(\St_2) |$.
Similarly, the azimuthal drift velocity ($v_{\phi}$) is given by
\begin{eqnarray}
	v_{\phi} &=& 
	\left(1- \frac{ (1+\rho_{\rm d}/\rho_{\rm g})^2 }{\left(1+\rho_{\rm d}/\rho_{\rm g} \right)^2 + \St^2 } \right)\eta v_{\rm K} .
	\label{eq:v_phi}
\end{eqnarray}
and $\Delta v_{\phi} = |v_{\phi}(\St_1) - v_{\phi}(\St_2) |$. 
The value of $v_{\phi}$ is almost equal to $0$ under $\St\ll\rho_{\rm d}/\rho_{\rm g}$. 
This is because aggregates and gas in the dense clump are in the near-Keplerian motion.

As described in Section \ref{sect:assump} (Assumption 3), we consider the two eddy models for $v_{\rm tur}$ and $\Delta v_{\rm tur}$.
There are three regimes in $v_{\rm tur}$, according to the relation between the turnover time and $t_{\rm s}$ \citep{Ormel&Cuzzi2007}.
The turnover time of the smallest eddies is $t_\eta = \Ret^{-1/2} t_{\rm L}$, where $\Ret$ is the turbulent Reynolds number, and $t_{\rm L}=\Omega_{\rm K}^{-1}$ is the turnover time of the largest eddy.
The turbulent Reynolds number is the ratio of the diffusion coefficient for the gas ($D_g=\delta v_g^2 t_{\rm L}$) to the molecular viscosity ($\nu_{\rm mol}$), 
where $\delta v_g^2=\alpha c_s^2$ is the mean-squared random velocity of the largest turbulent eddies.
Then, we can derive $t_\eta\simeq (\lambda_{\rm m}/\alpha h_{\rm g})^{1/2} \Omega_{\rm K}^{-1}\sim10^{-4} (\alpha/10^{-4})^{-1/2}\Omega_{\rm K}^{-1}$.
Consequently, $v_{\rm tur}$ is given by, 
\begin{eqnarray}
	&&(v_{\rm tur,L}^2+v_{\rm tur,S}^2)^{1/2} =
		\left\{
			\begin{array}{lr}
				\Ret^{1/4} \St \delta v_{\rm g},	& (\St < t_{\eta} \Omega_{\rm K}),\\
				1.7 \St^{1/2} \delta v_{\rm g},	& (t_{\eta} \Omega_{\rm K} \leq \St < 1),\\
				\left( 1+\frac{1}{1+\St} \right)^{1/2} \delta v_{\rm g},	& (1 \leq \St),
			\end{array}
		\right. 
		\label{eq:v_tur}
		\nonumber \\ \\
	&&v_{\rm tur,L} =
	\left\{
		\begin{array}{lr}
			\Ret^{1/4} \St \delta v_{\rm g},	& (\St < t_{\eta} \Omega_{\rm K}),\\
			\left( 1.6 + \frac{1}{2.6} \right)^{1/2} \St^{1/2} \delta v_{\rm g},	& (t_{\eta} \Omega_{\rm K} \leq \St < 1),\\
			0,	& (1 \leq \St).\\
		\end{array}
	\right. 
	\label{eq:v_tur1}
\end{eqnarray}
In the above equations, there is no difference between the whole and large eddy models when $\St < t_{\eta} \Omega_{\rm K}$.
This is simply because the turnover times of all eddies are longer than the stopping time and only $v_{\rm tur,L}$ is effective in the regime of $\St < t_{\eta} \Omega_{\rm K}$ under Assumption 3 in Section \ref{sect:assump}.
For the case that $t_{\eta} \Omega_{\rm K} \leq \St < 1$, both large and small eddies contribute to the turbulence-induced velocities,
which leads to different expressions between the whole and large eddy models.
For the case that $\St \geq1$ (i.e., $t_{\rm s}>t_{\rm L}$) there are no eddies whose turnover time is longer than $t_{\rm s}$.
In this regime, only $v_{\rm tur,S}$ is effective, and $v_{\rm tur,L}=0$ in the large eddy model.

The collision velocities induced by turbulence are 
\begin{eqnarray}
	\Delta v_{\rm tur} =
		\left\{
			\begin{array}{l}
				\Ret^{1/4} \delta v_{\rm g} (1-\epsilon)\St_1,	\hfill (\St_1 < t_{\eta} \Omega_{\rm K}),\\
				\left( 2.2 - \epsilon + \frac{2}{1+\epsilon} \left[ \frac{1}{2.6} + \frac{\epsilon^3}{1.6+\epsilon} \right]\right)^{1/2}
					\St_1^{1/2} \delta v_{\rm g},
					\\ \mbox{\hspace{0.4\hsize}} \hfill
					(t_{\eta} \Omega_{\rm K} \leq \St_1 < 1),\\
				\left( \frac{1}{1+\St_1} + \frac{1}{1+\St_1\epsilon} \right)^{1/2} \delta v_{\rm g},	\hfill (1 \leq \St_1),
			\end{array}
		\right.
		\label{eq:Delta_v_tur}
		\nonumber \\
\end{eqnarray}
in the whole eddy model, and 
\begin{eqnarray}
	\Delta v_{\rm tur} =
		\left\{
			\begin{array}{l}
				\Ret^{1/4} \delta v_{\rm g} (1-\epsilon)\St_1,	\hfill (\St_1 < t_{\eta} \Omega_{\rm K}),\\
				\left( \frac{1-\epsilon}{1+\epsilon} \right)^{1/2} \left( \frac1{2.6} -\frac{\epsilon^2}{1.6+\epsilon} \right)^{1/2} 
					\St_1^{1/2} \delta v_{\rm g},
					\\ \mbox{\hspace{0.4\hsize}} \hfill
					(t_{\eta} \Omega_{\rm K} \leq \St_1 < 1),\\
				0,	\hfill (1 \leq \St_1),
			\end{array}
		\right.
		\label{eq:Delta_v_tur1}
		\nonumber \\
\end{eqnarray}
in the large eddy model.
In the above equations, it is assumed that $\epsilon = \St_2/\St_1\leq1$ without loss of generality.
For CA-CA or MA-MA collisions, we also replace $1-\epsilon$ with 0.1 because of the internal density fluctuation of aggregates \citep{Okuzumi+2011}.

The whole eddy model is widely used in the previous studies of dust growth calculations.
Following \cite{Arakawa2017}, we adopt the minimum values of $v_{\rm tur}$ and $\Delta v_{\rm tur}$ among the above three regimes when computing the growth timescale in the whole eddy model.
\footnote{
This treatment would become important when aggregates move the above three regimes from one to another, where the expressions of $v_{\rm tur}$ and $\Delta v_{\rm tur}$ are not so accurate \citep{Ormel&Cuzzi2007}.
}

\subsection{Fragmentation velocity}
\label{sect:v_crit}

The outcome of collisions between two aggregates is important for the accretion mode.
We determine the collision outcome by comparing the collision velocity and the fragmentation velocity.

The fragmentation velocity can be derived from the energy balance between the impact energy ($E_{\rm imp}$) and the total contact energy of colliding aggregates.
Mathematically, $E_{\rm imp}=f_{\rm cr}N_{\rm tot} E_{\rm break}$, 
where $f_{\rm cr}=30$ is the factor for the critical breaking energy; $N_{\rm tot}$ is the total number of dust particles contained in two colliding aggregates; and $E_{\rm break}$ is the energy necessary to completely break the contact of particles in aggregates \citep{Wada+2007}. 
The value of $E_{\rm break}$ depends on the material properties and the size of monomer grains.
We briefly describe how $E_{\rm break}$ can be derived:
according to the elastic contact theory, $E_{\rm break}=1.54 F_{\rm c}\delta _{\rm c}$, 
where $F_{\rm c}=3\pi \gamma R$ is the maximum force needed to separate two contacting particles; and $\delta _{\rm c}=(9/16)^{1/3} (9\pi\gamma R^2/E^*)^{2/3}/(3R)$ is the critical separation. 
Considering collisions between aggregates composed of the same sized particles, $R=a_{\rm mat}/2$ 
and $E^{*-1}=(1-\nu^2)E^{-1}$, where $\nu=0.17$ and $E=54 \mbox{ G Pa}$ for silicate particles \citep{Dominik&Tielens1997}.

For MA-MA collisions, the impact energy is $E_{\rm imp}=0.5\mu_{\rm MA} v_{\rm frag,MA-MA}^2$, 
where $\mu_{\rm MA}=M_{\rm MA}/2$ is the reduced mass; and $v_{\rm frag,MA-MA}$ is the fragmentation velocity.
Then $v_{\rm frag,MA-MA}$ is written as
\begin{eqnarray}
	v_{\rm frag,MA-MA}
		&=& \sqrt{ \frac{2 f_{\rm cr}(2M_{\rm MA}/m_{\rm mat}) E_{\rm break}}{\mu_{\rm MA}} } \nonumber \\
		&=& 1.3\times10^4 \left( \frac{a_{\rm mat}}{2.5\times 10^{-7}\mbox{ cm}} \right)^{-5/6} \mbox{ cm s}^{-1},
		\label{eq:v_cr_ma-ma}
		\nonumber \\
\end{eqnarray}
where the dependence on $M_{\rm MA}$ is cancelled; and $v_{\rm frag,MA-MA}$ becomes a function only of the size of matrix grains.

For CA-CA collisions, both the experimental study \citep{Beitz+2012} and numerical study \citep{Gunkelmann+2017} suggest 
that matrix components in CAs can dissipate the collision energy and CAs can merge with much higher velocity collisions than that of pure chondrule-aggregates.
The fragmentation velocity for CA-CA collisions ($v_{\rm frag,CA-CA}$) is determined by matrix grains in CAs since $a_{\rm mat}/a_{\rm ch}\ll1$ and $(E_{\rm break}/M)^{1/2}\propto a^{-5/6}$.
Following \cite{Arakawa2017}, $v_{\rm frag,CA-CA}$ can be estimated by counting the number of matrix grains in CAs, which can be given as
\begin{eqnarray}
	v_{\rm frag,CA-CA}
		&=& \sqrt{ \frac{2 f_{\rm cr}(2(1-\chi) M_{\rm CA}/m_{\rm mat}) E_{\rm break}}{\mu_{\rm CA}} } \nonumber \\
		&=& 1.3\times10^4 (1-\chi)^{1/2}\left( \frac{a_{\rm mat}}{2.5\times 10^{-7}\mbox{ cm}} \right)^{-5/6} 
		\label{eq:v_cr_ca-ca}
		\nonumber \\&&
		\hfill\mbox{ cm s}^{-1},
\end{eqnarray}
where $\mu_{\rm CA}=M_{\rm CA}/2$.
This expression is applicable to collisions between the same compound aggregates that are composed of different materials with significantly different sizes of monomers.

The fragmentation velocity for CA-MA collisions can also be estimated in the same way as above
\begin{eqnarray}
	&&v_{\rm frag,CA-MA}
		\nonumber \\
		&=& \sqrt{ \frac{2 f_{\rm cr}((1-\chi) M_{\rm CA}/m_{\rm mat} + M_{\rm MA}/m_{\rm mat}) E_{\rm break}}{\mu_{\rm CM}} } \nonumber \\
		&=& 1.3\times10^4 \left(\frac{(1-\chi) M_{\rm CA}+M_{\rm MA}}{4\mu_{\rm CM}}\right)^{1/2}
		\nonumber \\&&\times 
		\left( \frac{a_{\rm mat}}{2.5\times 10^{-7}\mbox{ cm}} \right)^{-5/6} 
		\mbox{cm s}^{-1},
		\label{eq:v_cr_ca-ma}
\end{eqnarray}
where $\mu_{\rm CM}$ is the reduced mass of CAs and MAs.
This expression can be used for collisions between different compound aggregates that are composed of significantly different monomer sizes.

\section{Results of the whole eddy model}
\label{sect:resut_weddy}

\subsection{Overall features of aggregate growth}
\label{sect:fiducial}

\begin{figure*}[hbt]
	\plotone{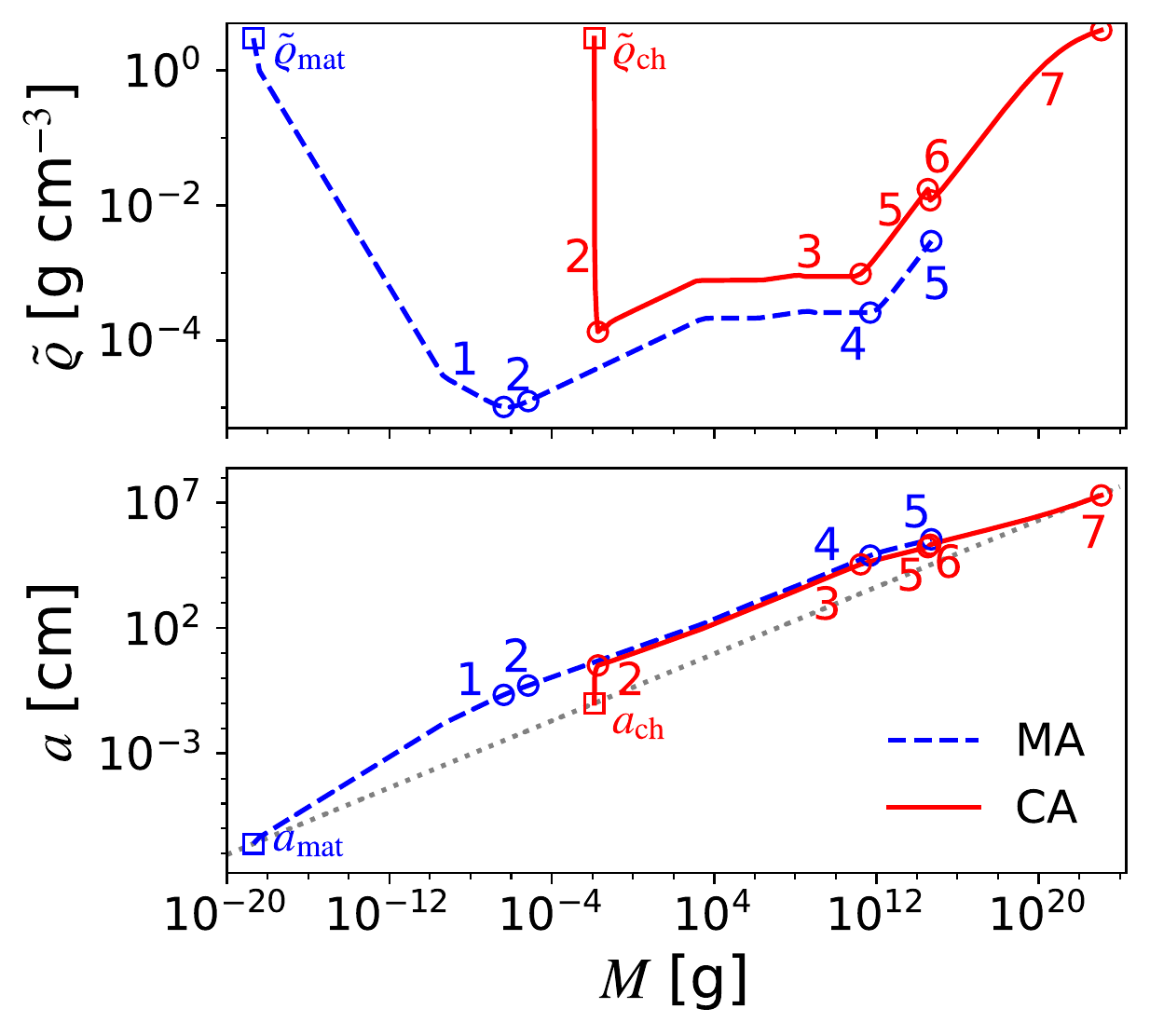}
	\caption{
		The growth of CAs and MAs on the $M$--$\rhoint$ (top) and $M$--$a$ (bottom) diagram for our fiducial case in the whole eddy model.
		The red solid lines denote the internal densities and radii of CAs and the blue dashed lines are for those of MAs.
		Aggregate growth can divide into 7 stages and proceeds in the order (from 1 to 7).
		The initial internal densities and radii of chondrules and matrix are plotted as the square points ($\rhoint_{\rm ch}=\rhoint_{\rm mat}=3~\mbox{g~cm}^{-3}$, $a_{\rm ch}=0.1$~cm and $a_{\rm mat}=2.5\times10^{-7}$~cm).
		In each stage, aggregates grow up to the circle points with the corresponding number.
		The dotted gray line in the bottom panel is the mass-radius relation calculated from the constant internal density of $\rhoint=3~\mbox{g~cm}^{-3}$.
		}
	\label{Fig:m_rho_a_v12}
\end{figure*}

\begin{figure*}[hbt]
	\plotone{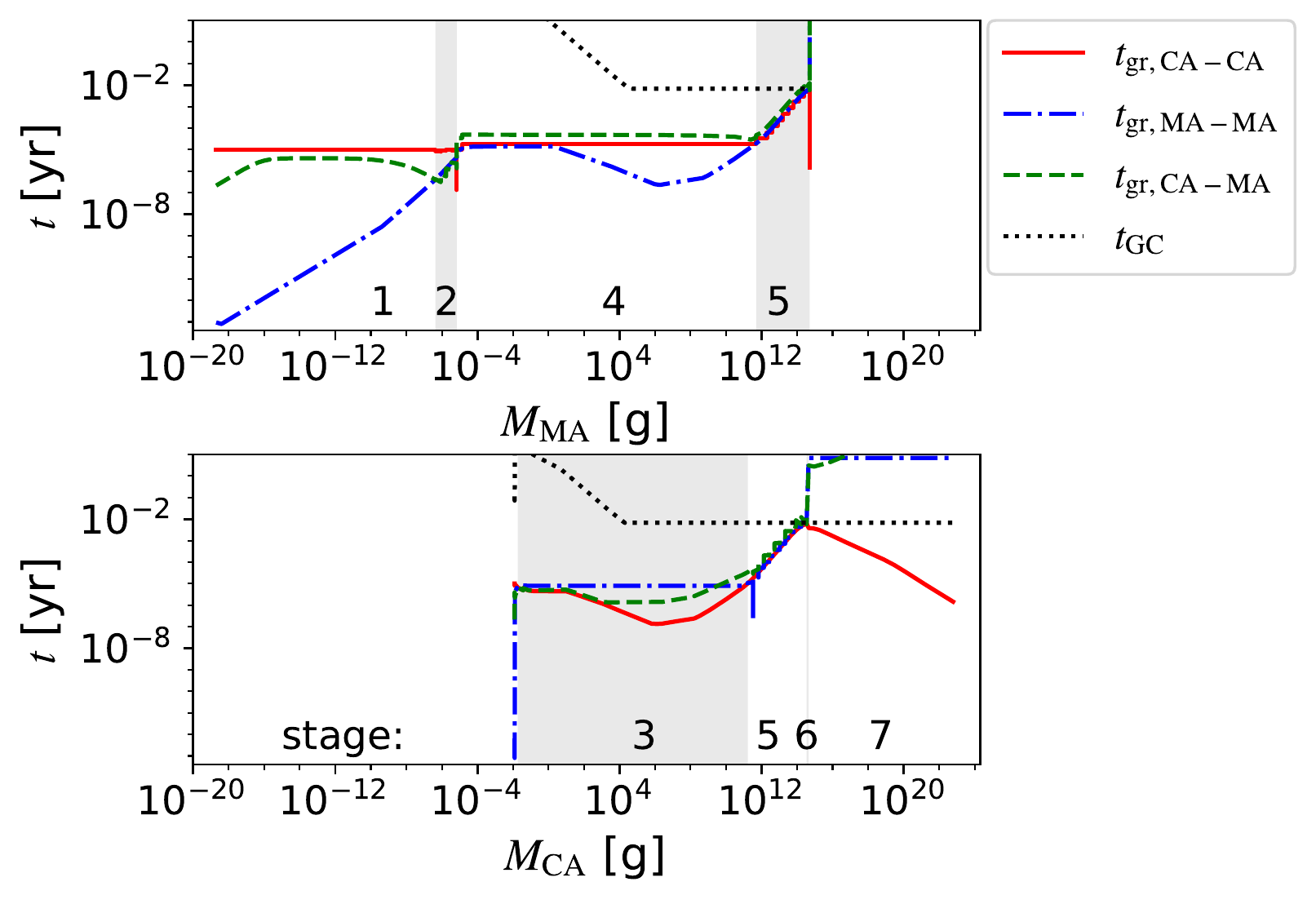}
	\caption{
		Timescales are shown as functions of $M_{\rm CA}$ and $M_{\rm MA}$ in our fiducial case of the whole eddy model. 
		The growth timescale between CA-CA collisions, MA-MA collisions, and CA-MA collisions are plotted as the red solid, blue dashed-dotted, and green dashed lines, respectively.
		The collapse timescale ($t_{\rm GC}$) is also plotted as the black dotted line.
		The evolution stages are distinguished by the white and gray regions with the stage numbers.
		Aggregates evolve in the order of the stage number.
	}
	\label{Fig:mCA_tgrow_tacc_v12}
\end{figure*}

We first present the results of aggregate growth calculations for our fiducial case 
($\rho_{\rm d,ch}=10^{-4} \mbox{~g~cm}^{-3},\ \chi_{\rm clump}=1/2$, $a_{\rm ch}=10^{-1}$~cm, and $a_{\rm mat}=2.5\times10^{-7}$~cm) in the whole eddy model.
The initial total mass of chondrules and matrix grains in the dense clump is $1.8\times10^{23}$ g.

Figure \ref{Fig:m_rho_a_v12} shows the internal density and radius evolutions of CAs and MAs as functions of their masses.
The evolution of MAs is similar to the results of \cite{Kataoka+2013, Arakawa&Nakamoto2016}.
The growth mode is accretion for this case.
Figure \ref{Fig:mCA_tgrow_tacc_v12} shows the timescales of aggregate growth and gravitational collapse as functions of $M_{\rm CA}$ and $M_{\rm MA}$.
We confirm that the collisional growth timescale is the shortest over the entire course of this simulation.
We find that the mass evolutions of CAs and MAs can divide into 7 stages.
These stages are distinguished by the shortest growth timescales, and the mass growth of CAs and MAs proceeds in their order (from stage 1 to stage 7).
The detailed explanation of aggregate growth at each stage is described in Appendix \ref{sec:results_evolution}.
Table \ref{table:collisions} summarizes the detailed properties of the dominant collisions at each stage.

\begin{figure}[ht]
	\plotone{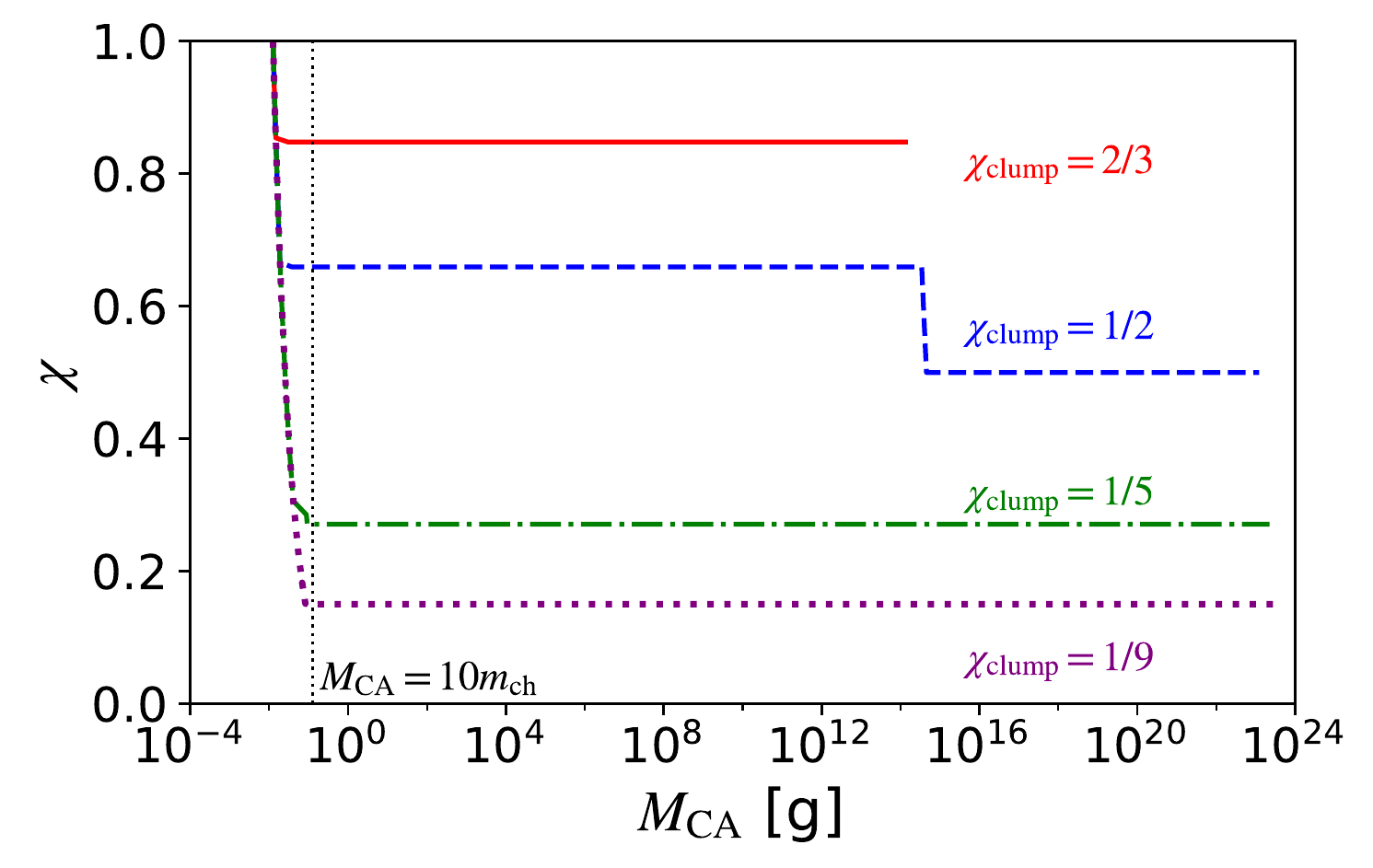}
	\caption{
		The evolution of the chondrule mass fraction in a CA ($\chi$) as a function of $M_{\rm CA}$ in the whole eddy model.
		We plot the case that $\rho_{\rm d,ch}=10^{-4} \mbox{~g~cm}^{-3}$.
		The red solid line represents the result for the case that $\chi_{\rm clump}=2/3$, 
		the blue dashed line is for the case that $\chi_{\rm clump}=1/2$, 
		the green dash-dotted line is for the case that $\chi_{\rm clump}=1/5$, and the purple dotted line is for the case that $\chi_{\rm clump}=1/9$.
	}
	\label{Fig:mCA_ch_ra_rho_ra_v12}
\end{figure}

We here briefly describe the evolution of aggregates in our fiducial case.
In stage 1, MAs form and grow
(Figure \ref{Fig:collisions} (a) and (b)).
While the mass of MAs is $4.3\times10^{-7}$~g at the end of this stage, 
the internal density decreases down to $\rhoint_{\rm MA}= 10^{-5}~\mbox{g~cm}^{-3}$.
The size of MAs becomes $a_{\rm MA}=$0.22~cm, which is larger than the size of chondrules (0.1~cm, Figure \ref{Fig:m_rho_a_v12}).
In stage 2, chondrules accrete MAs and become CAs
(Figure \ref{Fig:collisions} (c));
CAs are composed of single chondrules that are covered by the fluffy matrix components. 
The internal density of CAs decreases to $\rhoint_{\rm CA}= 1.3\times10^{-4}~\mbox{g~cm}^{-3}$ (Figure \ref{Fig:m_rho_a_v12}).
The size of CAs is $a_{\rm CA}=3.1$~cm, which is about 30 times larger than that of single chondrules.
At the end of this stage, CAs accrete 52~\% of MAs and $\chi$ of CAs becomes 0.66.
The evolution of $\chi$ is plotted in Figure \ref{Fig:mCA_ch_ra_rho_ra_v12} (see the blue dashed line).
In stage 3, CAs grow via CA-CA collisions
(Figure \ref{Fig:collisions} (d) and (e)).
This occurs because enough amount of matrix grains are accreted onto chondrules, and hence the merger of CAs becomes possible even when $\St_{\rm CA}\sim1$.
As CAs grow, their internal density increases due to compression via the ram pressure.
This compression depends on the relative velocity to gas and the stopping time (Equation (\ref{eq:rho_ram})).
This is why the internal densities of CAs do not increase monotonically (Figure \ref{Fig:m_rho_a_v12}).
In stage 4, MAs grow via MA-MA collisions.
This growth is the same as that of CAs in stage 3.
The mass of MAs becomes larger than that of CAs at the end of this stage (Table \ref{table:collisions}).
In stage 5, both CAs and MAs grow.
The compression becomes more efficient because the self-gravity of aggregates is now dominant.
In stage 6, CA-MA collisions occur again and CAs become more massive (Figure \ref{Fig:collisions} (f)).
We find that CAs accrete all MAs in this stage.
In stage 7, CAs undergo runaway growth via CA-CA collisions.
Note that stages 2 and 6 are the most critical for investigating the chondrule mass fraction ($\chi$) in CAs because of CA-MA collisions (see below).

\subsection{Effects of aggregate growth on $\chi$}\label{sect:chi_v12}

In this section, we present a possible relationship between aggregate growth and the chondrule mass fraction in CAs ($\chi$).

Figure \ref{Fig:mCA_ch_ra_rho_ra_v12} shows the results for the fiducial case (see the blue dashed line).
We find that the value of $\chi$ changes twice at $M_{\rm CA} \lesssim 0.1$ g and $M_{\rm CA} \sim 10^{15}$ g.
These two jumps correspond to CA-MA collisions (Table \ref{table:collisions}, see also stages 2 and 6 in Appendix \ref{sec:results_evolution})
and lead to dilution of the chondrule abundance in CAs.
This dilution would contain profound insights about the formation mechanisms and conditions of CPBs.
Furthermore, this would be a unique feature of the accretion mode.
We thus discuss this process in detail below.

We first refer to the internal structures set by stages 2 and 6 as the small and large scale distributions, respectively.
We then consider the small scale distribution.
This scale is the outcome of collisions between chondrules and MAs.
Accordingly, the resulting formed CAs are viewed as single chondrules covered by the fluffy matrix component.
Our results show that $M_{\rm CA}$ becomes $\lesssim 10m_{\rm ch}$ at the end of stage 2 for all the calculations (see Figure \ref{Fig:mCA_ch_ra_rho_ra_v12}).
This occurs because the total masses of chondrules and matrix grains are initially comparable for all the cases in our setup ($1/9\leq\chi_{\rm clump}\leq4/5$, see Table \ref{table:parameters}).
This mass estimate would be useful for characterizing the small scale distribution.
One can estimate the thickness of the matrix components ($a_{\rm th,mat}$) as
\begin{eqnarray}
	\label{eq:a_th_mat}
	a_{\rm th,mat} &\sim& \left(\frac{ m_{\rm ch}/\chi }{ 4\pi \rhoint_{\rm ch}/3 } \right)^{1/3} - a_{\rm ch}
	\nonumber \\
	&= & a_{\rm ch}(\chi^{-1/3}-1),
\end{eqnarray}
where the total mass of the CAs is given as $M_{\rm CA}=m_{\rm ch}/\chi$ (see Equation (\ref{eq:chi_def})); and $\chi$ is the value at $M_{\rm CA}\simeq 10m_{\rm ch}$, where $\chi_{\rm clump} \leq \chi \leq 1$.
It is also assumed in the above equation that the internal density of the matrix component becomes similar to that of chondrules due to compression.
This assumption would be justified because the CAs experience further growth (see Figures \ref{Fig:m_rho_a_v12} and \ref{Fig:mCA_tgrow_tacc_v12}).
Thus, we find that the corresponding thickness of the matrix surface layer is $1.4\times10^2~\mu$m for our fiducial case.
More importantly, our results imply that the spatial distribution of chondrules and matrix would be characterized by the matrix surface layer around chondrules and might be identical on the small scale.
Note that it would be reasonable to consider that this small scale distribution will be kept in the subsequent growth (Section \ref{sect:col_growth}, Panels (d) and (e) of Figure \ref{Fig:collisions}).

We now discuss the large scale distribution.
Our calculations show that this distribution is the outcome of collisions between CAs and MAs, both of which are similar in size (Panel (f) of Figure \ref{Fig:collisions}).
This suggests that the large scale distribution may be characterized by two regions;
one region is the collection of single chondrules covered by the matrix component (as in the small scale distribution).
We expect that this region should be chondrule-rich and is $\sim1$~km in size, 
which is estimated from the size of CAs at the end of stage 5 (Figure \ref{Fig:m_rho_a_v12}).
The other region is matrix-rich and has a size similar to the chondrule-rich region.
Thus, our results suggest that the spatial distribution of chondrules and matrix may become inhomogenous on the large scale. 

In summary, our calculations show that the {\it local} internal distributions of chondrules and matrix in the clump may be different from the homogenous one 
when the accretion mode is realized.
Note that inhomogeneity on the large scale distribution cannot be captured properly by our definition of $\chi$,
since it represents the {\it bulk} chondrule abundance.
In the following sections, we focus on the $\chi$ value of $M_{\rm CA}=10m_{\rm ch}$ and the final value of $M_{\rm CA}$ to discuss the small and large scale distributions, respectively.
For brevity, we call CAs that have the final value of $M_{\rm CA}$ as CPB.

\subsection{Parameter study}
\label{sec:param_study}

As discussed above, the accretion mode may generate two scale distributions of chondrules and matrix within the clump.
We here conduct a parameter study and examine how $\chi$ will vary,
by changing the values of $\rho_{\rm d,ch}$, $\chi_{\rm clump}$, $a_{\rm ch}$, and $a_{\rm mat}$ (Table \ref{table:parameters}).

\subsubsection{The dependence on the mass densities of chondrules and matrix}
\label{sect:dependence_rho_ratio}

\begin{figure}[ht]
	\plotone{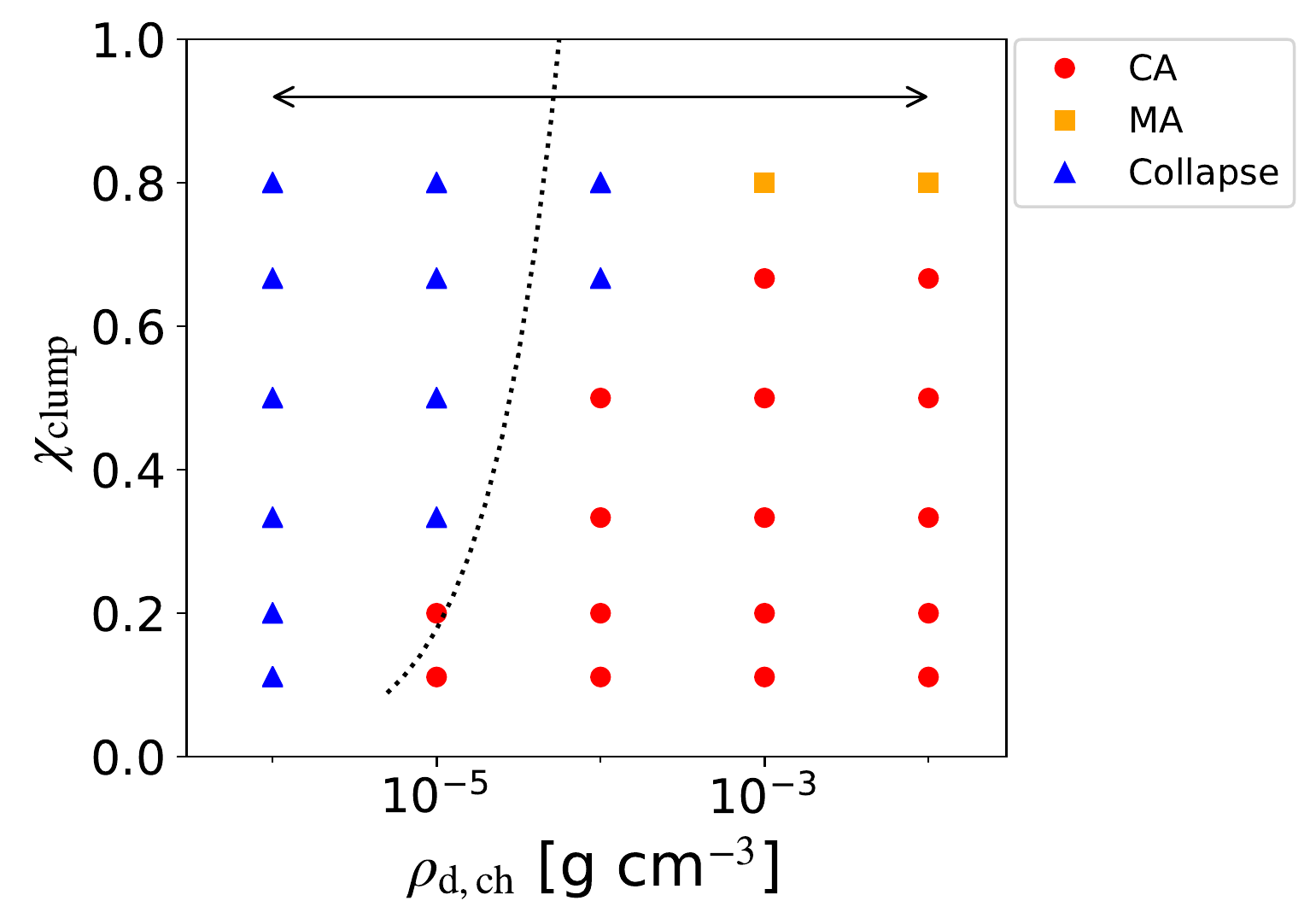}
	\caption{
		The growth modes on the $\rho_{\rm d,ch}$ -- $\chi_{\rm clump}$ diagram in the whole eddy model.
		The red circles denote the accretion mode of CAs, and the orange squares are for the accretion mode of MAs.
		The blue triangles are for the collapse mode.
		The dotted line is given by Equation (\ref{eq:t_ff_t_gr}) with $\chi=\chi_{\rm clump}$.
		The arrow indicates the density range inferred by \cite{Alexander+2008}.
	}
	\label{Fig:rho_ch_rho_mat}
\end{figure}

\begin{figure}[ht]
	\plotone{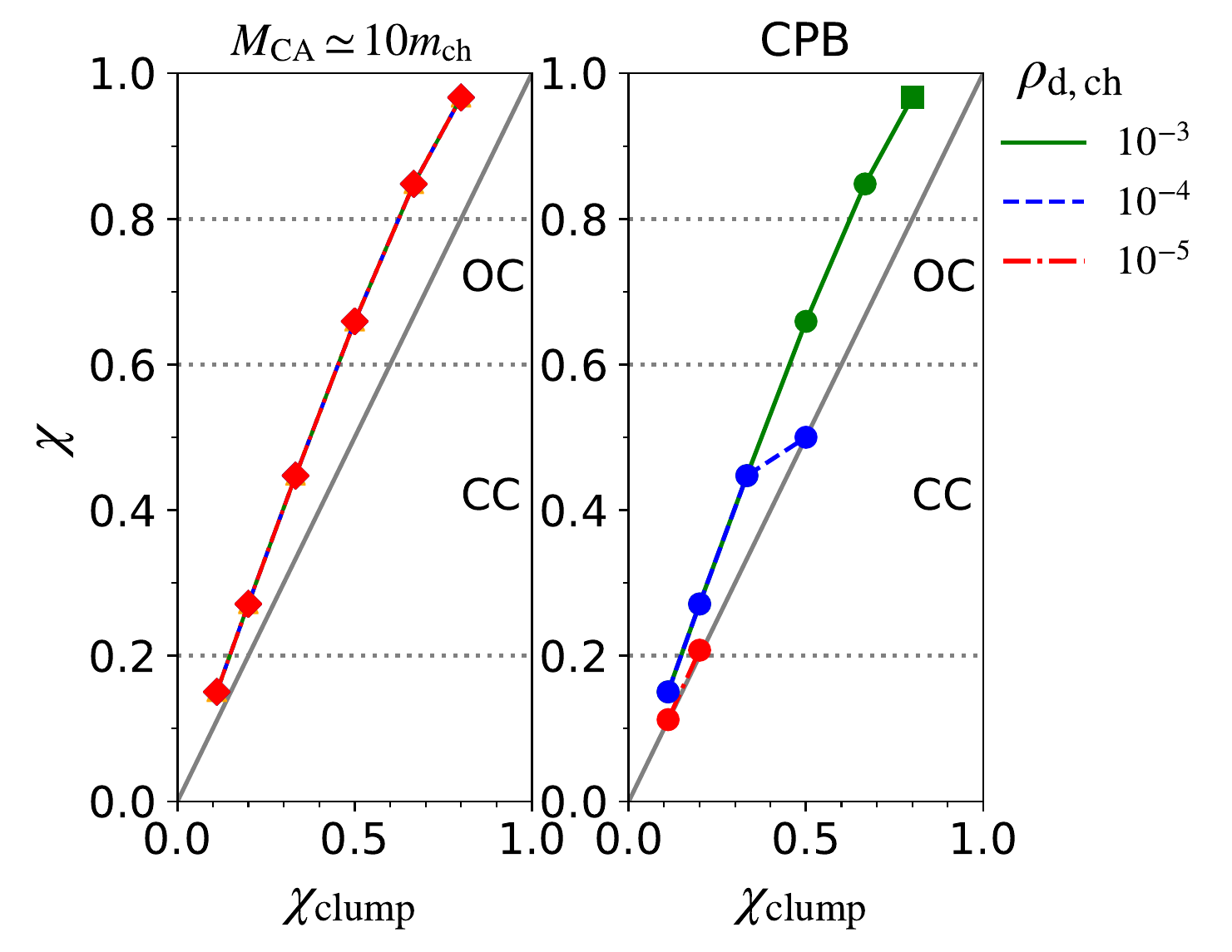}
	\caption{
		The chondrule mass fraction of CAs ($\chi$) as a function of that of the dense clump ($\chi_{\rm clump}$) in the whole eddy model.
		The left panel shows the fraction when their masses reach $\simeq 10m_{\rm ch}$, and the right one is the final chondrule fraction in CAs for the accretion mode.
		In the left panel, each point is plotted by diamonds, and all the lines match with each other.
		In the right panel, we use the circles for the CA accretion mode and the squares for the MA accretion mode.
		We do not plot the results that end up with the collapse mode.
		The green symbols and solid lines denote the results for the case that $\rho_{\rm d,ch}=10^{-3} \mbox{~g~cm}^{-3}$, 
		the blue ones are for the case that $\rho_{\rm d,ch}=10^{-4} \mbox{~g~cm}^{-3}$, and the red ones are for the case that $\rho_{\rm d,ch}=10^{-5} \mbox{~g~cm}^{-3}$.
		The gray solid lines denote that $\chi=\chi_{\rm clump}$.
		The horizontal dotted lines encompassing CC represent the typical range of the chondrule fraction in carbonaceous chondrites and those encompassing OC are for the typical range in ordinary chondrites.
	}
	\label{Fig:chi_rho_ra_v12}
\end{figure}

In this section, we change the values of $\rho_{\rm d,ch}$ and $\chi_{\rm clump}$.

We first discuss how the growth mode (accretion vs collapse) of the clump is determined as functions of these two parameters.
The results are plotted in Figure \ref{Fig:rho_ch_rho_mat}.
Our results show that the growth mode tends to be accretion when $\rho_{\rm d,ch}$ is high;
the accretion mode covers more than half part of the log-scaled density range inferred from Semarkona ordinary chondrite
\citep[][see the arrow range in Figure \ref{Fig:rho_ch_rho_mat}]{Alexander+2008}.
Thus, the accretion of chondrules and matrix grains is important if CPBs are born out of dense clumps in gas disks.

The transition from the accretion mode to the collapse one arises at $\rho_{\rm d,ch}\sim 10^{-5}$ -- $10^{-4}\mbox{~g~cm}^{-3}$ (Figure \ref{Fig:rho_ch_rho_mat});
while the accretion mode is realized at small $\chi_{\rm clump}$, 
the collapse one occurs for large values of $\chi_{\rm clump}$.
This can be understood as follows.
Based on our aggregate growth calculations, the growth timescale becomes the longest when runaway growth begins (see Figure \ref{Fig:mCA_tgrow_tacc_v12} and Appendix \ref{sec:results_evolution}).
Assuming that $\Theta_{\rm CA-CA}=1$ and that $\Delta v_{\rm CA-CA}$ is equal to the escape velocity of CAs ($v_{\rm esc}$), 
we can find out the value of $\rho_{\rm d,ch}$ that satisfies the condition that $t_{\rm GC}<t_{\rm gr,CA-CA}$:
\begin{eqnarray}
	t_{\rm ff} &<& \frac{ \chi M_{\rm CA} }{\rho_{\rm d,ch} \pi (4 a_{\rm CA}^2) 2 v_{\rm esc} } \nonumber \\
	\Leftrightarrow\ 
	\rho_{\rm d,ch} &<& 5.6\times10^{-5} \frac{\chi^2}{\chi_{\rm clump}} \left( \frac{\rhoint_{\rm CA}}{10^{-2} \mbox{~g~cm}^{-2}} \right) .\nonumber \\
	\label{eq:t_ff_t_gr}
\end{eqnarray}
Therefore, the critical values of $\rho_{\rm d,ch}$ are functions of $\chi_{\rm clump}$, $\chi$, and $\rhoint_{\rm CA}$.
Besides, our aggregate growth calculations suggest that $\chi$ and $\rhoint_{\rm CA}$ are functions of $\chi_{\rm clump}$.
As an example, Figure \ref{Fig:mCA_ch_ra_rho_ra_v12} shows that as $\chi_{\rm clump}$ decreases, the value of $\chi$ becomes smaller 
and CAs accrete more MAs efficiently.
Consequently, the critical value of $\rho_{\rm d,ch}$ varies with changing $\chi_{\rm clump}$.
This is the origin of the boundary feature at $\rho_{\rm d,ch}\sim 10^{-5}$ -- $10^{-4} \mbox{~g~cm}^{-3}$.
Furthermore, we can roughly estimate the location of the boundary by substituting $\chi=\chi_{\rm clump}$ in Equation (\ref{eq:t_ff_t_gr}),
although the relationship between $\chi$ and $\chi_{\rm clump}$ is complicated (Figure \ref{Fig:chi_rho_ra_v12}).
This estimation works well especially for a small value of $\chi_{\rm clump}$, where $\chi \simeq \chi_{\rm clump}$.
Note that the collapse mode is realized before CAs undergo runaway growth.
In collapse mode, the mass and size of CAs are roughly estimated by those values in the onset of CA runaway growth.
At that time, CAs have the mass of $4.5\times 10^{14}$~g and the radius of 2.1~km in our fiducial case (Figures \ref{Fig:m_rho_a_v12} and \ref{Fig:mCA_tgrow_tacc_v12}).
The mass and size of CAs in collapse mode decrease as $\rho_{\rm d,ch}$ and $\chi_{\rm clump}$ decrease.

Figure \ref{Fig:rho_ch_rho_mat} also shows that when $\chi_{\rm clump}=4/5$ and $\rho_{\rm d,ch}\geq 10^{-3} \mbox{~g~cm}^{-3}$, 
the growth mode becomes accretion of MAs (rather than CAs).
This is the outcome that MAs grow quickly before CA growth becomes effective.
More specifically, we find that CAs can contain only small fractions of matrix components in the high chondrule abundance environments (also see Figure \ref{Fig:chi_rho_ra_v12}).
This leads to the much higher internal densities of CAs than those of MAs, and CAs have a much smaller cross-section than MAs.
Eventually, $M_{\rm MA}$ grows faster than $M_{\rm CA}$ in stage 5 (Table \ref{table:collisions}).
MAs undergo runaway growth via MA-MA collisions for this case.

We now discuss the chondrule mass fraction in CAs ($\chi$) for the accretion mode.
We plot the values of $\chi$ at $M_{\rm CA}\simeq10m_{\rm ch}\simeq 0.1$ g and those at CPB on the left and right panels of Figure \ref{Fig:chi_rho_ra_v12}, respectively.
We here consider the cases that $\rho_{\rm d,ch}=10^{-3},\ 10^{-4},$ and $10^{-5}\mbox{~g~cm}^{-3}$.
Our results show that the value of $\chi$ at $M_{\rm CA}\simeq10m_{\rm ch}$ tends to be high for a high value of $\chi_{\rm clump}$ (see the left panel).
This is because while both MA-MA and CA-MA collisions occur at the early stage of aggregate growth, 
CAs become more chondrule-rich when the initial abundance of chondrules is higher in the clump.
Accordingly, the resulting slope becomes steeper than $\chi=\chi_{\rm clump}$.
We find that the maximum difference between $\chi_{\rm clump}$ and $\chi$ at $M_{\rm CA}\simeq10m_{\rm ch}$ is 18.1\% at $\chi_{\rm clump} =2/3$.

The mass fractions of chondrules at CPB exhibit some differences, depending on model parameters (see the right panel of Figure \ref{Fig:chi_rho_ra_v12}).
These fractions tend to be closer to $\chi_{\rm clump}$ than those at $M_{\rm CA}\simeq10m_{\rm ch}$.
This is the outcome of CA-MA collisions in stage 6, which leads to further dilution of the chondrule abundance in CPBs.
In fact, the chondrule mass fractions at CPBs become equal to $\chi_{\rm clump}$ 
for the case that $\rho_{\rm d,ch}=10^{-5} \mbox{~g~cm}^{-3}$ and the fiducial case ($\rho_{\rm d,ch}=10^{-4} \mbox{~g~cm}^{-3}$ and $\chi_{\rm clump}=1/2$). 
It is, however, interesting that the chondrule fractions at CPB do not change from the values at $M_{\rm CA}\simeq10m_{\rm ch}$ 
for the cases that $\rho_{\rm d,ch}=10^{-4} \mbox{~g~cm}^{-3}$ and $\chi_{\rm clump}<1/2$ and that $\rho_{\rm d,ch}=10^{-3} \mbox{~g~cm}^{-3}$.
In these cases, CA-MA collisions do not occur in stage 6 and CAs can keep the high value of the chondrule fractions at CPB.
The occurrence of CA-MA collisions is regulated both by the condition of stage 6 that depends on $\rho_{\rm d}$ (Appendix \ref{sect:s6}) and by the chondrule fraction in CAs through $t_{\rm gr,CA-MA}$.
When CAs do not experience CA-MA collisions in stage 6, 
the maximum difference between the chondrule fractions in CPBs and in clumps is 18.1\% at $\chi_{\rm clump} =2/3$ in $\rho_{\rm d,ch}=10^{-3} \mbox{~g~cm}^{-3}$.

We also change the location of clumps ($r$) from 1 au to 5 au as a separate parameter study.
We find that the chondrule fraction takes almost the same value even if $r$ is altered.
Since the $r$ dependence is minor, our results at $r$ = 2au can be applied to dense clumps at different locations.
Note that there is some difference in growth path when $r$ varies.
As $r$ increases, the gas mass density decreases and 
the evolution of the Stokes number of aggregates becomes different (see Equations (\ref{eq:t_s_Ep}), (\ref{eq:t_s_St}), (\ref{eq:t_s_Al}), (\ref{eq:t_s_Ne})).
Consequently, higher $\rho_{\rm d,ch}$ and smaller $\chi_{\rm clump}$ are needed to establish CA-MA collisions in stage 6 for a larger value of $r$ .

We now compare our results with the present chondrule fractions found in chondrites.
Our parameter study suggests that 
the chondrule fractions in ordinary chondrites can be reproduced with the range of $0.5\lesssim \chi_{\rm clump}\lesssim 0.67$ in both small and large scale distributions.
For carbonaceous chondrites, it is by $0.12< \chi_{\rm clump}<0.5$.

In summary, our calculations show that the accretion mode is realized when $\rho_{\rm d,ch}\gtrsim10^{-5}~\mbox{g~cm}^{-3}$.
Furthermore, chondrules and matrix grow via collisions in all cases.
This suggests that most chondrules in chondrites are surrounded by the matrix layer on the small scale.
The dependence of $\chi$ on $\chi_{\rm clump}$ at $10m_{\rm ch}$ is steeper than $\chi=\chi_{\rm clump}$.
When $\rho_{\rm d,ch}\gtrsim10^{-2}~\mbox{g~cm}^{-3}$, CA-MA collisions in stage 6 do not occur and the final CAs are the collection of chondrules covered by matrix surface layers.

\subsubsection{The dependence on chondrule size}\label{sect:dependence_chondrule_size}

\begin{figure}[ht]
	\plotone{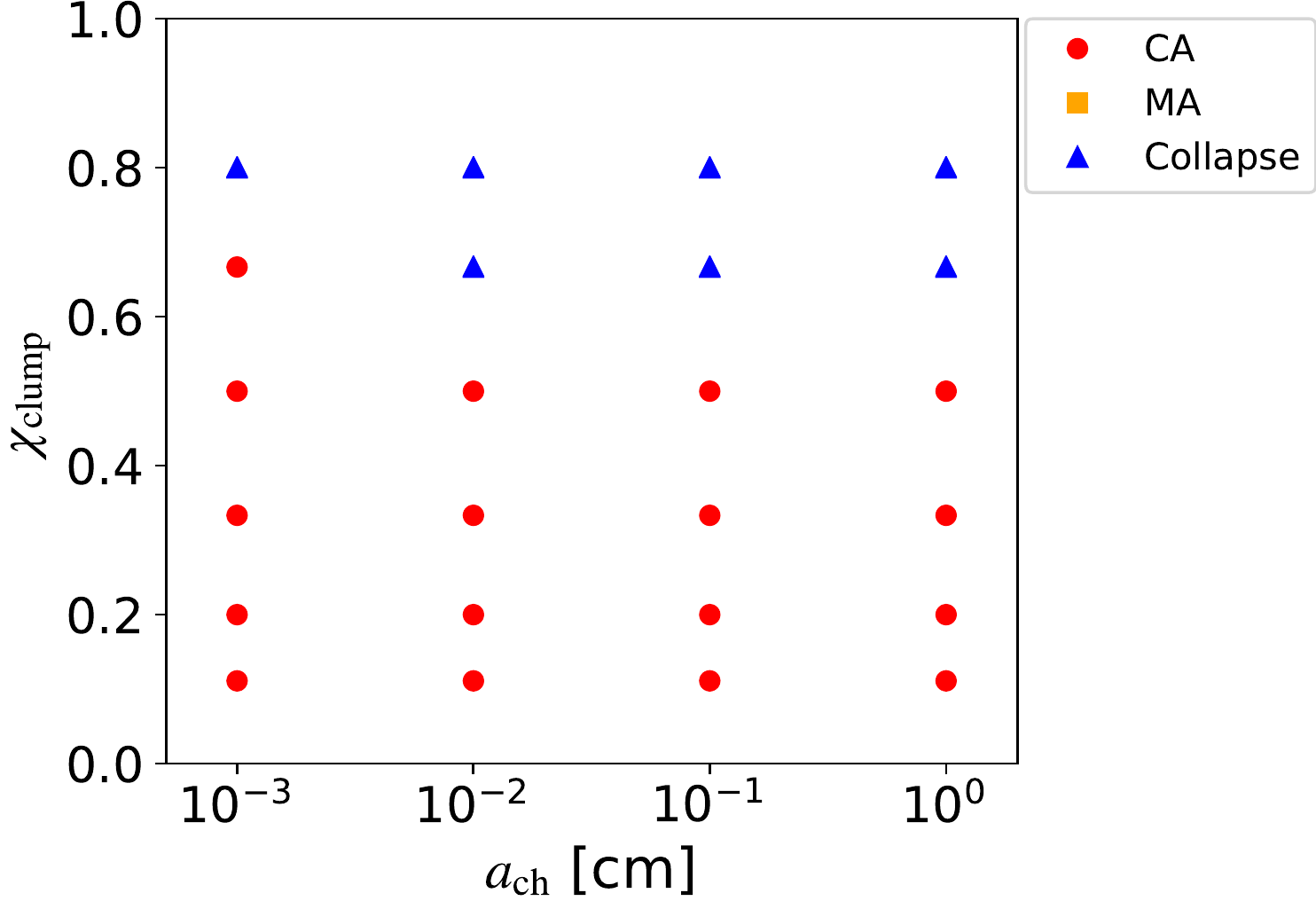}
	\caption{
		The growth modes on the $a_{\rm ch}$ -- $\chi_{\rm clump}$ diagram in the whole eddy model (as in Figure \ref{Fig:rho_ch_rho_mat}).
	}
	\label{Fig:ach_rho_mat_state_v12}
\end{figure}

In this section, we examine the effect of $a_{\rm ch}$ on the growth mode and the resulting value of $\chi$.
It can be expected that the variation of $a_{\rm ch}$ provides a considerable impact on the results.
This is because $t_{\rm gr,CA-MA}$ depends on $a_{\rm ch}$ in stage 2, where $\chi$ at $M_{\rm CA}\simeq10m_{\rm ch}$ is determined.
Here, we treat $a_{\rm ch}$ and $\chi_{\rm clump}$ as parameters while $\rho_{\rm d,ch}=10^{-4} \mbox{~g~cm}^{-3}$ and $a_{\rm mat}=2.5\times10^{-7}$ cm are fixed.

Figure \ref{Fig:ach_rho_mat_state_v12} shows the growth modes as a function of $a_{\rm ch}$. 
Our results indicate that most of the parameter space is covered by accretion.
The range of the accretion mode is the largest for the case that $a_{\rm ch}=10^{-3}$ cm (i.e., $m_{\rm ch}=1.26\times10^{-8}$~g).
This is because such small chondrules obey the Brownian motion, and the dependence of $t_{\rm gr, CA-CA}$ becomes the same as that of $t_{\rm gr,MA-MA}$.
Consequently, the CA-MA, CA-CA, and MA-MA collisions occur simultaneously after stage 1, and CAs accrete all MAs until $M_{\rm CA}\sim 10^{-6}$ g ($\simeq 100 m_{\rm ch}$).
This leads to $t_{\rm gr,CA-CA}$ that is shorter than $t_{\rm GC}$ for many cases.

\begin{figure}[ht]
	\plotone{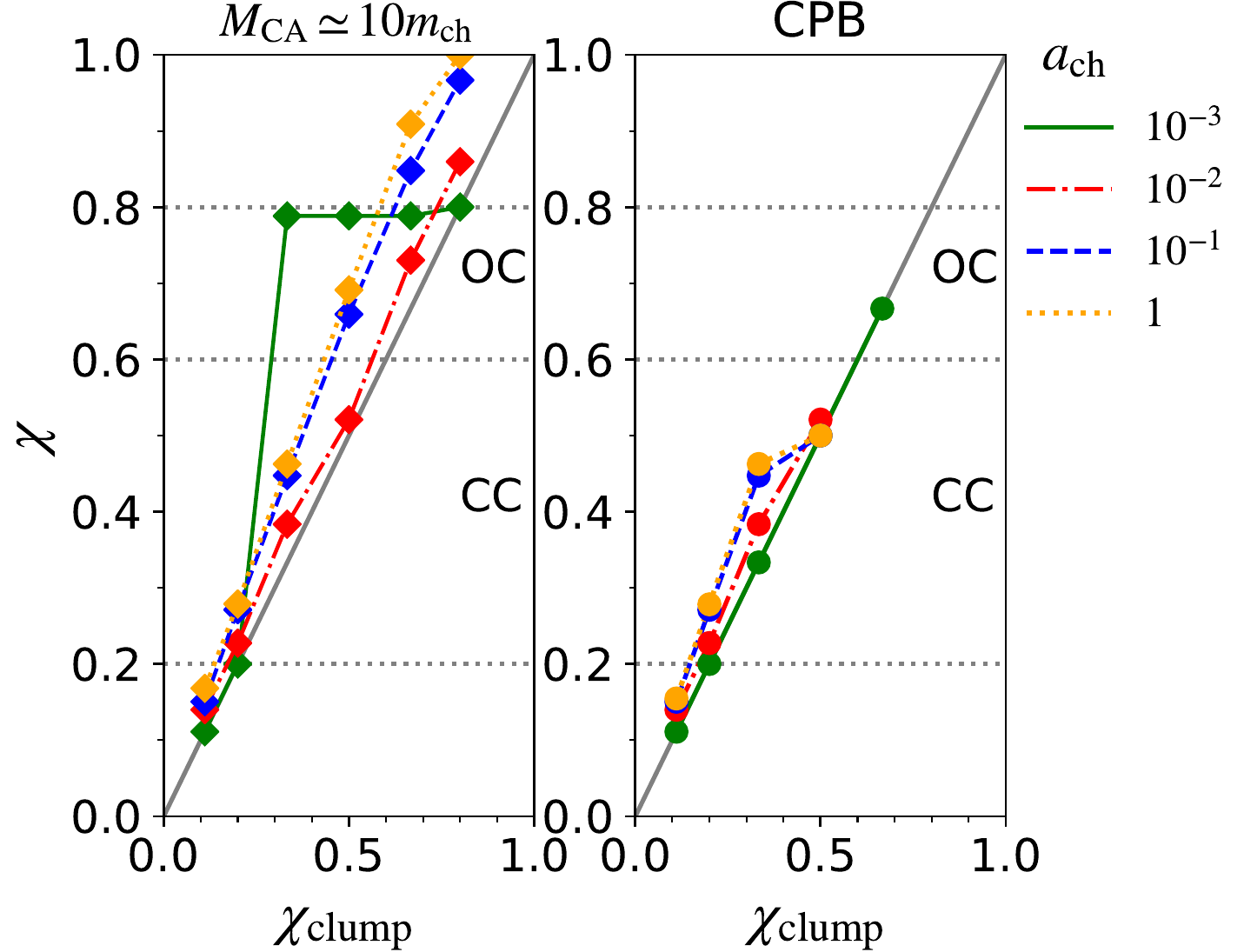}
	\caption{
		Same as Figure \ref{Fig:chi_rho_ra_v12}, but for the results changing $a_{\rm ch}$ from 1 cm to $10^{-3}$ cm.
	}
	\label{Fig:chi_a_ch_v12}
\end{figure}

Figure \ref{Fig:chi_a_ch_v12} shows the results of the chondrule mass fractions ($\chi$) with changing $a_{\rm ch}$.
We find that as $a_{\rm ch}$ increases, the value of $\chi$ increases at $M_{\rm CA}\simeq10 m_{\rm ch}$ except for the case that $a_{\rm ch} = 10^{-3}$ cm.
Namely, CAs become richer in chondrule abundance.
This arises because stage 2 begins when the sizes of MAs become almost equal to those of chondrules;
as $a_{\rm ch}$ increases, $\rhoint_{\rm MA}$ tends to be large and the cross-section of CAs becomes smaller.
Consequently, CAs have less chance to accrete MAs.
For the case with $a_{\rm ch}=10^{-3}$ cm, the value of $\chi$ at $10m_{\rm ch}$ is totally different from that of the other cases.
In addition, there is a large difference in $\chi$ even between the cases that $\chi_{\rm clump}\geq1/5$ and $\chi_{\rm clump}\leq1/5$ (see Figure \ref{Fig:chi_a_ch_v12}).
This can be understood as follows.
The evolution of such small chondrules is similar to that of MAs in stage 1 of the other cases. 
This is because the Brownian motion provides the dominant contribution to the relative velocity.
For the case that $\chi_{\rm clump}\geq1/5$, however, only a few CA-MA collisions occur in stage 2, and this stage quickly ends.
After that, MA-MA and CA-CA collisions occur, rather than only CA-CA collisions.
This leads to a higher abundance of chondrules in CAs.
On the contrary, for the case that $\chi_{\rm clump}\leq1/5$, 
the number densities of MAs are large enough for CA-MA collisions to occur efficiently before CAs grow.
As a result, $\chi$ becomes equal to $\chi_{\rm clump}$ at $10m_{\rm ch}$.

%

We now discuss the chondrule mass fraction at CPB (see the right panel of Figure \ref{Fig:chi_a_ch_v12}).
The resulting behaviors can be explained in the same way as that in our fiducial case ($a_{\rm ch}=10^{-1}$ cm, see Figure \ref{Fig:chi_rho_ra_v12}):
If CA-MA collisions occur in stage 6, then $\chi$ at CPB becomes equal to the initial value at the clump.
If not, then it becomes comparable to that at $10m_{\rm ch}$.
The case that $a_{\rm ch}=10^{-3}$ cm gives the exception that the chondrule fraction at CPB is equal to $\chi_{\rm clump}$ for a wide range of $\chi_{\rm clump}$.
This is because CA-MA collisions occur and CAs accrete all MAs eventually as both aggregates grow.

\begin{figure}[ht]
	\plotone{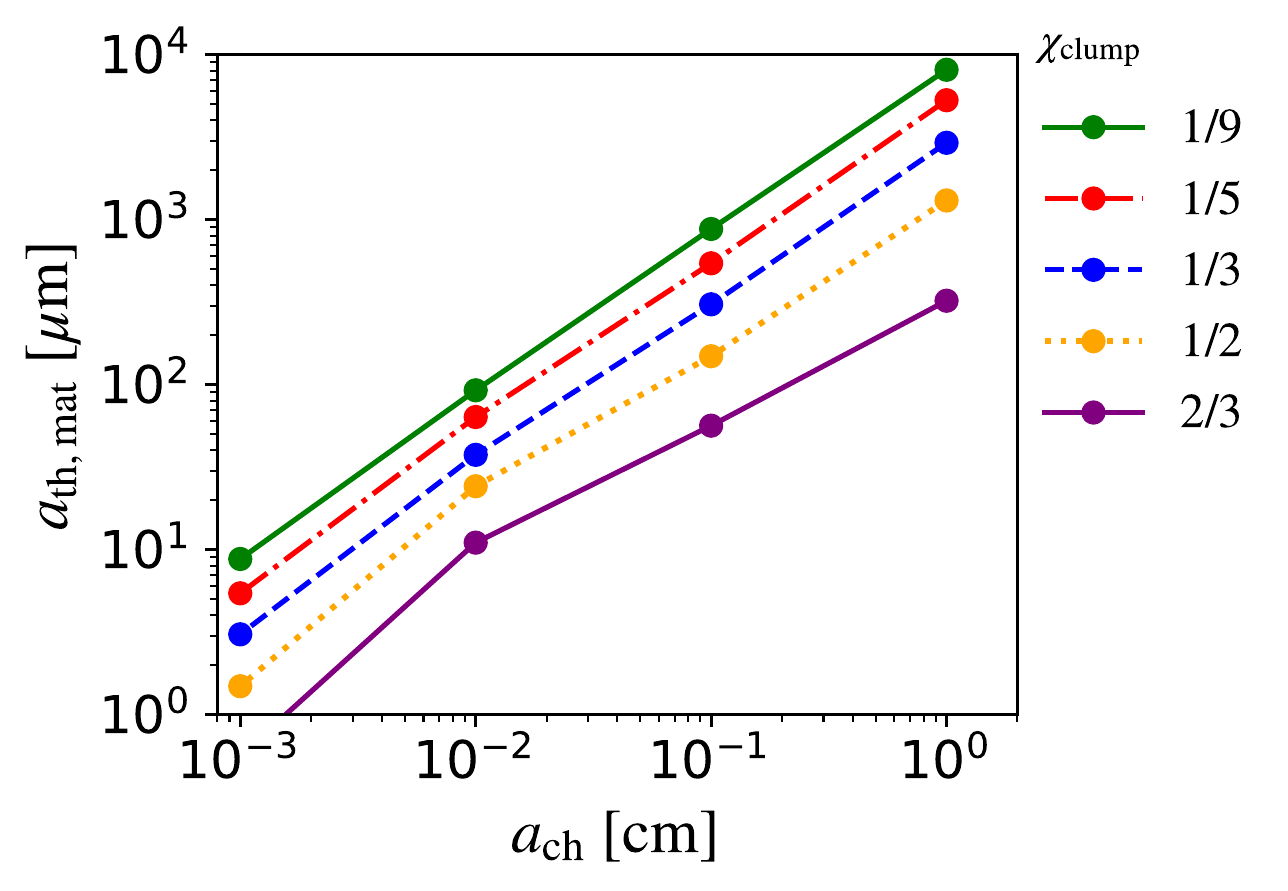}
	\caption{
	The thickness of the matrix components is shown as the function of the chondrule size.
	Each dotted line represents those for each $\chi_{\rm clump}$.
	}
	\label{Fig:ach_da_mat}
\end{figure}

In summary, as the chondrule size increases, CA-MA collisions occur less frequently and CAs become more chondrule-rich.
Note that the thickness of matrix components around chondrules ($a_{\rm th,mat}$) is an increasing function of $a_{\rm ch}$ (Figure \ref{Fig:ach_da_mat}).
This is because of the onset condition of stage 2 (see above, also see Equation (\ref{eq:a_th_mat})).
Thus, the effect of $a_{\rm ch}$ on $a_{\rm th,mat}$ is stronger than $\chi$.
We find that, for the case that $\chi_{\rm clump}=0.5$, 
$a_{\rm th,mat}= 1.4\times10^2~\mu$m if $a_{\rm ch}=10^{-1}$~cm, and $a_{\rm th,mat}= 24~\mu$m if $a_{\rm ch}=10^{-2}$~cm.
Interestingly, these sizes agree with the measurements of the rim thickness of chondrule in Allende carbonaceous chondrite \citep[][see also Section \ref{sect:disc_rim} for further discussion]{Simon+2018}.

\subsubsection{The dependence on matrix size}

\begin{figure}[ht]
	\plotone{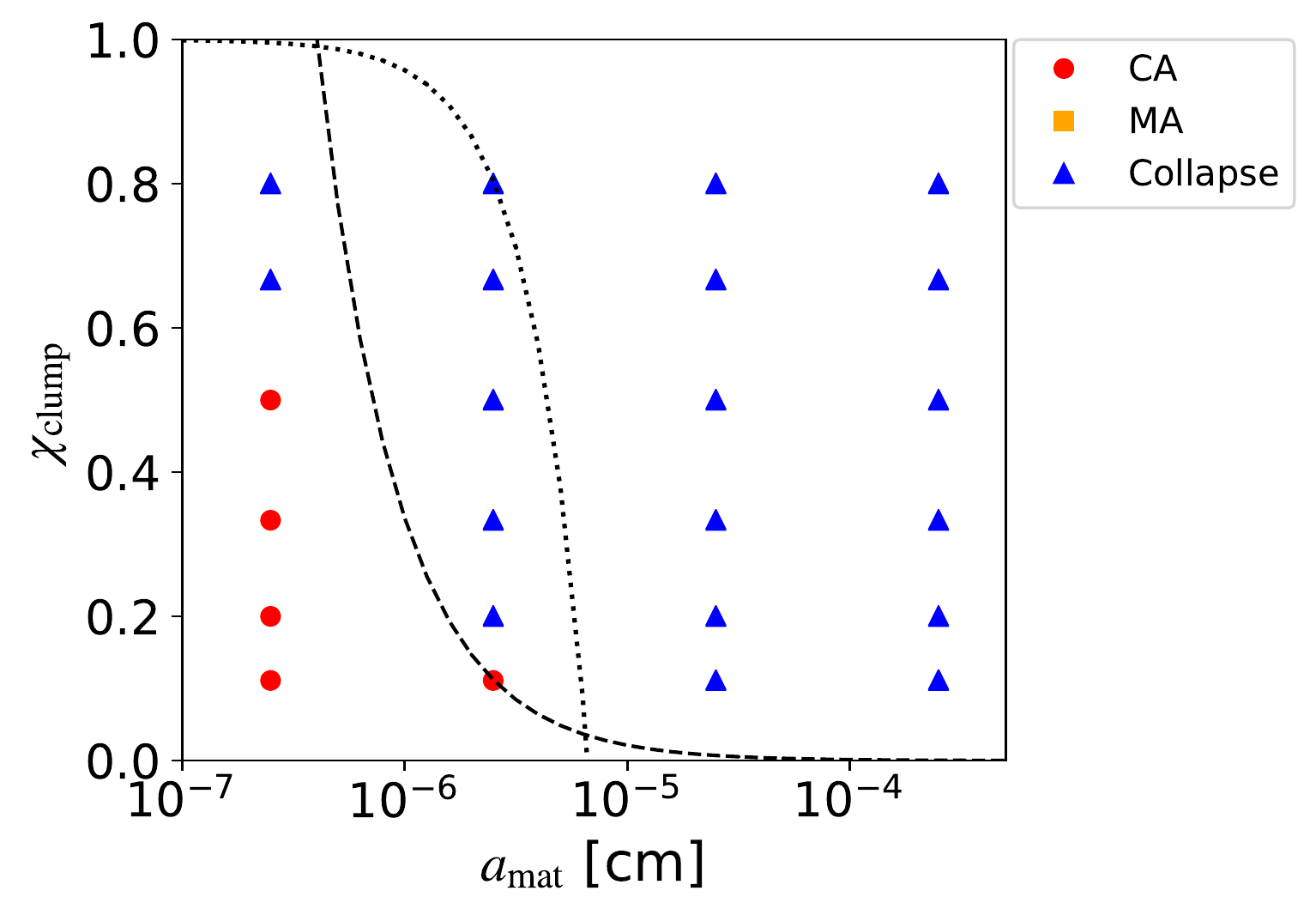}
	\caption{
		The growth modes on the $a_{\rm mat}$ -- $\chi_{\rm clump}$ diagram in the whole eddy model (as in Figure \ref{Fig:rho_ch_rho_mat}).
		The dashed line represents Equation (\ref{eq:t_ff_t_gr_amat}) with $\chi=\chi_{\rm clump}$, and the dotted line is for Equation (\ref{eq:v_cr_St_1}).
	}
	\label{Fig:a_mon_rho_mat_state_v12}
\end{figure}

In this section, we change the size of matrix grains ($a_{\rm mat}$), while the other parameters are the same as the fiducial case.
The variation of $a_{\rm mat}$ affects the number density of matrix grains when the value of $\chi_{\rm clump}$ is kept constant.
It also alters the fragmentation velocity ($v_{\rm frag}$, see Equations (\ref{eq:v_cr_ma-ma}), (\ref{eq:v_cr_ca-ca}), and (\ref{eq:v_cr_ca-ma})).
Here, we consider the range of $a_{\rm mat}$ from $2.5\times10^{-7}$ cm to $2.5\times10^{-4}$ cm (see Table \ref{table:parameters} and Section \ref{sect:params}).

Figure \ref{Fig:a_mon_rho_mat_state_v12} shows the results for the growth mode.
We find that as $a_{\rm mat}$ increases, the growth mode changes from accretion to collapse.
This can be understood by deriving a relationship between $a_{\rm mat}$ and $\chi_{\rm clump}$.
Given that gravitational collapse occurs when $t_{\rm GC}<t_{\rm gr,CA-CA}$ or $\Delta v_{\rm CA-CA}< v_{\rm frag,CA-CA}$,
we obtain for the former (see Equation (\ref{eq:t_ff_t_gr}))
\begin{eqnarray}
	a_{\rm mat} &>&
		4.0 \times 10^{-7} 
		\left(\frac{\rho_{\rm d,ch} }{ 10^{-4} \mbox{~g~cm}^{-3} } \right)^{5/6}
		\left( \frac{\chi_{\rm clump}}{\chi^{2}} \right)^{5/6} 
		\mbox{ cm},
	\nonumber \\
	\label{eq:t_ff_t_gr_amat}
\end{eqnarray}
where we have used that $\rhoint_{\rm CA} \propto a_{\rm mat}^{6/5}$ under the self-gravitational pressure regime (Equation (\ref{eq:rho_grav})), 
and for the latter,
\begin{eqnarray}
	&& 1- \left\{4.2\times10^{-3} \left( \frac{\alpha}{10^{-4}} \right) \left( \frac{c_s}{8.4\times10^4\mbox{ cm s}^{-1}} \right)^2 \right.\nonumber \\ && \left.\times
		\left( \frac{a_{\rm mat}}{2.5\times10^{-7}\mbox{ cm}} \right)^{5/3} \right\} 
		>\chi,
	\label{eq:v_cr_St_1}
\end{eqnarray}
where $\Delta v_{\rm CA-CA}$ is estimated with $\St_{\rm CA}=1$.
Equation (\ref{eq:v_cr_St_1}) suggests that CAs cannot grow through CA-CA collisions when $a_{\rm mat}\geq 6.7\times 10^{-6} \mbox{ cm}$ with the condition that $\chi \geq \chi_{\rm clump}$.
Our results broadly agree with Equation (\ref{eq:t_ff_t_gr_amat}) using $\chi=\chi_{\rm clump}$.
The fragmentation condition that $\Delta v_{\rm CA-CA}< v_{\rm frag,CA-CA}$ (Equation (\ref{eq:v_cr_St_1})) can be effective in the case of high $\alpha$.

\begin{figure}[hbt]
	\plotone{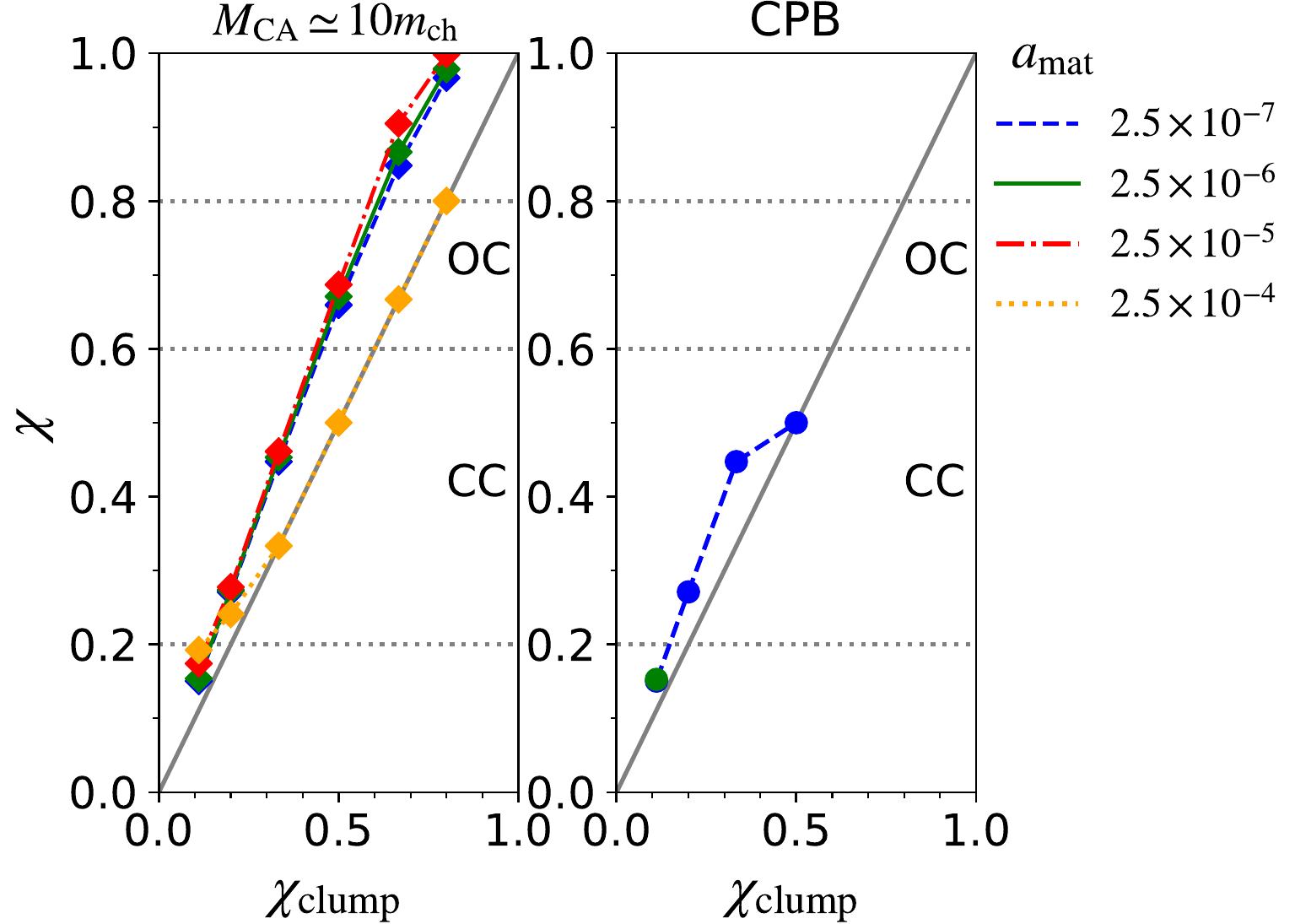}
	\caption{
		Same as Figure \ref{Fig:chi_rho_ra_v12}, but for the results changing $a_{\rm mat}$.
	}
	\label{Fig:chi_a_mat_v12}
\end{figure}

Figure \ref{Fig:chi_a_mat_v12} shows the chondrule fractions at $M_{\rm CA}\simeq10m_{\rm ch}$ and at CPB.
We find that at $M_{\rm CA}\simeq10m_{\rm ch}$, the larger $a_{\rm mat}$ makes slightly larger $\chi$ except for the case of $a_{\rm mat}=2.5\times10^{-4}$ cm (the left panel). 
This is because as $a_{\rm mat}$ increases, $\rhoint_{\rm MA}$ becomes larger (Equations (\ref{eq:rho_hit}) and (\ref{eq:rho_ram})).
Accordingly, CAs accrete MAs inefficiently and stage 2 ends quickly (Appendix \ref{sec:resu_stage2}).
As a result, the chondrule abundance in CAs increases slightly.
This slight change in turn suggests that the value of $a_{\rm th,mat}$ is similar even for the cases that $2.5\times10^{-7}~\mbox{ cm}\leq a_{\rm mat}\leq 2.5\times10^{-5}~\mbox{ cm}$.
On the contrary, for the case that $a_{\rm mat}=2.5\times 10^{-4}$ cm, 
the chondrule fraction at $M_{\rm CA}\simeq10m_{\rm ch}$ becomes equal to $\chi_{\rm clump}$ when $\chi_{\rm clump}\geq 1/3$.
Such a difference originates from the low abundance of matrix grains.
Our results show that CA-MA collisions serve as perfect mergers only after $M_{\rm MA}>M_{\rm CA}$, where CAs accrete all MAs.
The chondrule fractions at CPB are the same as those at $M_{\rm CA}\simeq10m_{\rm ch}$, 
except for our fiducial case ($a_{\rm mat}=2.5\times 10^{-7}$ cm and $\chi_{\rm clump}=1/2$, see the right panel).

\section{Results for the large eddy model}
\label{sect:resut_leddy}

\subsection{Overall features of aggregate growth}
\label{sect:fiducial_v1}

\begin{figure*}[hbt]
	\plotone{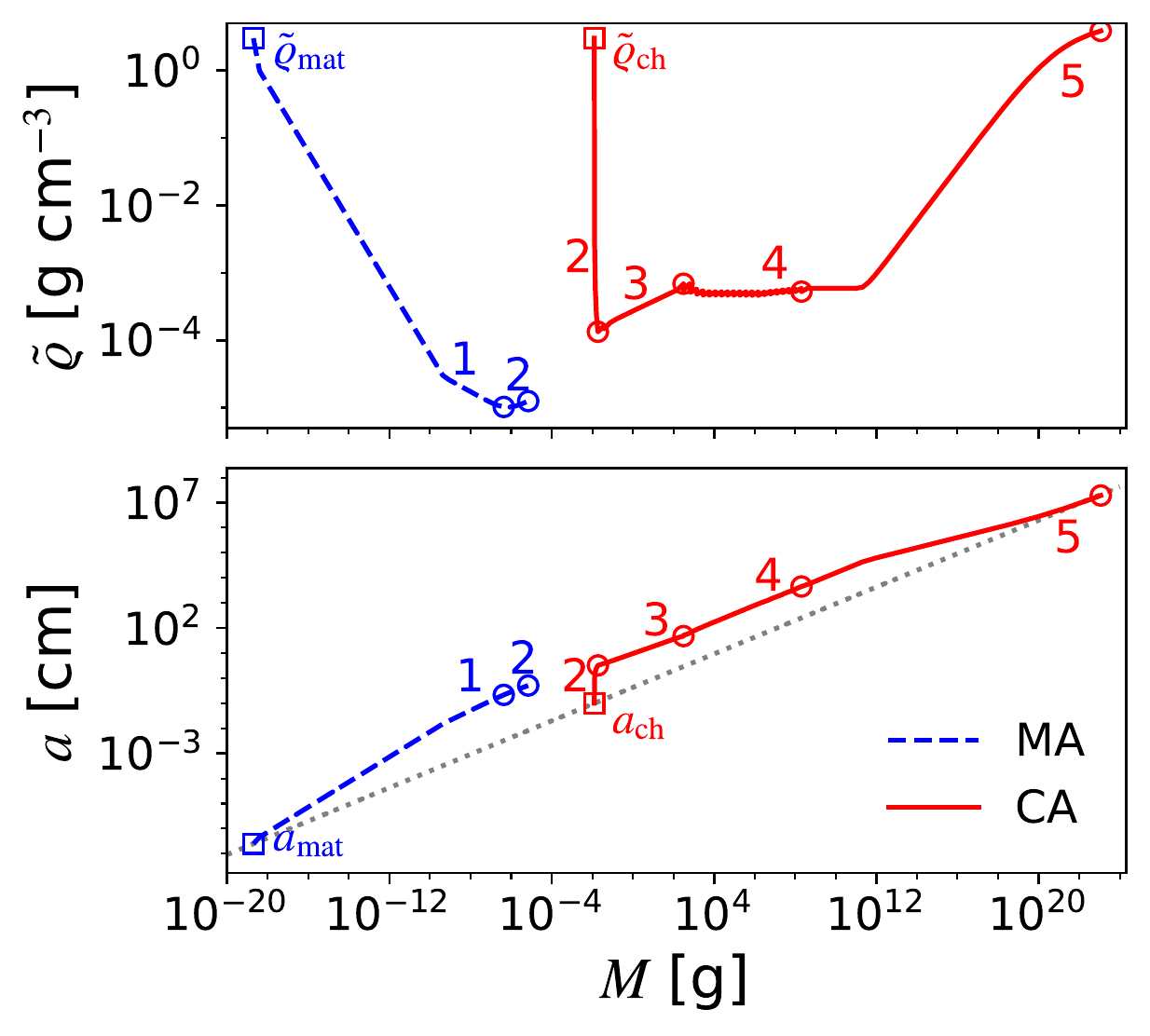}
	\caption{
		The same as Figure \ref{Fig:m_rho_a_v12}, but in the large eddy model.
	}
	\label{Fig:m_rho_a_v1}
\end{figure*}

\begin{figure*}[hbt]
	\plotone{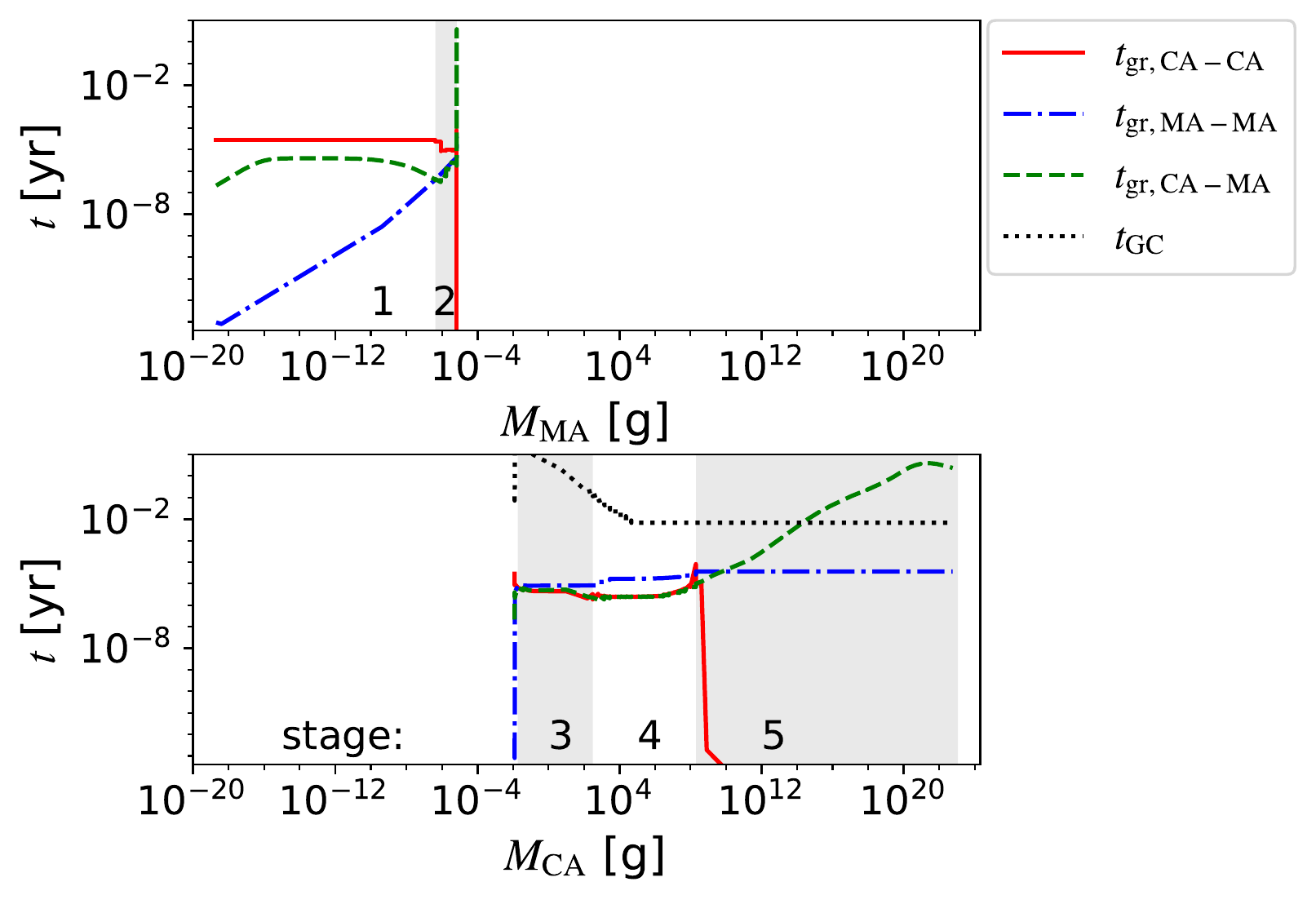}
	\caption{
		The same as Figure \ref{Fig:mCA_tgrow_tacc_v12}, but in the large eddy model.
	}
	\label{Fig:mCA_tgrow_tacc_v1}
\end{figure*}

In this section, we present the evolution of aggregates for our fiducial case in the large eddy model 
($\rho_{\rm d,ch}=10^{-4}\mbox{~g~cm}^{-3}$, $\chi_{\rm clump}=1/2$, $a_{\rm ch}=10^{-1}$ cm, and $a_{\rm mat}=2.5\times10^{-7}$ cm).

Figures \ref{Fig:m_rho_a_v1} and \ref{Fig:mCA_tgrow_tacc_v1} show the evolutions of the internal densities ($\rhoint$), radii ($a$), and growth timescales ($t_{\rm gr}$) of aggregates.
These evolutions are different between the whole and large eddy models.
These differences start from stage 3, where $v$ and $\Delta v$ are dominated by the turbulence.
The values of $v_{\rm tur}$ and $\Delta v_{\rm tur}$ in the large eddy model are smaller than those in the whole eddy model.
Consequently, the internal density of CAs becomes smaller in the large eddy model. 
In addition, the large eddy model can achieve that $t_{\rm gr,CA-CA} \simeq t_{\rm gr,CA-MA}$ in stage 4.
Accordingly, CAs accrete MAs efficiently and the chondrule abundance in CAs becomes lower.
At the end of stage 4, the mass density of MAs reaches about 10\% of the initial value.
In stage 5, the Stokes number of CAs exceeds unity.
As a result, CAs are no more affected by turbulence, and their collision velocities become significantly slower.
These slower collision velocities trigger CA-CA runaway collisions even at $M_{\rm CA}=2.0\times10^8$~g.
Finally, one large CA and small MAs are left in the clump.
The detailed explanation for the large eddy model is provided in Table \ref{table:collisions_v1} and Appendix \ref{sec:results_evolution_v1}.

\subsection{Evolution of the chondrule mass fraction in CAs}

\begin{figure}[ht]
	\plotone{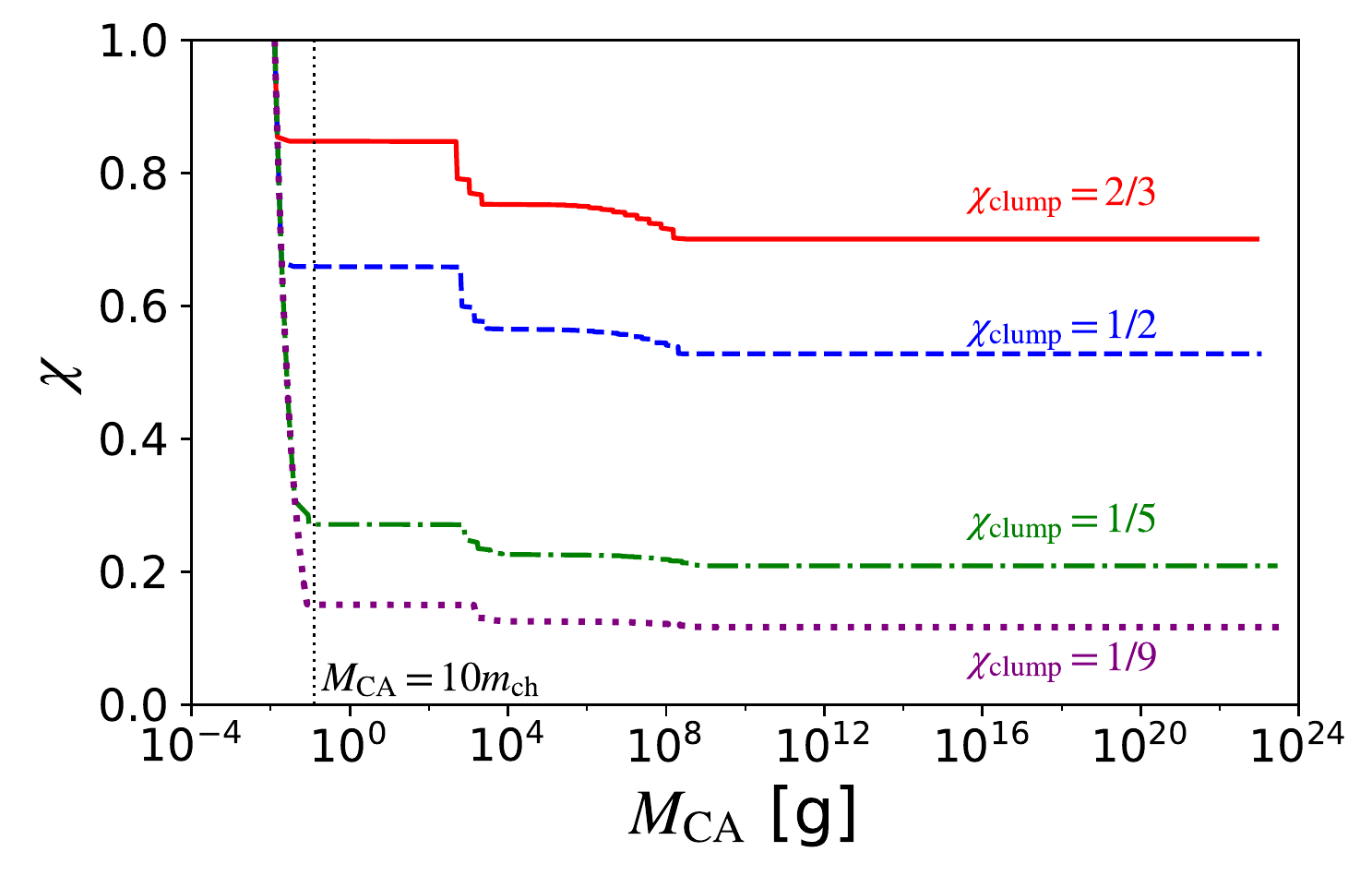}
	\caption{
		The same as Figure \ref{Fig:mCA_ch_ra_rho_ra_v12}, but in the large eddy model.
	}
	\label{Fig:mCA_ch_ra_rho_ra_v1}
\end{figure}

Figure \ref{Fig:mCA_ch_ra_rho_ra_v1} shows the evolution of $\chi$ in $\rho_{\rm d,ch}=10^{-4} \mbox{~g~cm}^{-3}$ cases.
In the large eddy model, the value of $\chi$ also changes twice at $M_{\rm CA} \lesssim 0.1$~g and at $10^2 \mbox{~g} \lesssim M_{\rm CA} \lesssim 10^8 \mbox{~g}$,
but these jumps correspond to stages 2 and 4, respectively.
The first reduction of $\chi$ in stage 2 is the same as that in the whole eddy model, so does the resulting small scale distribution (Sections \ref{sect:chi_v12} and \ref{sec:param_study}).
The second reduction occurs in phase 2 of stage 4 for the large eddy model (see Table \ref{table:collisions_v1} and Appendix \ref{sec:results_evolution_v1}).
In this phase, CAs accrete almost all MAs and hence $\chi$ becomes nearly equal to $\chi_{\rm clump}$.
This is the outcome of CA-MA collisions,
where MAs are smaller than CAs ($a_{\rm MA}=0.51$~cm, Figure \ref{Fig:collisions} (g)).
As a result, the large scale distribution of chondrules and matrix in CAs is characterized by matrix layers that encompass CAs.
Our results show that these features in the $\chi$ evolution are common 
for the case that $\rho_{\rm d,ch}=10^{-4} \mbox{~g~cm}^{-3}$ in the large eddy model (Figure \ref{Fig:mCA_ch_ra_rho_ra_v1}).

Thus, both the whole and large eddy models predict that, if CPBs form out of dense clumps and the accretion mode becomes important, there are certain internal distributions of chondrules within the CPBs.
There are two scales in these distributions.
The small scale distribution is characterized by single chondrules covered by matrix components.
Each chondrule has the same amount of the matrix grains.
The large scale distribution depends on the timing of CA-MA collisions and is determined by the masses of CAs and MAs at that time.
For the large eddy model, this scale is characterized by a matrix surface layer surrounding the small scale distribution.
This matrix rich layer should be composed of MAs with sizes of about 0.5~cm.

We also check the dependence on the orbital radius.
We find that in the large eddy model, the chondrule fraction is almost the same even if the orbital radius changes from 1~au to 5~au.

\subsection{Parameter study}\label{sect:parameter_v1}

\begin{figure}[ht]
	\plotone{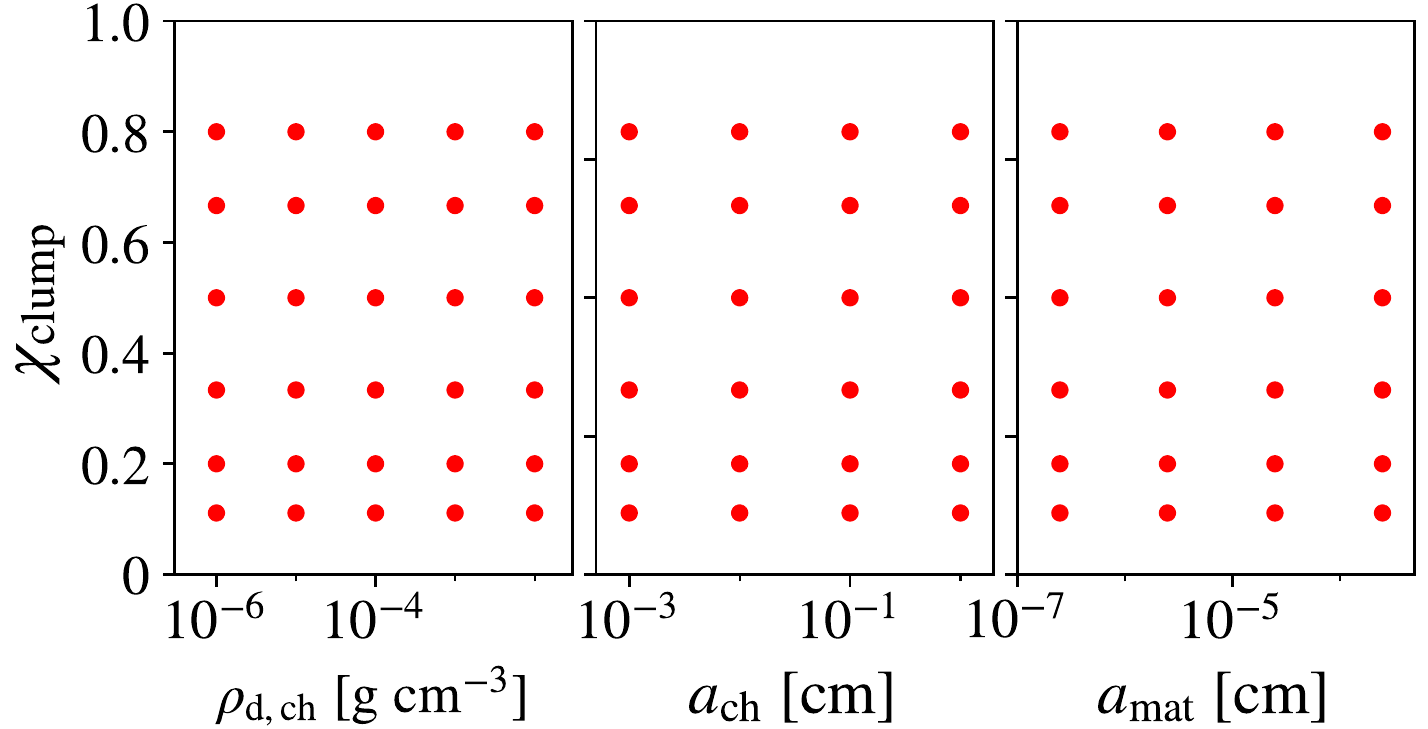}
	\caption{
		The growth modes in the large eddy model.
		Left panel is the $\rho_{\rm d,ch}$ -- $\chi_{\rm clump}$ diagram, 
		center one is the $a_{\rm ch}$ -- $\chi_{\rm clump}$ diagram, 
		and right one is the $a_{\rm mat}$ -- $\chi_{\rm clump}$ diagram.
		All simulations end up with the CA accretion mode.
	}
	\label{Fig:fs_all_v1}
\end{figure}

\begin{figure}[ht]
	\plotone{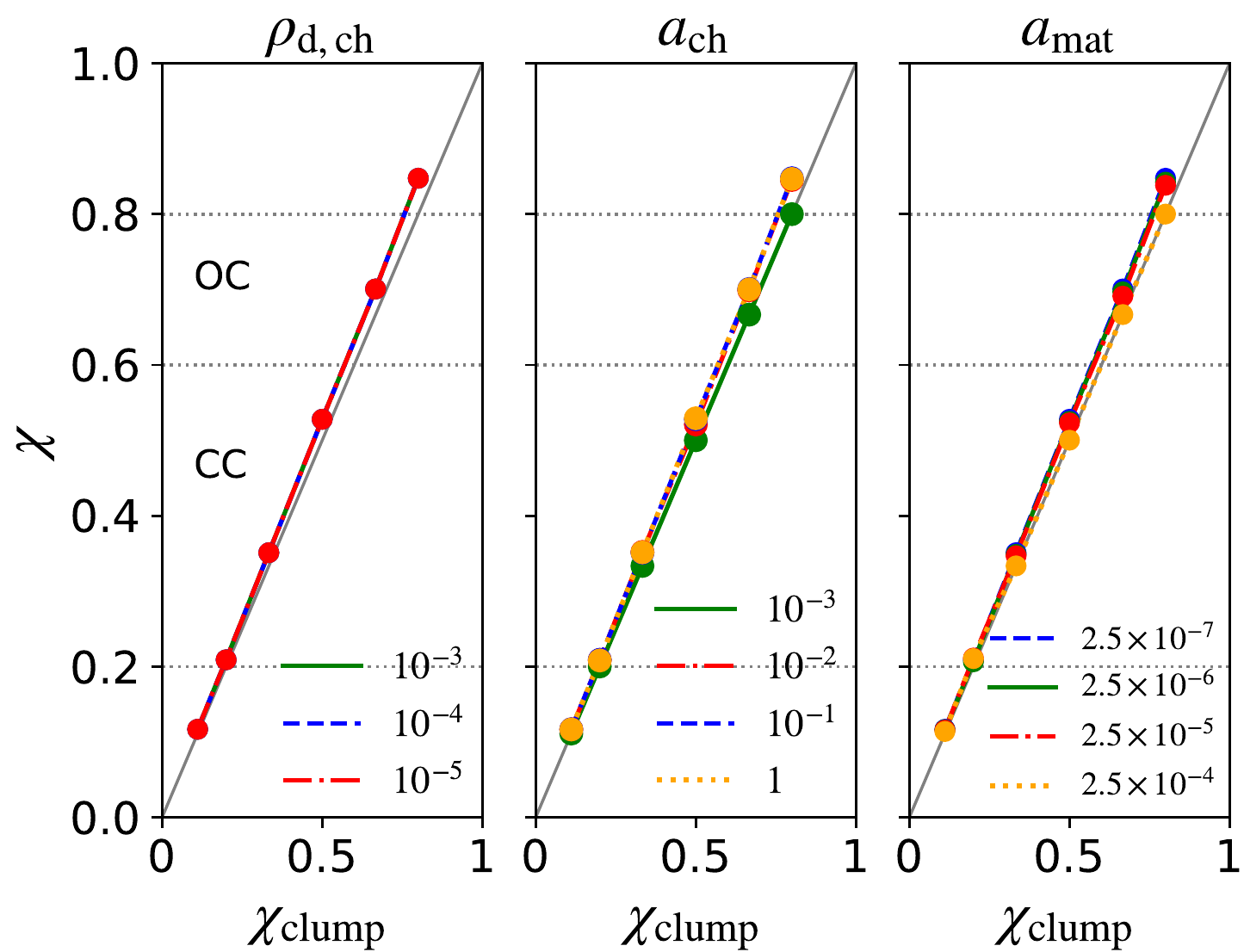}
	\caption{
		The chondrule mass fraction in a CA ($\chi$) as a function of that in the dense clump ($\chi_{\rm clump}$) in the large eddy model.
		The left panel is for the $\rho_{\rm d,ch}$ dependence, the center one is for the $a_{\rm ch}$ dependence, and the right one is for the $a_{\rm mat}$ dependence.
	}
	\label{Fig:chi_all_v1}
\end{figure}

As done in the whole eddy model (see Section \ref{sec:param_study}), we conduct the parameter study for the large eddy model, wherein $\rho_{\rm d,ch}$, $\chi_{\rm clump}$, $a_{\rm ch}$, and $a_{\rm mat}$ vary (Table \ref{table:parameters}).

Figure \ref{Fig:fs_all_v1} shows that only the CA accretion mode is realized in the large eddy model.
This is because the condition of $t_{\rm gr}> t_{\rm GC}$ is not satisfied.
Once CA-CA collisions occur, $t_{\rm gr,CA-CA}$ is always shorter than $t_{\rm GC}$ (Figure \ref{Fig:mCA_tgrow_tacc_v1}).
The other condition for avoiding accretion is collisional fragmentation.
In the large eddy model, however, CA-CA collisions do not end up with fragmentation even in the case that $a_{\rm mat}=2.5\times 10^{-4}$ cm.
This is simply because the turbulent induced velocity is smaller than the fragmentation velocity in the large eddy model, where only large eddies are taken into account.
The condition that $\Delta v_{\rm CA-CA}> v_{\rm cr,CA-CA}$ is rewritten as
\begin{eqnarray}
	&&1-\left\{
			2.2\times10^{-3} \left( \frac{\alpha}{10^{-4}} \right)
			\left( \frac{c_s}{8.4\times10^4\mbox{ cm s}^{-1}} \right)^2 \right.\nonumber \\ &&\left.\times 
			\left( \frac{a_{\rm mat}}{2.5\times10^{-4}\mbox{ cm}} \right)^{5/3} \right\}
		<\chi,
\end{eqnarray}
where the maximum value of $\Delta v_{\rm tur,CA-CA} (\simeq 0.06\sqrt{\alpha}c_{\rm s}$) is adopted (Equation (\ref{eq:Delta_v_tur1})).
This equation shows that CA-CA collisions result in fragmentation only when the chondrule fraction in the clump is $\chi_{\rm clump}>0.9978$.
Thus, the accretion mode always dominates in the large eddy model.

We now discuss the chondrule mass fractions in CPBs.
Since the fraction at $M_{\rm CA}\simeq10m_{\rm ch}$ is the same as that in the whole eddy model, we here focus on the value at CPB.
Figure \ref{Fig:chi_all_v1} shows the results.
As discussed above, CAs accrete almost all MAs at stage 4 (Figure \ref{Fig:mCA_ch_ra_rho_ra_v1}), and hence $\chi \simeq \chi_{\rm clump}$ for all the cases.

\section{Discussion}\label{sect:discussion}

We have demonstrated above that collisional growth of chondrules and matrix grains leads to the formation of aggregates in dense clumps.
This growth eventually produces the internal distribution of chondrules in the subsequently forming CPBs.
However, our results are derived from the fluffy aggregate growth calculations with a number of assumptions (Section \ref{sect:assump}).
In the following, we comment on our assumptions that may affect our finding.

\subsection{Effects of the collisional outcome}\label{sect:discuss_outcome}

We first discuss the outcome of collisions.
While we have effectively assumed perfect mergers and bouncing without the mass loss (Assumption 5 in Section \ref{sect:assump}), the realistic collisional outcome is more complicated.

One complexity in the collisional outcome is that aggregates may lose their components in collisions.
\cite{Gunkelmann+2017} have indeed shown that when aggregates composed of chondrules and smaller sized dust collide with each other, a fraction of dust particles can be lost while they are sticking together.
Based on the most porous dust shell case ($\phi=0.081$) in their study, 
CAs lose more than 10\% of their matrix components when the collisional velocity is $\gtrsim 20 \mbox{ cm s}^{-1}$.
In our calculations, CA-CA collision velocities exceed $20 \mbox{ cm s}^{-1}$ when $2.0 \times10^4\mbox{ g} < M_{\rm CA} < 1.7\times10^{14} \mbox{ g}$ in the whole eddy model, which corresponds to $3.9\times10^{-3}<\St_{\rm CA}<3.4\times10^3$.
The outer matrix components of CAs thus can be partially ejected in such velocity collisions ($\gtrsim 20 \mbox{ cm s}^{-1}$).
Note that the ejected mass and critical velocity for ejection would be affected by the size and filling factor of matrix grains.
In the large eddy model, the collision velocities in CA-CA collisions do not exceed $20 \mbox{ cm s}^{-1}$.
On the other hand, CA-MA collision velocities exceed $20 \mbox{ cm s}^{-1}$, while the effect of the partial ejection would be minor for such a collision.
It can be expected that the ejected fragments are quickly accreted onto CAs and MAs since their collision timescales are shorter than those of CA-CA and MA-MA collisions.
The balance between ejection and re-accretion would be important for determining whether the ejection of matrix components affects the evolution of CAs.
The collisions and subsequent ejections would affect the velocities of aggregates in the dense clump.

In addition, \cite{Arakawa2017} pointed out that chondrules themselves can also be ejected from CAs due to collisions.
The chondrule ejection process depends on the velocity difference between chondrules and surrounding matrix in CAs after collisions.
As the collision velocity between aggregates increases, the velocity difference becomes large, and the ejection of chondrules become important.
If chondrule ejection is effective, MAs can accrete ejected chondrules after MAs grow and satisfy $\St_{\rm MA}\gg1$. 
In such a case, MAs becomes CAs and these have the layer of re-accreted chondrules on their surfaces.

High-velocity collisions provide further diversity to their outcome.
One example is collisional compaction \citep[][see also Equation (\ref{eq:rho_col})]{Wada+2008, Meru+2013}.
CA-MA collisions would cause collisional compaction since the collision velocities of CA-MA collisions are higher than those of CA-CA and MA-MA collisions.
This compaction would be effective in $M_{\rm CA}<1$~g and affect the thickness of matrix layers around chondrules.

One may consider that the fragmentation velocity in our model is very high, which makes most collisions perfect mergers (see Section \ref{sect:v_crit}).
Numerical and experimental studies have shown that fluffy aggregates do not bounce but stick with each other when their filling factor $\phi\lesssim 0.10$ \citep{Langkowski+2008, Wada+2011,Seizinger&Kley2013,Kothe+2013}.
Equivalently, non-sticking events would occur only when the densities of CAs and MAs are more than $0.1\rhoint_{\rm mat}$ in our models.
Moreover, the fragmentation velocity is also affected by the filling factor \citep{Wada+2009, Meru+2013}.
Its dependence is still under the debate. 
We however find that most of our results do not change even if we reduce the fragmentation velocity by half or increase it.
Thus, our choice of fragmentation velocity does not alter our conclusions very much.

We only consider the growth of aggregates whose collision timescale is the shortest in the present calculations.
However, collisions whose timescale is the second or third shortest might not be negligible. 
When such collisions and the resulting growth would be taken into account, CAs may accrete MAs in earlier stages.
These CA-MA collisions can complicate the internal distribution of chondrules and matrix in CAs.

In our future work, we will consider the above effects and explore how the mass distribution of chondrules and matrix in CAs will be determined, 
according to the collision velocity and densities of colliding CAs and MAs.

\subsection{Rimmed and unrimmed chondrules} \label{sect:disc_rim}

Our results show that all chondrules have matrix surface layers.
This is the natural outcome that chondrules accrete matrix grains under the co-existence of them, and consistent with previous studies \citep{Ormel+2008, Arakawa2017}.
Note that these previous studies assume the standard nebular conditions, and do not consider dense clumps.
Thus, the presence of matrix surface layers around chondrules is very likely to be the robust results and insensitive to the surrounding environments.

We find that the layer thickness is approximately proportional to the chondrules size (Equation (\ref{eq:a_th_mat}) and Section \ref{sect:dependence_chondrule_size}), and there is no correlation to the high dust mass density (which we derived from Semarkona ordinary chondrite \citep{Alexander+2008}).
This finding agrees well with the analysis of the rimmed chondrules in carbonaceous chondrites \citep[e.g.,][]{Hanna+2018,Simon+2018}.
However, the meteoritic data show that the fraction of dust rims is minor;
$\sim$~15 -- 20~\% chondrules in Allende CV chondrite have rims, and those in NWA 5717 ordinary chondrite have almost none \citep{Simon+2018}.
This suggests that the growth of aggregates is not identical (Assumption 6) and/or that collisional outcomes (ejections and compaction) would be crucial for matrix components around chondrules (Section \ref{sect:discuss_outcome}).

The extension of this work may serve as an important step for understanding the origin of the dust rim.

\subsection{Effects of the spatial distribution of dust within clumps}

We have shown above that the internal distribution of chondrules within CPBs reflects the initial condition of the clumps and the dynamics of dust particles under the accretion mode.
This implies that the distribution might play an important role in exploring the origin of CPBs and planetesimals in general.
It is, however, important to recall that there is no spatial information about chondrules, matrix grains, CAs, and MAs within dense clumps in this work (Assumption 7 in Section \ref{sect:assump}).
In reality, these details should be crucial for accurately predicting the mass fraction and spatial distribution of chondrules in CPBs.
For instance, the inclusion of sedimentation of aggregates will enable specification of the spatial distribution of chondrules within CPBs and 
provide tighter constraints of which chondrule-rich layers within CPBs would trace their birth condition and planetesimals.
We will investigate these in our future work.

\subsection{Caveats on timescales}\label{sect:discuss_time}

We have adopted the timescale argument for determining either the collapse mode or the accretion one is realized in our calculations (Section \ref{sect:modes}).
The argument, however, contains some simplifications.
We here discuss them.

We first discuss the collapse timescale.
In our approach, the collapse timescale is estimated by simply comparing the free-fall and sedimentation timescales (Equation (\ref{eq:t_gc})).
While this approach broadly captures the basic picture of collapse,
the reality is much more complicated.
\cite{Shariff&Cuzzi2015} investigated this complexity by properly taking into account the interaction between gas and solid particles,
and provide a more detailed expression about the collapse timescale (see their equations (90) and (91)).
High dust mass densities in the clumps make the collapse timescale shorter and it corresponds to the free-fall timescale even in $\St<\St_{\rm ff}$ (Section \ref{sect:st_ff}).
We, however, emphasize that our results do not change at all even if the detailed expression is adopted.
This is because aggregates grow quickly in our setup, as discussed in Section \ref{sect:modes}.
Hence, our simplified approach works well.

We now discuss the growth timescale.
In our model, the growth timescale corresponds to the mass doubling timescale (Section \ref{sect:timescale}).
This indicates that the actual {\it total} growth timescale is longer than our growth timescale due to the cumulative effect.
In fact, \cite{Okuzumi+2012} showed that aggregate growth is limited by radial drift if the radial drift timescale becomes comparable to $\sim 30$ times the mass doubling timescale.
Similar consideration can be developed for our case;
the accretion mode will be realized if $\sim 30$ times the growth timescale is shorter than the collapse timescale (cf. Section \ref{sect:modes}).
Equivalently, our results overestimate the parameter range of the accretion mode.

In summary, better treatment of the timescales will improve our results.
However, it does not affect our conclusions very much.

\section{Conclusions}\label{sect:conclusion}

Chondrites are the common meteorites and composed mainly of chondrules and matrix.
The volume ratio of these two ingredients varies among the groups of chondrites.
Chondrite parent bodies very likely formed in dense clumps where the spatial dust densities were much larger than those of the solar nebula. 

As a first step, we have applied fluffy aggregate growth calculations to collisions among chondrules and matrix that are present in the surface regions of self-gravitating dense clumps.
We have assumed that there are two kinds of aggregates, which are aggregates composed of chondrules and matrix (called CAs) and aggregates composed purely of matrix grains (called MAs).
We have calculated the growth of these aggregates using their growth timescales and collision velocities.

Given that the interaction between aggregates and gas, especially turbulent gas motion in the dense clump is not well understood, we have considered two models for the turbulent induced velocity.
The effect of all eddies is taken into account in the whole eddy model.
In the large eddy model, it is assumed that only the eddies whose turnover time is longer than the stopping time of aggregates contribute to the relative velocity of aggregates.
Our results suggest that the growth path and distribution of chondrules in CA depend on these eddy models.

The results in the whole eddy model are summarized as follows:
\begin{enumerate}
	\item 
		In our fiducial case ($\rho_{\rm d,ch}=10^{-4}\mbox{~g~cm}^{-3}$, $\chi_{\rm clump}=1/2$, $a_{\rm ch}=10^{-1}$ cm, and $a_{\rm mat}=2.5\times10^{-7}$ cm), aggregates grow via collisions.
		We name this growth mode as the accretion mode.
	\item 
		In the accretion mode, CAs accrete MAs in one or two separated stages (Table \ref{table:collisions}).
		In the first CA-MA accretion stage, chondrules collide with MAs.
		This makes the matrix layer around chondrules, which may be the origin of the matrix rim around chondrules.
		After CAs and MAs grow due to CA-CA or MA-MA collisions, the second CA-MA collisions occur.		
	\item
		Growing aggregates may undergo sedimentation toward the clump center due to self-gravity in the middle of the growth calculations.
		We name this growth mode as the collapse mode.
	\item
		We have focused on the chondrule fraction of the aggregates in two scales.
		The small scale distribution of chondrules is the structure of the matrix layer around single chondrules.
		The large scale distribution originates from the second CA-MA accretion stage.
		Interestingly, the relation of chondrule and matrix rim sizes can be reproduced in our simulations.
\end{enumerate}

The results in the large eddy model are as follows: 
\begin{enumerate}
	\item 
		In the large eddy model, the velocities of aggregates are smaller than those in the whole eddy model. 
		The evolution of the densities of CAs becomes different from that in the whole eddy model since the static compression by the ram pressure depends on the relative velocity between gas and aggregates.
	\item 
		The growth modes are always the accretion mode in the large eddy model even when the size of matrix grains is $2.5~\mu$m due to small collision velocities.
		The internal structure of CPBs would be determined by collisions between aggregates.
	\item
		The matrix structure around single chondrules in the large eddy model is the same as that in the whole eddy model.
		However, the subsequent CA-MA collisions are different from that in the whole eddy model.
		In the large eddy model, more CA-MA collisions occur before MAs grow via MA-MA collisions.
		CAs accrete almost all MAs before their Stokes numbers reach unity.
\end{enumerate}

\acknowledgments

We thank Sota Arakawa and Satoshi Okuzumi for providing us with useful comments.
We also thank the anonymous referees for constructive comments.
Numerical simulations were in part carried out on the PC cluster at the Center for Computational Astrophysics, National Astronomical Observatory of Japan and at the Theoretical Institute for Advanced Research in Astrophysics (TIARA) in Academia Sinica Institute for Astronomy and Astrophysics (ASIAA).
Part of this research was carried out at the Jet Propulsion Laboratory, California Institute of Technology, under a contract with the National Aeronautics and Space Administration. Y.H. is supported by JPL/Caltech.

\appendix

\section{Mass evolution of aggregates in the fiducial case of the whole eddy model} \label{sec:results_evolution}

	\begin{table*}
		\begin{center}
		\tablenum{3}
		\caption{Summary of the dominant collisions at each stage for our fiducial case in the whole eddy model}
		\label{table:collisions}
		\setlength{\leftskip}{-.3in}
		\renewcommand{\arraystretch}{1.2}
		\scriptsize
		\begin{tabular}{c|c|c|c|c|c|c|c|c} 
		\hline \hline
			&	Collisions	&	Phase	&	Internal Density	&	$t_{\rm s}$	&	$v$	&	$\Delta v$	&	$t_{\rm gr}$	&	Transition mass	\\ \hline \hline
		Stage 1
			&	MA-MA	& 1	&	$\rhoint_{\rm hit}\propto M_{\rm MA}^{-1/2}$	&	($t_{\rm s}^{\rm Ep}$)	& ($v_{\rm B}$)	&	$\Delta v_{\rm B} \propto M_{\rm MA}^{-1/2}$	&	$t_{\rm gr} \propto M_{\rm MA}^{1/2}$	&	$M_{\rm MA} = 8.4\times10^{-10}$ g	\\
			&			& 2	&	$\rhoint_{\rm ram}\propto M_{\rm MA}^{-1/6}$	&	$t_{\rm s}^{\rm Ep}$	&	$v_{\rm B}$	&	$\Delta v_{\rm B} \propto M_{\rm MA}^{-1/2}$	&	$t_{\rm gr} \propto M_{\rm MA}^{13/18}$	&	$M_{\rm MA} = 4.3\times10^{-7}$ g	\\
			\hline
			&	CA-MA*	& 1 &	-	&	-	&	-	&	$\Delta v_{\rm B} \propto M_{\rm MA}^{-1/2}$	&	$t_{\rm gr} \propto M_{\rm MA}^{1/2}$	&	$M_{\rm MA} = 2.0\times10^{-16}$ g	\\
			&			& 2	&	-	&	-	&	-	&	$\Delta v_{\rm tur}\propto \St$	&	$t_{\rm gr} \propto M_{\rm MA}^{0}$	&	$M_{\rm MA} \sim 10^{10}$ g	\\ 
			&			& 3	&	-	&	-	&	-	&	$\Delta v_{\rm tur}\propto \St$	&	$t_{\rm gr} \propto M_{\rm MA}^{-2/3}\rhoint_{\rm MA}^{2/3}$	&	$M_{\rm MA} = 4.3\times10^{-7}$ g	\\ 
			\hline \hline
		Stage 2
			&	MA-MA	& - &	$\rhoint_{\rm ram}\propto M_{\rm MA}^{1/7}$	&	$t_{\rm s}^{\rm Ep}$	&	$v_{\rm tur} \propto$ St	&	$\Delta v_{\rm tur} \propto$ St $ \propto M_{\rm MA}^{3/7}$	&	$t_{\rm gr} \propto M_{\rm MA}^{0}$	&	$M_{\rm MA} = 6.9\times10^{-6}$ g	\\ 
			\hline
			&	CA-MA	& - & 	-	& -	&	-	&	$\Delta v_{\rm tur} \propto \St$	&	$t_{\rm gr} \propto M_{\rm MA}^{1/2}$	&	$M_{\rm CA} = 1.9\times10^{-2}$ g ($\chi = 0.66$)	\\
			\hline \hline
		Stage 3 
			&	CA-CA	& 1 &	$\rhoint_{\rm ram}\propto M_{\rm CA}^{1/7}$		&	$t_{\rm s}^{\rm Ep}$	&	$v_{\rm tur} \propto \St$		&	$\Delta v_{\rm tur} \propto\St \propto M_{\rm CA}^{3/7}$		&	$t_{\rm gr} \propto M_{\rm CA}^{0}$ 	&	$M_{\rm CA} = 20$ g	\\
			&			& 2	&	$\rhoint_{\rm ram}\propto M_{\rm CA}^{1/7}$		&	$t_{\rm s}^{\rm St}$	&	$v_{\rm tur} \propto \St$		&	$\Delta v_{\rm tur} \propto\St \propto M_{\rm CA}^{5/7}$		&	$t_{\rm gr} \propto M_{\rm CA}^{-2/7}$	&	$M_{\rm CA} = 2.5 \times10^{3}$ g	\\
			&			& 3	&	$\rhoint_{\rm ram}\propto M_{\rm CA}^{0}$		&	$t_{\rm s}^{\rm St}$	&	$v_{\rm tur} \propto \St^{1/2}$	&	$\Delta v_{\rm tur} \propto\St \propto M_{\rm CA}^{2/3}$		&	$t_{\rm gr} \propto M_{\rm CA}^{-1/3}$	&	$M_{\rm CA} = 2.6 \times10^{6}$ g	\\
			&			& 4	&	$\rhoint_{\rm ram}\propto M_{\rm CA}^{4/91}$	&	$t_{\rm s}^{\rm Al}$	&	$v_{\rm tur} \propto \St^{1/2}$	&	$\Delta v_{\rm tur} \propto \St^{1/2} \propto M_{\rm CA}^{3/13}$&	$t_{\rm gr} \propto M_{\rm CA}^{12/91}$ &	$M_{\rm CA} = 1.6 \times10^{8}$ g	\\
			&			& 5	&	$\rhoint_{\rm ram}\propto M_{\rm CA}^{0}$(fix)	&	$t_{\rm s}^{\rm Al}$	&	$v_{\rm tur} \propto \St^{-1/2}$&	$\Delta v_{\rm tur} \propto\St^{-1/2} \propto M_{\rm CA}^{-1/3}$&	$t_{\rm gr} \propto M_{\rm CA}^{2/3}$	&	$M_{\rm CA} = 1.7 \times10^{11}$ g	\\
			\hline \hline 
		Stage 4
			&	MA-MA	& 1	&	$\rhoint_{\rm ram}\propto M_{\rm MA}^{1/7}$		&	$t_{\rm s}^{\rm Ep}$	&	$v_{\rm tur} \propto \St$		&	$\Delta v_{\rm tur} \propto\St\propto M_{\rm MA}^{3/7}$			&	$t_{\rm gr} \propto M_{\rm CA}^{0}$		&	$M_{\rm MA} = 3.6$ g	\\
			&			& 2	&	$\rhoint_{\rm ram}\propto M_{\rm MA}^{1/7}$		&	$t_{\rm s}^{\rm St}$	&	$v_{\rm tur} \propto \St$		&	$\Delta v_{\rm tur} \propto\St\propto M_{\rm MA}^{5/7}$			&	$t_{\rm gr} \propto M_{\rm CA}^{-2/7}$	&	$M_{\rm MA} = 3.7 \times10^{3}$ g	\\
			&			& 3	&	$\rhoint_{\rm ram}\propto M_{\rm MA}^{0}$		&	$t_{\rm s}^{\rm St}$	&	$v_{\rm tur} \propto \St^{1/2}$	&	$\Delta v_{\rm tur} \propto\St^{1/2} \propto M_{\rm MA}^{2/3}$	&	$t_{\rm gr} \propto M_{\rm CA}^{-1/3}$	&	$M_{\rm MA} = 1.9 \times10^{6}$ g	\\
			&			& 4	&	$\rhoint_{\rm ram}\propto M_{\rm MA}^{4/91}$	&	$t_{\rm s}^{\rm Al}$	&	$v_{\rm tur} \propto \St^{1/2}$	&	$\Delta v_{\rm tur} \propto\St^{-1/2} \propto M_{\rm MA}^{3/13}$&	$t_{\rm gr} \propto M_{\rm CA}^{12/91}$ &	$M_{\rm MA} = 4.8 \times10^{8}$ g	\\
			&			& 5	&	$\rhoint_{\rm ram}\propto M_{\rm MA}^{0}$(fix)	&	$t_{\rm s}^{\rm Al}$	&	$v_{\rm tur} \propto \St^{-1/2}$&	$\Delta v_{\rm tur} \propto\St^{-1/2} \propto M_{\rm MA}^{-1/3}$&	$t_{\rm gr} \propto M_{\rm CA}^{2/3}$	&	$M_{\rm MA} = 5.0 \times10^{11}$ g	\\
			\hline \hline 
		Stage 5
			&	CA-CA	& -	&	$\rhoint_{\rm grav}\propto M_{\rm CA}^{2/5}$	&	$t_{\rm s}^{\rm Al}$	&	$v_{\rm tur} \propto \St^{-1/2}$	&	$\Delta v_{\rm tur} \propto\St^{-1/2} \propto M_{\rm CA}^{-9/20}$	&	$t_{\rm gr} \propto M_{\rm CA}^{3/4}$	&	$M_{\rm CA} = 3.4 \times10^{14}$ g ($\chi = 0.66$)	\\
			\hline
			&	MA-MA	& 1 &	$\rhoint_{\rm grav}\propto M_{\rm MA}^{2/5}$	&	$t_{\rm s}^{\rm Al}$	&	$v_{\rm tur} \propto \St^{-1/2}$	&	$\Delta v_{\rm tur} \propto\St^{-1/2} \propto M_{\rm MA}^{-9/20}$	&	$t_{\rm gr} \propto M_{\rm MA}^{3/4}$	&	$M_{\rm MA} = 1.3 \times10^{14}$ g	\\
			&	MA-MA	& 2 &	$\rhoint_{\rm grav}\propto M_{\rm MA}^{2/5}$	&	$t_{\rm s}^{\rm Ne}$	&	$v_{\rm tur} \propto \St^{-1/2}$	&	$\Delta v_{\rm tur} \propto\St^{-1/2} \propto M_{\rm MA}^{-3/5}$	&	$t_{\rm gr} \propto M_{\rm MA}^{6/5}$	&	$M_{\rm MA} = 5.1 \times10^{14}$ g	\\
			\hline \hline 
		Stage 6
			&	CA-MA	&	-	& 	$\rhoint_{\rm grav}\propto M_{\rm MA}^{2/5}$	& $t_{\rm s}^{\rm Ne}$	&	$v_{\rm tur} \propto \St^{-1/2}$	&	$\Delta v_{\rm r} \propto\St$	&	-	&	$M_{\rm CA} = 4.5\times10^{14}$ g ($\chi=0.5$)	\\ 
			\hline \hline 
		Stage 7
			&	CA-CA	&	1	&	$\rhoint_{\rm grav}\propto M_{\rm CA}^{2/5}$	&	$t_{\rm s}^{\rm Ne}$	&	$v_{\rm tur} \propto \St^{-1/2}$&	$\Delta v_{\rm tur} \propto$ St$^{-1/2} $	&	$t_{\rm gr} \propto M_{\rm CA}^{-17/20}$	&	$M_{\rm CA} = 1.4\times10^{16}$ g ($\chi = 0.5$)	\\ 
			&	CA-CA	&	2	&	$\rhoint_{\rm grav}\propto M_{\rm CA}^{2/5}$	&	$t_{\rm s}^{\rm Ne}$	&	$v_{r}\propto \St$				&	$\Delta v_{\rm tur} \propto$ St$^{-1/2} $	&	$t_{\rm gr} \propto M_{\rm CA}^{-23/35}$	&	$M_{\rm CA} = 5.9\times10^{19}$ g ($\chi = 0.5$)	\\ 
			&	CA-CA	&	3	&	$\rhoint_{\rm grav}\propto M_{\rm CA}^{2/5}$	&	$t_{\rm s}^{\rm Ne}$	&	$v_{\phi}\propto \St^{0}$		&	$\Delta v_{\rm tur} \propto$ St$^{-1/2} $	&	$t_{\rm gr} \propto M_{\rm CA}^{-19/25}$	&	$M_{\rm CA} = 1.2\times10^{23}$ g ($\chi = 0.5$)	\\ 
			\hline \hline 
		\end{tabular}
		\end{center}
		\tablecomments{
			The columns of $v$ and $\Delta v$ show the dominant components in each collision.
			Since the actual value of $v$ is affected by the other components, the dependences of $\rhoint$ and $t_{\rm gr}$ are rough estimations and their behaviors in Figures \ref{Fig:m_rho_a_v12} and \ref{Fig:mCA_tgrow_tacc_v12} are somewhat different from the values in this table for certain cases.
			We put brackets when the corresponding quantities are not involved with the computation of the growth timescale.
			We put * when the corresponding collisions do not occur at the stage (see CA-MA collision at stage 1). 
			This is the reason why the transition mass of MAs (not CAs) is labeled for this case.
			When the internal density is fixed to constant as the aggregate grows up, we put (fix).
			The columns of internal density, $t_{\rm s}$, $v$ for CA-MA collisions are not filled out because CA-MA collisions do not change the regimes of CAs and MAs.
			The dependences of $\Delta v$ and $t_{\rm gr}$ for CA-MA collisions are derived from the values of $\rhoint$, $t_{\rm s}$ $v$ for CAs and MAs at the corresponding transition masses.
		}
	\end{table*}

We explain how CAs and MAs grow at each stage, focusing on their growth timescales (see Figure \ref{Fig:mCA_tgrow_tacc_v12}).
Since the growth timescale is determined by $\rhoint$, $\Delta v$, and $M$ (see Equations (\ref{eq:t_gr_MA-MA}), (\ref{eq:t_gr_CA-CA}), and (\ref{eq:t_gr_CA-MA})), we derive $\rhoint$ and $\Delta v$ as a function of $M$.
These dependences are summarized in Table \ref{table:collisions}.

\subsection{Stage 1} \label{sec:resu_stage1}

In stage 1, MAs form and grow via MA-MA collisions (see Figure \ref{Fig:m_rho_a_v12}). 
This occurs because $t_{\rm gr, MA-MA}$ becomes the shortest due to the larger number density of matrix grains than that of chondrules (see Figure \ref{Fig:mCA_tgrow_tacc_v12}).

We find that this stage can divide into two phases (see Table \ref{table:collisions}).
At phase 1, matrix grains first collide together to form MAs, and these MAs grow up to bigger ones by MA-MA collisions, subsequently.
Collisions between matrix grains and MAs end up with hit-and-stick, and hence the internal density of MAs is determined by $\rhoint_{\rm hit}$ (see Equation (\ref{eq:rho_hit})).
We apply $\rhoint_{\rm hit}$ for MA growth after the first matrix-matrix collisions occur.
This is why $\rhoint_{\rm MA}$ decreases more rapidly than $\rhoint_{\rm hit}\propto M^{-1/2}$ at the initial collisions.
As the mass of MAs increases, their internal densities decrease and their sizes enlarge (Figure \ref{Fig:m_rho_a_v12}).
Since the size of growing MAs is still small, their stopping time and relative velocity are given by $t_{\rm s}^{\rm Ep}$ and $\Delta v_{\rm B}$ (see Equations (\ref{eq:t_s_Ep}) and (\ref{eq:v_B})), respectively.
As a result, the MA-MA growth timescale is written as 
\begin{eqnarray}
	t_{\rm gr,MA-MA}
	&\propto & M_{\rm MA}^{1/3} \rhoint_{\rm MA}^{\,2/3} \Delta v_{\rm MA-MA}^{-1} 
	\ \propto M_{\rm MA}^{1/2}.
\end{eqnarray} 
Thus, $t_{\rm gr,MA-MA}$ becomes longer as MAs grow in mass.
Our results show that MAs grow up in this phase until $M_{\rm MA}<8.4\times10^{-10} \mbox{ g}$.

After $M_{\rm MA}$ exceeds $8.4\times10^{-10} \mbox{ g}$, the ram pressure becomes more important for determining the internal density of MAs (phase 2, see Equation (\ref{eq:rho_ram})).
The decreasing rate of the internal densities of MAs is smaller than the hit-and-stick regime, and then their internal densities start to increase as MAs grow up.
The relative velocity between MAs and the disk gas ($v_{\rm MA}$) also affects $t_{\rm gr,MA-MA}$ through $\Delta v_{\rm MA-MA}$ and $\rhoint_{\rm ram}$.
The dependence of $t_{\rm gr,MA-MA}$ on $M_{\rm MA}$ changes from $M_{\rm MA}^{1/2}$ to $M_{\rm MA}^{13/18}$.
This phase lasts until $M_{\rm MA}$ reaches $4.3\times10^{-7}$ g.
Stage 1 ends when $t_{\rm gr,MA-MA}$ exceeds $t_{\rm gr,CA-MA}$.
Thus, MAs with the mass of $4.3\times10^{-7}$ g, the internal density of $\rhoint_{\rm MA}=1.0\times10^{-5} \mbox{~g~cm}^{-3}$, and the radius of $a_{\rm MA}=0.22$~cm are formed from matrix grains with the initial mass of $2.0\times10^{-19}$ g in stage 1.

While CA-MA collisions do not occur at stage 1 in our simulations since $t_{\rm gr,CA-MA} > t_{\rm gr,MA-MA}$ (see Figure \ref{Fig:mCA_tgrow_tacc_v12}), we here explain how MA growth changes the timescale of CA-MA collisions for reference.
In stage 1, the CA-MA growth timescale is computed as $M_{\rm CA}t_{\rm col,CA-MA}/M_{\rm MA}$ since $M_{\rm CA}\gg M_{\rm MA}$.
The collisional cross-section $\sigma_{\rm CA-MA}$ is given by $\pi a_{\rm CA}^2$.
At the end, $t_{\rm gr,CA-MA}$ is determined mainly by $\Delta v_{\rm CA-MA}$.
At the early stage of MA growth, $\Delta v_{\rm CA-MA}$ is given by the Brownian motion, and hence $t_{\rm gr,CA-MA}$ depends on $M_{\rm MA}^{1/2}$.
This growth mode of $t_{\rm gr,CA-MA}$ on $M_{\rm MA}$ stops when the relative velocity is determined by $\Delta v_{\rm tur,CA-MA}$ (rather than $\Delta v_{\rm B,CA-MA}$).
This switching occurs at $M_{\rm MA}\simeq 2.0\times10^{-16}$ g. 
The turbulence-induced velocity is given by $\St$ (Equation (\ref{eq:v_tur})).
Considering $\Delta v_{\rm tur,CA-MA}$ is proportional to $\St$, the constant $\St_{\rm MA}$ under the hit-and-stick regime ($\rhoint_{\rm MA}=\rhoint_{\rm hit}$) give the constant $t_{\rm gr,CA-MA}$ (Figure \ref{Fig:mCA_tgrow_tacc_v12}).
Once $M_{\rm MA}$ becomes larger than $10^{-10}$ g (equivalently $a_{\rm MA}\gtrsim 0.1$ mm = 0.1 $a_{\rm ch}$), the collisional cross-section ($\sigma_{\rm CA-MA}$) increases as $M_{\rm MA}$ increases, and the CA-MA growth timescale becomes a decreasing function of $M_{\rm MA}$.

\subsection{Stage 2}\label{sec:resu_stage2}

In stage 2, both CA-MA and MA-MA collisions occur.
This arises because the values of $t_{\rm gr,MA-MA}$ and $t_{\rm gr,CA-MA}$ increase together and compete with each other.
At the beginning of this stage, the size of the MAs is about twice larger than that of chondrules (Figure \ref{Fig:m_rho_a_v12}).

The MA-MA growth timescale is initially given by $\rhoint_{\rm MA}$, $t_{\rm s}^{\rm Ep}$, $v_{\rm B}$, and $\Delta v_{\rm B}$ as in phase 2 of stage 1.
Following the increase of $M_{\rm MA}$, however, $v_{\rm B}$ decreases and $v_{\rm tur}$ becomes more effective.
The switching of the dominant velocity component occurs at $M_{\rm MA}\simeq 8.6\times 10^{-7}\mbox{ g}$.

CAs are formed through collisions between chondrules and MAs.
Chondrules become to be covered by matrix (see Figure \ref{Fig:collisions} (c)).
These collisions decrease the chondrule fraction in CAs ($\chi$) and hence decrease $\rhoint_{\rm CA}$ (see Equation (\ref{eq:rho_CA}) and Figure \ref{Fig:m_rho_a_v12}).
Given that the internal density of a MA is $\rhoint_{\rm MA}\sim10^{-5} \mbox{~g~cm}^{-3}$, $\rhoint_{\rm CA}$ decreases down to $1.3\times10^{-4} \mbox{~g~cm}^{-3}$.
In contrast, $M_{\rm CA}$ does not increase significantly, and $M_{\rm CA}$ becomes $1.9\times 10^{-2}$ g at the end of this stage.
Through such evolution of $\rhoint_{\rm CA}$ and $M_{\rm CA}$, $t_{\rm gr,CA-MA}$ increases.
This is because $\Delta v_{\rm tur,CA-MA}$ is proportional to $\St_{\rm CA}$ under $\St_{\rm CA}\gg\St_{\rm MA}$, and $\St_{\rm CA}$ decreases as $\rhoint_{\rm CA}$ decreases ($\St_{\rm CA}\propto t_{\rm s}^{\rm Ep}\propto M_{\rm CA}^{1/3} \rhoint_{\rm CA}^{2/3}$).

Competition between $t_{\rm gr,MA-MA}$ and $t_{\rm gr,CA-MA}$ 
is an outcome that $t_{\rm gr,CA-MA}$ becomes shorter via MA-MA collisions (see Section \ref{sec:resu_stage1}),
while $t_{\rm gr,CA-MA}$ becomes longer via CA-MA collisions.
Switching of the shortest timescale between CA-MA and MA-MA collisions is realized when the net change in $t_{\rm gr,CA-MA}$ is comparable with $t_{\rm gr,MA-MA}$ that increases due to MA-MA collisions.

The growth timescale between CA-CA collisions does not change through CA-MA collisions.
When $\Delta v_{\rm CA-CA}$ is given by the turbulence-induced velocity and $\St_{\rm CA}$ is given by the Epstein regime, 
\begin{eqnarray}
	t_{\rm gr,CA-CA} 
		\propto M_{\rm CA}^{1/3}\rhoint_{\rm CA}^{\,2/3} \Delta v_{\rm tur,CA-CA}^{-1}
		= {\rm const.},
\end{eqnarray}
and then $t_{\rm gr,CA-CA}$ becomes independent of $M_{\rm CA}$.

At the end of this stage, CAs have $\chi=0.67$ that is larger than the value of $\chi_{\rm clump}=0.5$.
This is the consequence of CA-MA collisions, which makes $\rho_{\rm d,MA}$ smaller as well.
Our results show that $\rho_{\rm d,MA}$ becomes less than half of $\rho_{\rm d,MA,init}$ (Equation (\ref{eq:rhod_MA})).
Through stage 2, $M_{\rm MA}$ and $\rhoint_{\rm MA}$ grow up to $6.9\times 10^{-6}$ g and $1.2\times 10^{-5} \mbox{~g~cm}^{-3}$, respectively (see Table \ref{table:collisions}).
The size of a CA becomes almost 10 times larger than that of a MA (Figure \ref{Fig:m_rho_a_v12}).
These mean that CAs are chondrules covered by fluffy matrix layers, which are thick and have a half of a chondrule mass.

\subsection{Stage 3}\label{sect:s3}

In stage 3, CAs grow due to CA-CA collisions (see Figure \ref{Fig:collisions} (d) and (e)).
We find that $M_{\rm CA}$ increases from $1.9\times 10^{-2}$ g to $1.7\times10^{11}$ g.
This growth becomes possible due to the inclusion of matrix grains in CAs which can dissipate the collision energy efficiently.
In this stage, the evolution of $\rhoint_{\rm CA}$ behaves similarly to that of $\rhoint_{\rm MA}$ due to the constant $\chi$ (Equation (\ref{eq:rho_CA})).
We confirm that $\rhoint_{\rm CA}$ is given by $\rhoint_{\rm ram}$ as for MA-MA collisions in stage 2.
Due to significant growth of $M_{\rm CA}$, $\St_{\rm CA}$ increases and the law of $\St_{\rm CA}$ is changed in this stage.
Such change leads to a different dependence of $t_{\rm gr,CA-CA}$ on $M_{\rm CA}$ through $v_{\rm CA}$ and $\Delta v_{\rm CA-CA}$.
We explain the dependence of $t_{\rm gr,CA-CA}$ and $\rhoint_{\rm ram}$ using five phases below (see Table \ref{table:collisions}).

At phase 1, CAs obey the Epstein's law and $\St_{\rm CA}<t_{\eta}\Omega_{\rm K}$.
As shown in stage 2, $t_{\rm gr,CA-CA}$ does not depend on $M_{\rm CA}$ in this phase.
At phase 2, CAs are under the Stokes' law which begins at $M_{\rm CA}\simeq 20$ g.
Then, the CA-CA growth timescale is 
\begin{eqnarray}
	t_{\rm gr,CA-CA}
	\propto M_{\rm CA}^{1/3}\rhoint_{\rm CA}^{2/3} \St_{\rm CA}^{-1} 
	\propto M_{\rm CA}^{-2/7}, 
\end{eqnarray}
where $\rhoint_{\rm CA}$ is given by $\rhoint_{\rm ram}$
	$(\propto M_{\rm CA}^{1/7} v_{\rm CA}^{3/7} \St^{-3/7}
	\propto M_{\rm CA}^{1/7} 
	$),
when the turbulent induced velocity is proportional to $\St_{\rm CA}$ ($\St_{\rm CA}<t_{\eta}\Omega_{\rm K}$).

Phase 3 begins when the regime of the turbulence-induced velocity is changed ($t_{\eta}\Omega_{\rm K}< \St_{\rm CA}$, which corresponds to $2.5\times 10^{3} \mbox{ g}<M_{\rm CA}$ in our fiducial case).
The dependence of $v_{\rm CA}$ on $\St_{\rm CA}$ changes to $\St_{\rm CA}^{\ 1/2}$ (Equation (\ref{eq:v_tur})), while $\St_{\rm CA}$ is still in the Stokes' law.
Under this dependence, $\rhoint_{\rm ram}$ becomes independent of $M_{\rm CA}$ and the growth of $\rhoint_{\rm CA}$ stalls.
The growth timescale decreases as $t_{\rm gr,CA-CA}$ is proportional to $M_{\rm CA}^{-1/3}$.
Thus, CA-CA collisions accelerate as CAs grow.
Phase 3 lasts for $\St_{\rm CA}<9.5\times 10^{-2}$ ($M_{\rm CA}<2.6\times 10^{6} \mbox{ g}$).

At phase 4, CAs obey the Allen's law. 
The relative velocity between CAs and gas continues to be given by $v_{\rm tur}\propto \St^{1/2}$. 
The CA-CA growth timescale eventually becomes an increasing function of $M_{\rm CA}$,
\begin{eqnarray}
	t_{\rm gr,CA-CA} 
	\propto M_{\rm CA}^{12/91}.
\end{eqnarray}
We find that the minimum value of $t_{\rm gr,CA-CA}$ is obtained at the transition from phase 3 to phase 4 (see Figure \ref{Fig:mCA_tgrow_tacc_v12}).

After $\St_{\rm CA}$ exceeds unity, phase 5 begins.
This is the last phase in Stage 3 and corresponds to $M_{\rm CA}>1.6\times10^8$ g.
The dependence of $\Delta v_{\rm tur,CA-CA}$ on $\St_{\rm CA}$ is changed to $(1+\St_{\rm CA})^{-1/2}$.
Substituting $\Delta v_{\rm CA-CA}$ and $\St\propto t_{\rm s}^{\rm Al}$, $\rhoint_{\rm CA}$ becomes a decreasing function of $M_{\rm CA}$.
The power-law index of $\rhoint_{\rm CA}$ is fixed at zero to avoid unphysical expansion of CA volume, and then $t_{\rm gr,CA-CA}$ can be given as a proportional to $M_{\rm CA}^{2/3}$.
This stage lasts until CAs become $M_{\rm CA}=1.7\times10^{11}$~g, $\rhoint_{\rm CA}=9.6\times10^{-4} \mbox{~g~cm}^{-3}$, and $a_{\rm CA}=3.5\times10^4$~cm.

\subsection{Stage 4}

In stage 4, the growth of MAs comes back (see the top panel of Figure \ref{Fig:mCA_tgrow_tacc_v12}).
In this stage, MAs grow similarly to the CA growth in stage 3 (see Table \ref{table:collisions}).
The mass and internal density of MAs become $M_{\rm MA}=5.0\times 10^{11} \mbox{ g}$ and $\rhoint_{\rm MA}=2.6\times10^{-4}\mbox{~g~cm}^{-3}$, respectively, at the end of this stage.
These values are comparable to those of $M_{\rm CA}$ and $\rhoint_{\rm CA}$ at the end of stage 3.

We find that the values of $M_{\rm MA}$ and $a_{\rm MA}$ are slightly larger than those of $M_{\rm CA}$ and $a_{\rm CA}$ at the end of stage 4 (Figure \ref{Fig:m_rho_a_v12}).
Stage 4 ends when the MA-MA growth timescale exceeds $t_{\rm gr,CA-CA}$.
Substituting $\Delta v$ and $\St$, the ratio of the growth timescales can be written as the function of the mass ratio ($M_{\rm CA}/M_{\rm MA}$), internal density ratio ($\rhoint_{\rm CA}/\rhoint_{\rm MA}$) and mass density ratio ($\rho_{\rm d,ch}/\rho_{\rm d,MA}$) of aggregates.
These density ratios are affected by $\chi_{\rm clump}$.
Then, the mass ratio at the end of stage 4 is also affected by $\chi_{\rm clump}$.

\subsection{Stage 5}

We find that CA-CA and MA-MA collisions occur with keeping $t_{\rm gr,CA-CA}\simeq t_{\rm gr,MA-MA}\lesssim t_{\rm gr,CA-MA}$ in stage 5.
The compression regime changes from the ram pressure to the self-gravity ($\rhoint_{\rm grav}$).
The internal densities of CAs and MAs increase more rapidly than those in previous stages (Figure \ref{Fig:m_rho_a_v12}).
Then, the mass relationship of $M_{\rm CA}<M_{\rm MA}$ is also kept in this stage.
The growth timescales of both CA-CA and MA-MA collisions are proportional to $M^{3/4}$ (see Table \ref{table:collisions}).
The stopping time of MAs begins to obey the Newton's law before that of CAs does so.
After $\St_{\rm MA}$ obeys the Newton's law, the dependence of $t_{\rm gr,MA-MA}$ on $M_{\rm MA}$ becomes steeper.
This leads to $t_{\rm gr,MA-MA}$ that is longer than $t_{\rm gr,CA-CA}$ and $t_{\rm gr,CA-MA}$.

\subsection{Stage 6}\label{sect:s6}

In stage 6, CAs collide with MAs.
Before this stage, CAs are composed of the assemblage of chondrules with matrix layers, and CAs and MAs have comparable sizes ($\sim1$~km, Figure \ref{Fig:m_rho_a_v12}).
The CA-MA collisions in this stage make a partial matrix excess in the internal distribution of CAs (Figure \ref{Fig:collisions} (f)).
CA-MA collisions in this stage occur because $\Delta v_{\rm CA-MA}$ becomes high.
The Stokes numbers of CAs and MAs becomes $\sim \rho_{\rm d}/\rho_{\rm g}$, and $\Delta v_{r}$ and $\Delta v_{\phi}$ are effective (Equations (\ref{eq:v_r}) and (\ref{eq:v_phi})).
The CA-MA growth timescale becomes the shortest. 
All matrix grains are accreted by CAs in this stage ($\chi=\chi_{\rm clump}=0.5$).

\subsection{Stage 7}\label{sec:res_stage7_v12}

In stage 7, CA-CA runaway collisions occur.
The CA-CA collisions become runaway collisions when the Safronov parameter ($\Theta_{\rm CA-CA}$) exceeds unity.
The CA-CA runaway collisions begins at $M_{\rm CA}=4.5\times10^{14} \mbox{ g}$ (see Table \ref{table:collisions}).
The growth timescale by CA-CA collisions takes the longest value in the CA growth when the runaway growth begins, and then $t_{\rm gr,CA-CA}$ becomes shorter due to the increase of $\Theta_{\rm CA-CA}$ (see Figure \ref{Fig:mCA_tgrow_tacc_v12}).

There are three phases in stage 7.
These phases are divided by the dominant component of $v_{\rm CA}$.
As $\St_{\rm CA}$ increases, the dominant component of $v_{\rm CA}$ changes from the turbulence-induced velocity ($v_{\rm tur}$, phase 1) to $v_r$ (phase 2) and $v_{\phi}$ (phase 3).
While $\Delta v_{\rm CA-CA}$ is given by the turbulence-induced velocity at any phases, the dominant component of $v_{\rm CA}$ affects $t_{\rm gr,CA-CA}$ through $\St_{\rm CA}$, which is in the Newton's regime (Equation (\ref{eq:t_s_Ne})).
The growth timescale becomes 
\begin{eqnarray}
	t_{\rm gr,CA-CA}
		&\propto& M_{\rm CA}^{1/3} \rhoint_{\rm CA}^{\,2/3} \Delta v_{\rm CA-CA}^{-1} (\Theta_{\rm CA-CA})^{-1} \nonumber \\
		&\propto& M^{-1/3} \rhoint^{1/3} \Delta v 
		\propto M^{-1/5} \Delta v.
		\label{eq:t_gr_runaway}
\end{eqnarray}

The value of $t_{\rm gr,CA-CA}$ takes the longest in the growth of CA, and it is shorter than the timescales of gravitational collapse.
CAs grow up by their collisions before their collapse occurs.

\section{Mass evolution of aggregates in the fiducial case of the large eddy model} 
\label{sec:results_evolution_v1}

\begin{table*}
	\begin{center}
	\tablenum{4}
	\caption{Summary of the dominant collisions at each stage for our fiducial case in the large eddy model.}
	\label{table:collisions_v1}
	\setlength{\leftskip}{-.3in}
	\renewcommand{\arraystretch}{1.2}
	\scriptsize
	\begin{tabular}{c|c|c|c|c|c|c|c|c} 
	\hline \hline
		&	Collisions	&	Phase	&	Internal Density	&	$t_{\rm s}$	&	$v$	&	$\Delta v$	&	$t_{\rm gr}$	&	Transition mass	\\ \hline \hline
		%
	Stage 3 
		&	CA-CA	& 1 &	$\rhoint_{\rm ram}\propto M_{\rm CA}^{1/7}$		&	$t_{\rm s}^{\rm Ep}$	&	$v_{\rm tur} \propto \St$		&	$\Delta v_{\rm tur} \propto\St \propto M_{\rm CA}^{3/7}$		&	$t_{\rm gr} \propto M_{\rm CA}^{0}$ 	&	$M_{\rm CA} = 20$ g ($\chi = 0.66$)\\
		&			& 2	&	$\rhoint_{\rm ram}\propto M_{\rm CA}^{1/7}$		&	$t_{\rm s}^{\rm St}$	&	$v_{\rm tur} \propto \St$		&	$\Delta v_{\rm tur} \propto\St \propto M_{\rm CA}^{5/7}$		&	$t_{\rm gr} \propto M_{\rm CA}^{-2/7}$	&	$M_{\rm CA} = 3.1 \times10^{2}$ g \\
		\hline \hline 
	Stage 4
		&	CA-CA	& 1	&	$\rhoint_{\rm ram}\propto M_{\rm CA}^{0}$		&	$t_{\rm s}^{\rm St}$	&	$v_{\rm tur} \propto \St^{1/2}$	&	$\Delta v_{\rm tur} \propto\St^{1/2}\propto M_{\rm CA}^{1/3}$	&	$t_{\rm gr} \propto M_{\rm CA}^{0}$		&	$M_{\rm CA} = 1.5\times10^6$ g	\\
		&			& 2	&	$\rhoint_{\rm ram}\propto M_{\rm CA}^{4/91}$	&	$t_{\rm s}^{\rm Al}$	&	$v_{\rm tur} \propto \St^{1/2}$	&	$\Delta v_{\rm tur} \propto\St^{1/2}\propto M_{\rm CA}^{3/13}$	&	$t_{\rm gr} \propto M_{\rm CA}^{12/91}$	&	$M_{\rm CA} = 2.0 \times10^{8}$ g ($\chi = 0.53$)	\\
		\hline
		&	CA-MA	& 1 & 	-	& -	&	-	&	$\Delta v_{\rm tur} \propto \St^{1/2}$	&	$t_{\rm gr} \propto M_{\rm CA}^{0}$	&	$M_{\rm CA} = 1.5\times10^6$ g	\\
		&			& 2 & 	-	& -	&	-	&	$\Delta v_{\rm tur} \propto \St^{1/2}$	&	$t_{\rm gr} \propto M_{\rm CA}^{12/91}$	&	$M_{\rm CA} = 2.0 \times10^{8}$ g	\\
		\hline \hline 
	Stage 5
		&	CA-CA	& 1	&	$\rhoint_{\rm ram}\propto M_{\rm CA}^0$ (fix)	&	$t_{\rm s}^{\rm St}$	&	$v_{\rm r} \propto \St$		&	$\Delta v_{\rm B} \propto M_{\rm CA}^{-1/2}$	&	$t_{\rm gr} \propto M_{\rm CA}^{-6/5}$	&	$M_{\rm CA} = 2.1 \times10^{11}$ g ($\chi = 0.53$)	\\
		&			& 2	&	$\rhoint_{\rm grav}\propto M_{\rm CA}^{2/5}$	&	$t_{\rm s}^{\rm St}$	&	$v_{\rm r} \propto \St$		&	$\Delta v_{\rm B} \propto M_{\rm CA}^{-1/2}$	&	$t_{\rm gr} \propto M_{\rm CA}^{-7/10}$	&	$M_{\rm CA} = 8.4 \times10^{11}$ g ($\chi = 0.53$)	\\
		&			& 3	&	$\rhoint_{\rm grav}\propto M_{\rm CA}^{2/5}$	&	$t_{\rm s}^{\rm Al}$	&	$v_{\rm r} \propto \St$		&	$\Delta v_{\rm B} \propto M_{\rm CA}^{-1/2}$	&	$t_{\rm gr} \propto M_{\rm CA}^{-7/10}$	&	$M_{\rm CA} = 6.9 \times10^{15}$ g ($\chi = 0.53$)	\\
		&			& 4	&	$\rhoint_{\rm grav}\propto M_{\rm CA}^{2/5}$	&	$t_{\rm s}^{\rm Ne}$	&	$v_{\rm r} \propto \St$		&	$\Delta v_{\rm B} \propto M_{\rm CA}^{-1/2}$	&	$t_{\rm gr} \propto M_{\rm CA}^{-7/10}$	&	$M_{\rm CA} = 1.1 \times10^{20}$ g ($\chi = 0.53$)	\\
		&			& 5	&	$\rhoint_{\rm grav}\propto M_{\rm CA}^{2/5}$	&	$t_{\rm s}^{\rm Ne}$	&	$v_{\phi} \propto \St^{0}$	&	$\Delta v_{\rm B} \propto M_{\rm CA}^{-1/2}$	&	$t_{\rm gr} \propto M_{\rm CA}^{-7/10}$	&	$M_{\rm CA} = 1.2 \times10^{23}$ g ($\chi = 0.53$)	\\
		\hline \hline 
	\end{tabular}
	\end{center}
\end{table*}

The typical evolution of aggregates for our fiducial case in the large eddy model is shown in this section.
The evolution of aggregates is the same as that in the whole eddy model until phase 1 of stage 3 (see Figures \ref{Fig:m_rho_a_v12} and \ref{Fig:mCA_tgrow_tacc_v12}).
We, therefore, focus on stages 4 and 5.

\subsection{Stage 4} 

The evolution of aggregates becomes different from that in the whole eddy model when $\St_{\rm CA}$ exceeds $t_{\rm \eta}\Omega_{\rm K}$, which corresponds to $M_{\rm CA}=3.1\times10^2$~g and $a_{\rm CA}=$48~cm.
In the large eddy model, $v_{\rm tur,CA}$ and $\Delta v_{\rm tur,CA-CA}$ become smaller than those in the whole eddy model.
As a result, $\rhoint_{\rm CA}$ keeps smaller value (Figure \ref{Fig:m_rho_a_v1}).
The CA-CA growth timescale in the large eddy model is longer than that in the whole eddy model, and $t_{\rm gr,CA-CA}$ is comparable to $t_{\rm gr,CA-MA}$ (Figure \ref{Fig:mCA_tgrow_tacc_v1}).
Both CA-CA and CA-MA collisions occur in stage 4.

CA-CA and CA-MA collisions produce a jagged structure on the evolution of $\rhoint_{\rm CA}$ (Figure \ref{Fig:m_rho_a_v1}).
CAs grow up by CA-CA collisions with keeping $\rhoint_{\rm CA}$ constant (Table \ref{table:collisions_v1}).
For CA-MA collisions, $\rhoint_{\rm CA}$ becomes smaller with keeping $M_{\rm CA}$ value almost constant since $M_{\rm CA}\gg M_{\rm MA}$.
This CA-MA growth occurs before the growth of MAs via MA-MA collisions.
The mass and size of MAs do not grow up from stage 2, $M_{\rm MA}=6.9\times10^{-6}$~g and $a_{\rm MA}=$0.51~cm.
The small MAs are continuously accreted by the large CAs ($\geq48$~cm) between CA-CA collisions.
These CA-MA collisions also make a partial matrix excess in the internal distribution of CAs (Figure \ref{Fig:collisions} (g)).
However, this partial matrix excess would result in thin matrix layers.
We find that the value of $\chi$ decreases from 0.66 to 0.53 in this stage.
The spatial density of MAs in the clump becomes $1.1\times10^{-5}~\mbox{g cm}^{-3}$, which is about 10\% of the initial value.
This stage lasts until $\St_{\rm CA}$ exceeds 1.

\subsection{Stage 5} \label{sec:res_stage5_v1}

After $\St_{\rm CA}$ exceeds unity, which corresponds to $M_{\rm CA}=2.0\times 10^8$ g, $v_{\rm CA}$ and $\Delta v_{\rm CA-CA}$ become extremely small since CAs are no longer affected by turbulence (Table \ref{table:collisions_v1}).
Such small $\Delta v_{\rm CA-CA}$ makes $\Theta_{\rm CA-CA}>1$, and the runaway CA-CA accretion begins.
After the runaway collisions begins, $t_{\rm gr,CA-CA}$ becomes proportional to $\Delta v_{\rm CA-CA}$ (Equation (\ref{eq:t_gr_runaway})).
This is why $t_{\rm gr,CA-CA}$ also becomes extremely small and CA-CA collisions are the dominant process in the large eddy model.

This stage holds 5 phases, which is divided by the regime of the internal density ($\rhoint_{\rm CA}$), stopping time ($t_{\rm s}$), and relative velocity ($v_{\rm CA}$).
At phase 1, $\rhoint_{\rm CA}$ is fixed to $\rhoint_{\rm CA}(\St_{\rm CA}=1)$ to avoid unphysical expansion of CAs.
This expansion is because $\rhoint_{\rm CA}$ is given by $\rhoint_{\rm ram}$, which depends on $v_{\rm CA}$ (Equation (\ref{eq:rho_ram})).
Except for the initial expansion, CAs are compressed by the ram and self-gravitational pressure (Table \ref{table:collisions_v1}).

\end{document}